\RequirePackage{fix-cm}
\documentclass[smallextended]{svjour3}       
\smartqed  
\usepackage[numbers,sort&compress]{natbib}
\usepackage{textcomp}
\usepackage{mathtools}
\usepackage{amsmath,amssymb,amsfonts}
\usepackage{graphicx}
\usepackage{textcomp}
\usepackage{xcolor}
\usepackage{soul}
\usepackage{subfigure}
\usepackage{tikz}
\usepackage{tabu}
\usepackage{threeparttable}
\usepackage{balance}
\usepackage{colortbl}
\usepackage{amsmath}
\usepackage{amssymb}
\usepackage{multirow}
\usepackage{adjustbox}
\usepackage{xcolor}

\definecolor{codegreen}{rgb}{0,0.6,0}
\definecolor{codegray}{rgb}{0.5,0.5,0.5}
\definecolor{codepurple}{rgb}{0.58,0,0.82}
\definecolor{backcolour}{rgb}{0.95,0.95,0.92}

\usepackage{listings}
\lstdefinestyle{mystyle}{
    float=tp,
    abovecaptionskip=-5pt,
    backgroundcolor=\color{backcolour},   
    commentstyle=\color{codegreen},
    keywordstyle=\color{magenta},
    numberstyle=\tiny\color{codegray},
    stringstyle=\color{codepurple},
    basicstyle=\ttfamily\footnotesize,
    breakatwhitespace=false,         
    breaklines=true,                 
    captionpos=b,                    
    keepspaces=true,                 
    numbers=left,                    
    numbersep=5pt,                  
    showspaces=false,                
    showstringspaces=false,
    showtabs=false,                  
    tabsize=2
}

\usepackage[utf8]{inputenc}
\usepackage{float}
\usepackage{amsmath}
\usepackage[english]{babel}
\usepackage[edges]{forest}

\usepackage{float}
\usepackage{chngpage}
\usepackage{tcolorbox}
\usepackage{ulem} 
\usepackage[ruled,vlined,linesnumbered]{algorithm2e}
\usepackage[hyphens]{url}
\usepackage{hyperref}
\usepackage{graphics}
\SetKwInOut{Input}{Input}
\newcommand*\circled[1]{\tikz[baseline=(char.base)]{
            \node[shape=circle,draw=red!60,inner sep=2pt] (char) {#1};}}
\newtheorem{defn}[theorem]{Definition}

\newcommand{\basicalert}[2]{\fbox{\bfseries\sffamily\scriptsize\color{blue} #1}{\sf\small$\blacktriangleright$\textit{\color{red} #2}$\blacktriangleleft$} }
\newcommand{\app}{$^\dag\,$}
\newcommand{\Foutse}[1]{\basicalert{From Foutse}{#1}}

\newcommand{\Amin}[1]{\textcolor{orange}{{\it [Amin: #1]}}}

\newcommand{\Paulina}[1]{\textcolor{blue}{{\it [Paulina: #1]}}}

\newcommand\myeq{\stackrel{\mathclap{\normalfont\mbox{def}}}{=}}
\usepackage[figuresleft]{rotating}
\journalname{Empirical Software Engineering}
\begin{document}

\title{Harnessing Pre-trained Generalist Agents for Software Engineering Tasks}


\author{Paulina Stevia Nouwou Mindom \and Amin Nikanjam \and Foutse Khomh} 

\authorrunning{Nouwou Mindom et al.} 

\institute{Paulina Stevia Nouwou Mindom \and Amin Nikanjam \and Foutse Khomh \at  
               Polytechnique Montréal, Québec, Canada \\
              \email{\{paulina-stevia.nouwou-mindom, amin.nikanjam, foutse.khomh\}@polymtl.ca} 
}

\date{Received: date / Accepted: date}

\maketitle
\begin{abstract}
Nowadays, we are witnessing an increasing adoption of Artificial Intelligence (AI) to develop techniques aimed at improving the reliability, effectiveness, and overall quality of software systems. 
Deep reinforcement learning (DRL) has recently been successfully used for automation in complex tasks such as game testing and solving the job-shop scheduling problem, as well as learning efficient and cost-effective behaviors in various environments. However, these specialized DRL agents, trained from scratch on specific tasks, suffer from a lack of generalizability to other tasks and they need substantial time to be developed and re-trained effectively. Recently, DRL researchers have begun to develop generalist agents, able to learn a policy from various environments (often Atari game environments) and capable of achieving performances similar to or better than specialist agents in new tasks. In the Natural Language Processing or Computer Vision domain, these generalist agents are showing promising adaptation capabilities to never-before-seen tasks after a light fine-tuning phase and achieving high performance.
To the best of our knowledge, no study has investigated the applicability of 
these generalist agents to SE tasks. 
This paper investigates the potential of generalist agents for solving SE tasks. Specifically, we conduct 
an empirical study aimed at assessing the performance of two generalist agents 
on two important SE tasks: the detection of bugs in games (for two games) and the minimization of makespan in a scheduling task, to solve the job-shop scheduling problem (for two instances). 
Our results show that the generalist agents outperform the specialist agents with very little effort for fine-tuning, achieving a 20\% reduction of the makespan over specialized agent performance on task-based scheduling. In the context of game testing, some generalist agent configurations detect 85\% more bugs than the specialist agents. Building on our analysis, we provide recommendations for researchers and practitioners looking to select generalist agents for SE tasks, to ensure that they perform effectively.
\end{abstract}

\section{Introduction}\label{Introduction}
Researchers and practitioners invest a significant amount of time in developing techniques aimed at enhancing the reliability, effectiveness, and overall quality of software systems. Techniques such as coverage-based testing \cite{zhu1997software}, and search-based testing  \cite{harman2015achievements} have been proposed to ensure that software products behave as expected. Munro et al.\cite{munro2005product}, Singh et al. \cite{singh2011effectiveness} studied how to detect bad smells in code-based projects. Gao et al. \cite{gao2007hybrid}, Kaur et al. \cite{kaur2012efficient} proposed task scheduling algorithms to minimize time loss and maximize performance in cloud computing systems. Recently, DRL has been increasingly leveraged in Software Engineering (SE) tasks, thanks to the availability of DRL algorithms such as Proximal Policy Optimization (PPO) \cite{schulman2017proximal}, Advantage Actor Critic (A2C) \cite{mnih2016asynchronous}, Deep Q-Networks (DQN) \cite{mnih2013playing} that can train agents to accomplish the task at hand. For example, Bagherzadeh et al. \cite{bagherzadeh2021reinforcement}, trained DRL agents to prioritize test cases, Zheng et al. \cite{zheng2019wuji}, Tufano et al. \cite{tufano2022using} trained DRL agents to detect bugs in games, and Ahmadi et al. \cite{ahmadi2022dqn} leveraged DRL for automatic refactorings in code-based projects.

Specialist DRL agents (i.e., agents trained from scratch on a specific task) have shown promising performance in SE tasks, but that often comes with a long training time due to the large amount of training data \cite{bagherzadeh2021reinforcement, zhang2020learning}. Moreover, researchers have observed that specialized DRL agents trained to excel at specific tasks often struggle to generalize to never-before-seen tasks \cite{wang2019generalization,lazaric2012transfer}, which can leave them unable to solve real-world tasks, as these tasks are likely to evolve constantly because of changing environments \cite{zhu2023transfer}. Developing a generalist agent trained on multiple tasks that can solve never-before-seen tasks is a goal for the AI community \cite{mccarthy2006proposal}. In other domains, like NLP \cite{devlin2018bert} and Computer Vision \cite{arnab2021vivit}, researchers have built generalist models that can adapt to never-before-seen tasks and achieve high performance, by training them on large task-agnostic datasets with a light fine-tuning step. In the field of DRL, researchers such as Espeholt et al. \cite{espeholt2018impala}, Lee et al. \cite{lee2022multi} proposed generalist agents that are trained on the Atari suite environments \cite{mnih2013playing}, and able to perform almost as good as specialist agents on never-before-seen tasks with affordable effort for fine-tuning. The high performance of these generalist agents is due to their scalable model architecture, the continuous improvements of their learning policy in high-performance computing infrastructure, and the diversity of the training data \cite{lee2022multi}.

While generalist agents, with prior knowledge/skills, can facilitate efficient exploration \cite{argall2009survey} on the task at hand, transfer prior skills with their rapid adaptation capability \cite{lee2022multi}, and enable efficient use of resources in never-before-seen tasks, given the broad usage of DRL in SE, whether or not such generalist agents can be beneficial is unclear. Hence, in this paper, we examine the applicability of generalist agents for SE tasks. Specifically, we fine-tune pre-trained generalist agents on different data budgets (i.e., size of the data used to fine-tune the generalist agents) to automate bug detection in games and minimize the makespan of scheduling operations (also called the Job-shop scheduling problem \cite{zhang2020learning}) via a heuristic approach called the priority dispatching rule. Automatic bug detection in games is essential and has been the focus of research due to the frequent changes that can occur during the game development process \cite{Santos18}. Pfau et al. \cite{pfau2017automated} studied the detection of crashes and blocker bugs in games while the agent is playing. Minimizing makespan in scheduling is essential in many areas such as cloud computing, to foster systems that complete tasks as quickly as possible against a rising demand for cloud resources. Guo et al. \cite{guo2018energy} and Tong et al. \cite{tong2020scheduling} proposed a scheduling algorithm that leverages DRL to minimize makespan in cloud systems. Hence, we believe the study of these tasks can be beneficial for the SE community.

In this paper, we leverage generalist agents, namely a Multi-Game Decision Transformer (MGDT) \cite{lee2022multi} and a Scalable Distributed Deep-RL with Importance Weighted Actor-Learner Architectures (IMPALA) \cite{espeholt2018impala}. We employ the pre-trained models of these generalist agents and empirically compare their performance against the performance of specialist agents \cite{zheng2019wuji,tufano2022using,zhang2020learning} on two SE tasks: the detection of bugs in games and the minimization of makespan of scheduling operations. We investigate which pre-trained generalist agents (or which of its different configurations) with regards to the fine-tuning budget allowed, perform better than or are close to the specialist agents' performance. Results show that MGDT performs close to a specialist agent with a few fine-tuning steps. IMPALA agent outperforms the specialist agent in minimizing the makespan of scheduling operations. 
Our results also show that prior knowledge, the fine-tuning budget, and the scalable architecture of generalist agents positively impact their performance on the studied SE tasks. To summarize, our work makes the following contributions:
\begin{itemize}
    \item We propose the first study that leverages generalist agents for SE tasks.
    \item To evaluate the usefulness of generalist agents on SE tasks, we utilized two pre-trained generalist agents: MGDT and IMPALA. We carefully fine-tuned them (on zero-shot, 1\%, and 2\% data budgets) on the Blockmaze and MsPacman games for bug detection and collected the number of bugs, the cumulative reward, and the average training and testing times using model-free DRL algorithms. Similarly, we fine-tuned them on a scheduling-based task and collected the makespan, the cumulative reward, and the average training and testing times. We have evaluated a total of 45 configurations and some of them perform close to or better than the baseline specialist agents.
    \item We provide recommendations for researchers looking to leverage generalist agents on SE tasks, as we found that generalist agents offer good transferability performance at low fine-tuning costs. 
\end{itemize}

The rest of this paper is organized as follows. In Section \ref{Background}, we review the necessary background knowledge on the bug detection task, the scheduling task, and DRL. The methodology followed in our study is described in Section \ref{studydesign}. We discuss the obtained results in Section \ref{sec:Experimental results}. Some recommendations for future work are presented in Section \ref{Recommendations about generalist agents selection}. We review related work in Section \ref{Related work}. Threats to the validity of our study are discussed in Section \ref{threats}. Finally, we conclude the paper and highlight some avenues for future works in Section \ref{Conclusion}.

\section{Background}\label{Background}
In this section, we first introduce DRL and then describe the terms and notations used to define the bug detection task and the job-shop scheduling problem. 
\subsection{Deep Reinforcement Learning} \label{sec:Reinforcement Learning}
The interaction between a Reinforcement Learning (RL) agent and its environment can be modeled as a Markov decision process $(\mathcal{S, A, P,\gamma})$ with the following components:

\textbf{State of the environment:}  A state $ s \in \mathcal{S} = \mathbb{R}^n $ represents the perception that the agent has of the environment. 

\textbf{Action:} Based on the state of the environment, the agent chooses among available actions in $\mathcal{A}$ to be executed in the environment and move to the next state.

\textbf{State transition distribution:} $\mathcal{P}(s_{t+1},r_{t}|s_{t},a_{t})$ defines the probability of moving to the next state $s_{t+1}$ given $a_{t}$. When performing an action $a_{t}$ in state $s_{t}$, the agent also receives a reward as $r_{t}$. The goal of the agent is to maximize the expected rewards discounted by $\gamma$. 

\textbf{Episode:} An episode is a sequence of environment states, actions performed by an agent and rewards that ends as the agent reaches a terminal state or a maximum number of steps.

\textbf{Policy.} A policy $\pi$ is defined as a function $\pi:S \rightarrow A$ mapping each state $s \in S$ to an action $a \in A$ performs by an agent. The policy defines the agent's decision in each state of the underlined environment. It can be either a strategy developed by a human expert or one drawn from experience.

DRL algorithms use Deep Neural Networks (DNNs), in an RL problem, to estimate the value function or model (state transition function and reward function) to estimate the policy and therefore tend to offer a more manageable solution space in large, and complex environments. DRL algorithms can be classified based on the following properties:

\textbf{Model-based and Model-free.} In model-based DRL, the agent knows in advance the outcome of the environment to possible actions and the rewards it will be getting by taking those actions. During training, the goal is to learn the optimal policy by running the selected action based on the current state, observing the next state, and the immediate reward. Model-based techniques can sometimes be referred to as offline DRL methods. In model-free DRL, the agent learns the optimal behavior by interacting directly with the environment, which includes taking actions within it. Model-free methods can sometimes be referred to as online DRL methods. In this work, we are interested in model-free DRL as the location of faults in real games is unknown beforehand as well as the processing time of a job in the context of job shop scheduling.

\textbf{Value-based, policy-based, and actor-critic learning.} Value-based methods estimate at every state the Q-value and select the action that has the best Q-value. A Q-value indicates how good an action is, given the current state of the environment. Policy-based methods parameterize an initial policy, and then that policy is updated during training using gradient-based/gradient-free optimization techniques. Actor-critic methods combined value-based and policy-based techniques: the policy function (actor) selects the action and the value function (critic) estimates the Q-values based on the action selected by the policy. In this work, we fine-tune our generalist agents with DQN \cite{mnih2013playing} and PPO \cite{schulman2017proximal} which are respectively value-based and actor-critic methods.

\textbf{On-policy vs Off-policy.} With On-policy methods the policy used to generate trajectories during training is the same used to evaluate and improve the target policy and take future actions. With Off-policy methods the policy (called behavior policy) used to generate trajectories during training is different from the target policy. DQN and PPO are respectively Off-policy and On-policy methods.

\textbf{Decision transformers.} 
In this work, one of our studied generalist agents follows an architecture similar to a Decision Transformer (DT). A DT is an architecture that tackles DRL as a sequential modeling problem \cite{chen2021decision}. It leverages transformer \cite{vaswani2017attention} architecture to predict future actions. Specifically, a trajectory is represented as a sequence of states, actions and return-to-go: \[ T= g_{1},s_{1},a_{1} ... g_{\mid T \mid},s_{\mid T \mid},a_{\mid T \mid} \]  A DT learns a deterministic policy: 
\[ \pi (a_t \mid s_{-K,t},g_{-K,t})\] where \textit{K} is the context length of the transformer, and $s_{-K,t}$ denotes the sequence of the last K states, which is the case for $g_{-K,t}$ as well. DT policy is parameterized through a Generative Pre-trained Transformer (GPT) \cite{radford2018improving} architecture to predict the next action sequence. 

\subsection{Bugs detection in a game} \label{sec:Bugs detection in a game}
\begin{figure}
    \centering
    \includegraphics[width=0.8\textwidth]{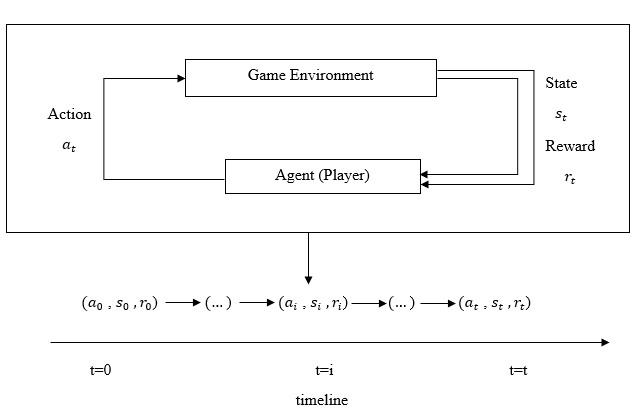}
    \caption{The interaction between an agent and a game environment \cite{nouwou2023comparison}.}
    \label{fig:GameEnvMDP}
\end{figure}

The process of detecting bugs in a game is an essential activity before its release. Given the complexity of the activity, researchers have investigated various ways to automate it \cite{tufano2022using,zheng2019wuji}. In the following, we introduce some concepts that are important for understanding the automatic detection of bugs in games. 
\begin{defn}
\label{gamedefinition}: \textbf{Game.} A game $G$ is defined as a function $G: A^n \rightarrow (S \times R)^n$, where $A$ is the set of actions that can be performed in the game by the agent playing it, $S$ is the set of states of the game, $R$ represents the set of rewards that come from the game, and $n$ is the number of steps in the game. An agent undertakes a sequence of actions ($n$ actions) based on the observations he has received up to the end of the game. If we view the game as an environment with which the agent interacts, each state refers to the observations of the environment the agent perceives at each time step. An action is a decision that is taken by the agent and that can be positively or negatively rewarded by the environment. 
\end{defn}
Figure \ref{fig:GameEnvMDP} depicts an overview of the interaction between an agent and a game. Given a state $s_t$ at time step $t$ the agent takes an action $a_t$ to interact with the game environment and receives a reward $s_t$ from the game environment. The environment moves into a new state $s_{t+1}$, affecting the selection of the next action. 

\begin{defn}
\label{gamestatedefinition}: \textbf{Game state.} A state in a game refers to its current status and can be represented as a fixed-length vector $(v_0, v_1, ..., v_n)$ \cite{nouwou2023comparison}. Each element $v_i$ of the vector represents an aspect of the state of the game such as the agent's position, and the location of the gold trophy in case of a Blockmaze game. 
\end{defn}

\begin{defn}
\label{gametestingdefinition}: \textbf{Game tester.} Given a game $G$, a set of policies $\Pi$ to play $G$, a set of states $S$ of $G$, and a set of bugs $B$ on $G$, a game tester $T$ can be defined as a function $T_G: \Pi \rightarrow S \times B$ \cite{nouwou2023comparison}. 
\end{defn}
A test case for a game is a sequence of actions. As a game might be non-deterministic (i.e., G is a stochastic function), each test case may result in various distinct states. A game tester \cite{nouwou2023comparison} implements different strategies to explore the different states of the game to detect bugs. The game tester, in this paper, acts as an oracle indicating the existence of a bug in an output state. A game tester, therefore, generates a series of valid actions, as a test case, that leads to a state in which there might be a bug. In this work, we consider the detection of bugs by one agent similar to our baselines \cite{zheng2019wuji,tufano2022using}.

\subsection{Job-Shop Scheduling Problem} \label{sec: Job-Shop Scheduling Problem}
The Job-Shop Scheduling Problem (JSSP) is a well-known optimization problem in manufacturing, computer science, and operations research \cite{kan2012machine}. This problem consists of minimizing the total time taken to process (i.e., makespan) a number of jobs by a number of machines. Since it is an NP-hard problem, multiple heuristics have been proposed to find approximate solutions \cite{gromicho2012solving,zhang2008effective}. In the rest of this section, we adopt the Priority Dispatching Rule (PDR) as a heuristic rule to solve the JSSP, as it is the same technique adopted by our baseline \cite{zhang2020learning}. In the following, we define some concepts that are important for understanding JSSP and 
PDR.
\begin{defn}
\label{JSSPdefinition}: \textbf{Job-Shop Scheduling Problem.} A JSSP instance consists of a set of jobs $J$ and a set of machines $M$. A JSSP instance is denoted as $\mid J \mid \times \mid M \mid $, where each job $J_i \in J$ goes through $m_i \in M$ machines in a constraint order as follows:
\[ O_{i1} \rightarrow O_{i2} \rightarrow ... \rightarrow O_{ij}\rightarrow ...\rightarrow O_{im_i}\] where $1 \leq j \leq m_i$, and $O_{ij}$ is an operation of $J_i \in J$ with a processing time of $t_{ij}$. Solving a JSSP instance means finding a schedule that starts at a startime $ST_{ij}$ for operation $O_{ij}$, such that the makespan: \[ C_{max}=max_{i,j}(ST_{ij}+t_{ij})\] is minimized. 
\end{defn}

\begin{defn}
\label{disjunctivedefinition}: \textbf{Disjunctive graph.} A JSSP instance can be represented as a disjunctive graph \cite{blazewicz2000disjunctive} 
$ G=(\mathcal{O}, \mathcal{C}, \mathcal{D}) $
with 
\[ \mathcal{O}= \{O_{ij} \mid 1 \leq j \leq m_i \}\cup \{ \textit{start},\textit{end} \}\]
$\mathcal{O}$ is the set of all operations and $i$ is the number of jobs; \textit{start} and \textit{end} are dummy operations with no processing time representing the start and the end of the schedule. $\mathcal{C} $ is a set of directed arcs between operations of the same job and $\mathcal{D} $ is a set of undirected arcs between operations requiring the same machine for processing. Consequently, finding a solution to a JSSP instance is equivalent to finding a direction for each undirected arc. 
\end{defn}

\begin{defn}
\label{Heuristicdefinition}: \textbf{Priority Dispatching Rule for JSSP.} 
A PDR is a heuristic technique to solve the JSSP and is widely used in scheduling systems \cite{zhang2020learning}. Given a set of operations, a PDR-based method can solve a JSSP instance by computing a priority index for each operation, then the operation with the highest priority is selected for dispatching. In this work, DRL is used to automatically generate PDRs. 
\end{defn}

\label{sec:Decision transformers}
\section{Study design}\label{studydesign}
In this section, we describe the methodology of our study which aims to leverage pre-trained generalist DRL agents on SE tasks. 

\subsection{Research questions}
The \emph{goal} of our work is to leverage generalist DRL agents on SE tasks by carefully fine-tuning their pre-trained models for automatically detecting bugs in games and solving JSSP on two scheduling instances. To achieve this goal, we focus on answering the following Research Questions (RQs). 
\begin{itemize}
    \item \textbf{RQ1:} How can we leverage generalist DRL agents for SE tasks? 
    \item \textbf{RQ2:} How do different pre-trained DRL generalist agents perform on SE tasks compared to DRL specialist agents?
   \item \textbf{RQ3:} How do different model-free DRL algorithms affect the performance of pre-trained generalist agents on SE tasks? 
\end{itemize}

\subsection{Methodology} \label{sec:Method}

We employ pre-trained generalist agents in several SE tasks. More specifically, the pre-trained agents are fine-tuned on selected SE tasks in order to accomplish the task at hand. In our work, we leverage two generalist agents IMPALA \cite{espeholt2018impala} and MGDT \cite{lee2022multi}.

\subsubsection{IMPALA}\label{sec:impala}
\begin{figure}
    \centering
    \includegraphics[width=0.6\textwidth]{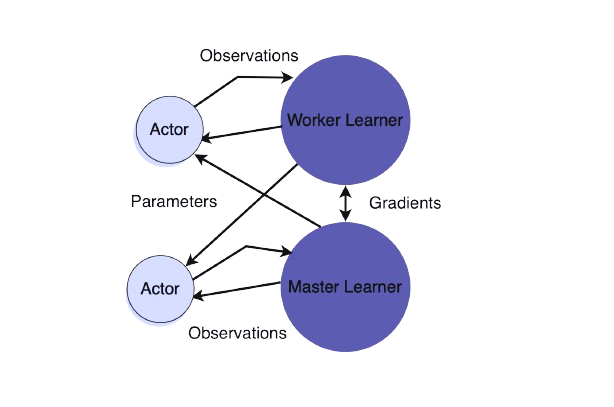}
    \caption{IMPALA architecture \cite{espeholt2018impala}.}
    \label{fig:impalaarch}
\end{figure}

IMPALA (Figure \ref{fig:impalaarch}) \cite{espeholt2018impala} is based on an actor-critic architecture with a set of actors and a set of learners. The actors generate trajectories of experiences and the learners use the experiences to learn an off-policy $\pi$. At the beginning of the training, each actor updates its own policy $\mu$ to the latest learner policy $\pi$ and runs it for $n$ steps in its environment. At the end of $n$ steps, each actor sends the collected observations, actions, rewards, as well as its policy distribution $\mu$ and an initial Long Short-Term Memory (LSTM) state to the learners via a queue. The learners update their own policy $\pi$ based on experiences collected from the actors. An IMPALA model architecture consists of multiple convolutional networks for feature extraction, followed by an LSTM \cite{hochreiter1997long} and a fully connected output layer. This architecture of IMPALA allows the learners to be accelerated on GPUs and the actors distributed across multiple machines. Given that pre-trained version of IMPALA are not available, we pre-trained it on 57 games of the Atari learning environment \cite{bellemare2013arcade} using V-trace \cite{espeholt2018impala}, an off-policy actor-critic algorithm. 

\textbf{V-trace} is an off-policy actor-critic algorithm used to pre-train IMPALA, with the ability to correct the policy lag between actors and learners. As stated in Section \ref{Background} off-policy algorithms use the behavior policy $\mu$ to generate trajectories in the form of

\[ (x_{t},a_{t},r_{t})_{t=s}^{t=s+n} \]
then, use those trajectories to learn the value function $V^{\pi}$ of another policy called target policy. V-trace target policy is formulated as:

\begin{equation}
    v_s ~ \myeq ~ V(x_s) + \sum _{t=s}^{s+n-1} \gamma ^{t-s} (\prod _{i=s}^{t-1} c_i ) \delta _t V
\end{equation}
where a temporal difference for $V$ is defined as \[\delta _t V ~ \myeq ~ \rho_{t}(r_t + \gamma V(x_{t+1}) - V(x_{t}))\] 
where $\rho_{t}$ and $c_i$ are truncated Importance Sampling (IS) weights defined as:
\[\rho_{t} ~ \myeq ~ min(\bar{\rho}, \frac{\pi(a_{t} \mid x_{t})}{\mu(a_{t} \mid x_{t})}),\] \[c_i ~ \myeq ~ min(\bar{c}, \frac{\pi(a_{i} \mid x_{i})}{\mu(a_{i} \mid x_{i})})\] 

At training time $s$, the value parameters $\theta$ are updated by the gradient descent on $l2$ loss to the target function $v_s$ in the direction of \[(v_s - V_{\theta}(x_{s}))\Delta_{\theta}V_{\theta}(x_{s}),\] and the policy parameters $\omega$ are updated in the direction of the policy gradient: 
\[\rho_s \Delta_{\omega} log\pi_{\omega}(a_{s} \mid x_{s})(r_s + \gamma v_{s+1} - V_{\theta}(x_{s}))\] 
We pre-trained an IMPALA agent following an online policy on 57 games of the Atari suite environments \cite{bellemare2013arcade} for 50M steps. Please note that the MsPacman game is excluded from the pre-training since it is among our evaluation tasks.

\subsubsection{Multi-Game Decision Transformer (MGDT)}\label{sec:Multi-Game Decision Transformers}
\begin{figure}
    \centering
    \includegraphics[width=0.5\textwidth]{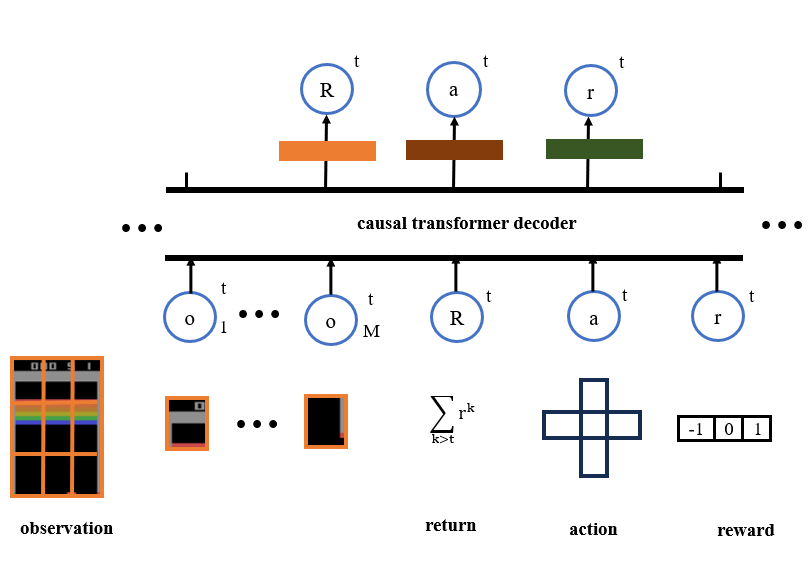}
    \caption{Multigame Decision Transformer Architecture (MGDT)\cite{lee2022multi}.}
    \label{fig:mgdtarch}
\end{figure}

MGDT, as illustrated in Figure \ref{fig:mgdtarch}, is a decision-making agent similar to a DT that receives at time $t$ an observation $o_{t}$, chooses an action $a_{t}$, and receives a reward $r_{t}$. 
The MGDT is an offline pre-trained RL agent with the goal of learning a single optimal policy distribution $P_{\theta}^{*}(a_{t} \mid o_{\leq t},a_{< t},r_{< t})$ with parameters $\theta$ that maximizes the agent's total future return $R_t= \sum _{k>t}r^k$. Further, during training, the trajectories are modeled in the form of a sequence of tokens: 
\[ s= \langle ...,o_{t} ^{1},...,x_{t} ^{M},\hat{R}_t,a_{t},r_{t},...\rangle \] 
where $t$ represents a time-step, $M$ is the number of image patches per observation and $\hat{R}_t$ is the MGDT agent target return for the rest of the sequence. The returns, actions and rewards are converted to discrete tokens after being generated via multinomial distributions. The scalar rewards are converted to ternary quantities $\{-1,0,+1\}$, and the returns are uniformly quantized into a discrete range shared by the environment considered by Lee et al.\cite{lee2022multi}. We did not pre-train the MGDT agent as the checkpoint of the pre-training done by Lee et al. \cite{lee2022multi} is available online\footnote{https://github.com/google-research/google-research/tree/master/multi\_game\_dt}. For pre-training, Lee et al.\cite{lee2022multi} used data from the Atari trajectories of 41 environments introduced by Agarwal et al. \cite{agarwal2020optimistic} on which a DQN agent was trained \cite{mnih2015human}. The data used for pre-training are from two training runs, each with roll-outs from 50 policy checkpoints for a total of $4.1$ billion steps. Similar to the case of IMPALA, the MsPacman game is excluded from the pre-training since it is among the evaluation tasks \cite{lee2022multi}.

\subsubsection{Online fine-tuning on SE tasks}\label{sec:Online fine-tuning on software testing tasks}
In our study, we fine-tune pre-trained generalist agents via online interactions (i.e., with model-free DRL algorithms) to perform SE tasks. Regarding the MGDT agent, we employ Max-entropy sequence modelling (MAENT) \cite{zheng2022online}, PPO\cite{schulman2017proximal} and DQN\cite{mnih2013playing} as the learning policy of the DRL agent. Regarding the IMPALA agent, in addition to V-trace, we leverage PPO as the agent's learning policy. We did not use DQN with IMPALA because IMPALA has an actor-critic architecture incompatible with DQN's value-based learning policy.

MAENT was introduced by Zheng et al. \cite{zheng2022online} to fine-tune a pre-trained decision transformer via online interactions. The goal of the MAENT is to learn a stochastic policy that maximizes the likelihood of the dataset with its ability to balance the tradeoffs between exploration and exploitation. During fine-tuning, the agent has access via a replay buffer to a data distribution $\mathcal{T}$. Let $\tau$ denotes a trajectory with the length of $\mid \tau \mid$, where $a = (a_1,..., a_{\mid \tau \mid}), ~ x=(x_1,..., x_{\mid \tau \mid})$, and $~ R =(R_1,..., R_{\mid \tau \mid})$ denote the sequence of actions, states and returns of $\tau$ respectively. The constrained function to solve is formulated as follows:

\begin{equation}
\underset{\theta}{min} ~ J(\theta) ~ subject ~ to ~ H^{\mathcal{T}} _{\theta} [a \mid x,R ] \ge \beta
\end{equation}
where $\beta$ is a predefined hyperparameter, $J(\theta)$ denotes the objective function used to minimize the negative log-likelihood loss of the trajectories of the dataset: \[J(\theta)= \frac{1}{K} \mathbb{E}_{(a,x,R) \sim \mathcal{T}}[-log \pi_{\theta}(a \mid x,R)]\]  and $H^{\mathcal{T}} _{\theta} [a \mid x,R ]$ is the entropy of policy used to quantify exploration and is formulated as follows: 
\begin{equation}
H^{\mathcal{T}} _{\theta} [a \mid x,R ]=\frac{1}{K} \mathbb{E}_{(a,R) \sim \mathcal{T}} [H[\pi_{\theta}(a \mid x,R)]]
\end{equation}
where $H[\pi_{\theta}(a_k)]$ denotes the Shannon entropy to calculate the entropy of the distribution $\pi_{\theta}(a_k)$. 

PPO \cite{schulman2017proximal} and DQN \cite{mnih2013playing} are the traditional model-free DRL algorithms respectively actor-critic and value-based. PPO follows an on-policy learning behavior while DQN follows an off-policy learning behavior. Algorithms \ref{alg:Online fine-tuning of generalist agents} and \ref{alg:Online fine-tuning of the IMPALA agent} summarize the overall fine-tuning procedure we implement to leverage the pre-trained generalist agents on our SE tasks. 

\begin{algorithm}[t]
\caption{Online fine-tuning of the MGDT agent}\label{alg:Online fine-tuning of generalist agents}
\Input{model parameters $\theta$, replay buffer $\mathcal{R}$, number of iterations $\mathcal{I}$, context length $\mathcal{K}$, batch size $\mathcal{B}$}

 \For{t=1,..., $\mathcal{I}$}{
 Generate trajectories and fill $\mathcal{R}$ \\
 Sample $\mathcal{B}$ trajectories out of $\mathcal{R}$\\
  \For{ each sampled trajectory $\tau$}{
    $(a,x,R) \gets$ a length $\mathcal{K}$ sub-trajectory sampled uniformly from $\tau$
  }
  $\theta \gets $  one gradient update using the sampled sub-trajectory $\{(a,x,R)\}s.$
  
 }
\end{algorithm}

\begin{algorithm}
\caption{Online fine-tuning of the IMPALA agent}\label{alg:Online fine-tuning of the IMPALA agent}
\Input{model parameters $\theta$, replay buffer $\mathcal{R}$, number of iterations $\mathcal{I}$,  $\{Actor_1,...,Actor_N\}$, Learner $\mathcal{L}$}

 \For{t=1,..., $\mathcal{I}$}{
 In parallel $Actor_i$ generates trajectories  $\{(x_1, a_1, r_1),...,(x_n, a_n, r_n)\}$\\
    $\mathcal{R}$ $\gets$ $\{(x_1, a_1, r_1),...,(x_n, a_n, r_n)\}$ \\
    $Actor_i$ retrieves policy parameters from $\mathcal{L}$\\
  $\theta \gets $  one gradient update of $\mathcal{L}$ using trajectories in $\mathcal{R}$
 }
\end{algorithm}
With Algorithm \ref{alg:Online fine-tuning of generalist agents}, at each iteration we generate trajectories to fill up the replay buffer. Then, we sample $\mathcal{K}$ trajectories to perform one gradient update of the policy of either one of the online training methods we used (i.e., MAENT, PPO, and DQN). With Algorithm \ref{alg:Online fine-tuning of the IMPALA agent}, each actor generates trajectories that are sent to the learner (PPO or V-trace) to update its policy gradients.

\subsubsection{Creation of the DRL tasks}\label{sec:Creation of the DRL algorithms}
In this paper, we consider two tasks: The detection of bugs in two games, Blockmaze \cite{zheng2019wuji} and MsPacman \cite{MsPacman}, and a PDR-based scheduling. We consider these tasks because they have been leveraged by previous studies \cite{zheng2019wuji, tufano2022using, zhang2020learning}, which we consider as baseline studies in this work. 

\textbf{1)} The Blockmaze game can be mapped into a DRL process by defining the states (i.e., observations), actions, reward function, end of an episode, and the information related to the bug. The goal is to navigate the game until reaching the goal position. 

\textbf{Observation space:} An observation is a set of features describing the state of the game. In our case, the observation space of the Blockmaze game has the size of a $20 \times 20$ matrix.
 
\textbf{Action space:} The action space describes the available moves that can be made in the Blockmaze by the agent. We consider the Blockmaze game with four discrete actions: north, south, east, and west.
 
\textbf{Reward function:} The reward function is the feedback from the environment regarding the agent's actions. It should be designed so that the agent can accomplish its mission. In this paper, the agent is rewarded negatively when it reaches a non-valid position in the game or any other position that is not the goal position of the game. In all other cases, it receives a positive reward.

\textbf{2)} The MsPacman game is based on the classic Pac-man game in which the goal is to eat all the dots without touching the ghosts \cite{rohlfshagen2017pac}. The environment of the MsPacman game is integrated into the gym Python toolkit \cite{1606.01540}. 

\textbf{Observation space:} The observation space of the MsPacman game has the size of an 84 $\times$ 84 matrix.

\textbf{Action space:} The action space describes the available moves that can be made in the MsPacman game by the agent. It is a set of 5 discrete actions.

\textbf{Reward function:} The reward function provides +1 when the agent eats one dot and 0 otherwise \cite{1606.01540}.

\textbf{3)} The last task we consider is a PDR task to solve JSSP. As mentioned in Definition \ref{Heuristicdefinition}, the goal of the PDR task is to dispatch all operations by minimizing the makespan. Similarly to the Blockmaze and MsPacman games, a PDR task can be mapped into a DRL process by defining the observations, actions, reward function, and end of an episode. 

\begin{figure*}[t]
     \centering
     \begin{minipage}{0.45\textwidth}
         \centering
         \includegraphics[width=\textwidth]{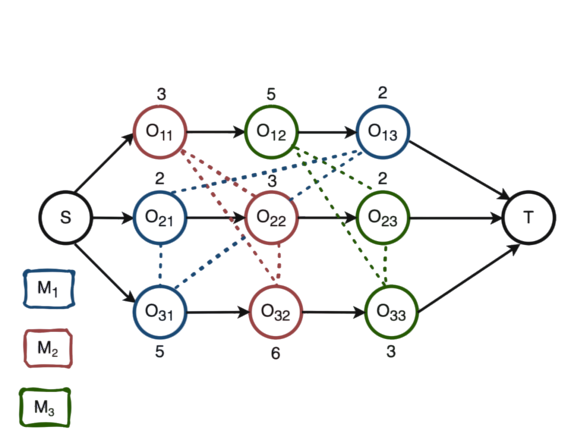}
         \label{fig:JSSP instance}
     \end{minipage}
     \hfill
     \begin{minipage}{0.45\textwidth}
         \centering
         \includegraphics[width=\textwidth]{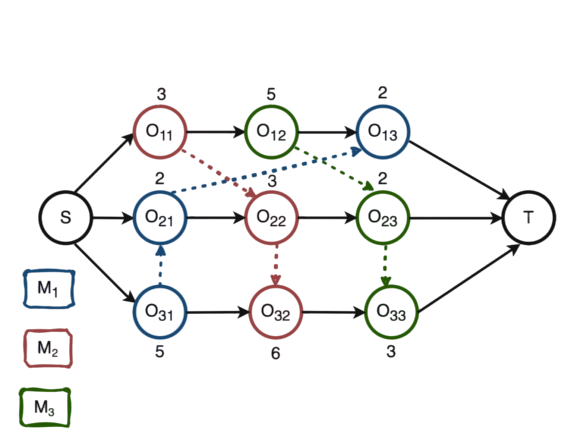}
         \label{fig: Complete solution}
     \end{minipage}
        \caption{Example of a PDR task observation. The left figure represents a 3 × 3 JSSP instance. The black arrows represent conjunctive arcs and the dotted lines are disjunctive arcs grouped into machines of different colors. The right figure is a complete solution, where all disjunctive arcs are assigned with directions \cite{zhang2020learning}.}
        \label{fig:JSSPobservation}
\end{figure*}

\textbf{Observation space:} The observation at each step \textit{t} of the agent in the environment is a disjunctive graph: \[G(t)=(O, C \cup D_u(\textit{t}), D(\textit{t}))\] that represents the current status of solution for the task instance. $D_u(\textit{t}) \subseteq D$ contains disjunctive arcs that have been assigned a direction at step \textit{t} while $D(\textit{t}) \subseteq D$ has the remaining arcs. At the beginning of an episode, the observation is the original JSSP instance, while at the end, it is the complete solution where all disjunctive arcs have been assigned a direction. An example of a $3 \times 3$ scheduling instance observation is depicted in Figure \ref{fig:JSSPobservation}. 
 
\textbf{Action space:} The action space describes operations that can be performed at each step \textit{t}. Each job has a set of operations (as defined in Section \ref{sec: Job-Shop Scheduling Problem}), therefore the size of action space is $\mid J \mid$. As an episode evolves, the action space becomes smaller as more jobs are completed. An episode ends when all jobs have been completed.
 
\textbf{Reward function:} The reward function is the quality difference between the solution of each two consecutive states: \[R=H(s_t)-H(s_{t+1})\] where \[H(s_t)=max_{i,j}\{C_{LB}(O_{i,j},s_t)\}\] is the lower bound of the makespan $C_{max}$. The goal is to dispatch operations step by step so that the makespan is minimized. Therefore, the cumulative reward obtained by the agent at each episode coincides with the minimum makespan \cite{zhang2020learning}.

\subsubsection{Baselines studies}\label{sec:Baselines studies}
In order to assess the performance of our fine-tuned generalist agents on SE tasks, we consider three baseline studies that we replicate, collect the performance of the specialist DRL agents being used and compare them against the pre-trained generalist. We picked these baseline studies because they target challenging problems and the source code of their proposed approaches is available. 

\textbf{1)} Zheng et al. \cite{zheng2019wuji} proposed Wuji, an automated game testing framework that combines Evolutionary Multi-Objective Optimization (EMOO) and DRL to detect bugs in a game. \\
\textbf{Approach:} Wuji randomly initializes a population of policies (represented by DNNs), adopts EMOO to diversify the state's exploration, and then uses DRL to improve the capability of the agent to accomplish its mission which is to play the game. \\
Zheng et al. \cite{zheng2019wuji} evaluated their approach on three games namely a Blockmaze game, L10 a Chinese ghost story\footnote{https://xqn.163.com/} and NSH a Treacherous Water Online\footnote{http://n.163.com/}. Given that the two latest are commercial games, with closed source, we do not consider them in our study. We replicated Wuji and collected the number of bugs it detected, and its training and testing times performance on the Blockmaze game.

\textbf{2)} Tufano et al. \cite{tufano2022using} proposed RELINE, an approach that leverages DRL algorithms to detect performance bugs. Specifically, the authors injected artificial performance bugs in two games, Cartpole \cite{Cartpole}, and MsPacman \cite{MsPacman}, and investigated whether or not the DRL agents can detect the bugs. Further, the authors studied a 3D kart racing game, Supertuxkart\footnote{https://github.com/supertuxkart/stk-code} and investigated whether or not their approach can find parts of the game resulting in a drop in the number of frames per second.\\
\textbf{Approach:} RELINE leverages DQN \cite{mnih2013playing} to train a DRL agent. Specifically, DQN adopts a Convolutional Neural Network (CNN) architecture that takes as input screenshots of the game and returns one of the possible actions of the game. Moreover, to incentivize the agent to look for bugs, an additional +50 is added to the reward earned by the agent every time it finds a bug during an episode. \\
In our study, we consider MsPacman as one of our evaluation studies. We did not consider Cartpole because our generalist agents are only compatible with 2D observation space. We did not consider the Supertuxkart game due to the unavailability of the source code of the environment. We replicated RELINE and collected the number of bugs it detected, and its training and testing times performance on the MsPacman game.

\textbf{3)} Zhang et al. \cite{zhang2020learning} leveraged DRL to automatically learn a PDR-based scheduling task for solving JSSP. They tested their approach on scheduling instances of various sizes $(6 \times 6), (10 \times 10), (15 \times 15), (20 \times 20), (30 \times 20), (50 \times 20), (100 \times 20)$ as well as on JSSP benchmarks and collected the makespan metric.\\
\textbf{Approach:} Zhang et al. \cite{zhang2020learning} used PPO algorithm to train a DRL agent that solves the JSSP. They proposed a Graph Neural Network (GNN) based architecture to encode the nodes of the disjunctive graphs of each JSSP instance. Particularly, the GNN captures the feature embedding of each node in a non-linear fashion.\\
In our study, we consider two instances $((6 \times 6),(30 \times 20) )$ to compare the performance of our fine-tuned generalist agents against a PPO-based specialist agent utilized by Zhang et al \cite{zhang2020learning}. We picked two scheduling instances each among the small and medium sizes. 
We replicated Zhang et al.'s \cite{zhang2020learning} approach and collected the makespan metric, the cumulative reward earned, as well as the training and testing times performance. Since the authors of Wuji\cite{zheng2019wuji} and RELINE\cite{tufano2022using} did not consider the cumulative reward as a metric in the original work, we did not report it here. The reason is that we would not have any baselines to compare our results.

\subsubsection{Datasets}\label{sec:Datasets}

In this section, we describe the training dataset used to fine-tune the selected pre-trained generalist agents for each task. We fine-tuned the pre-trained generalist agents on trajectories collected from the Blockmaze game, MsPacman game, and a PDR-based scheduling environment.

\begin{figure}
    \centering
    \includegraphics[width=0.4\textwidth]{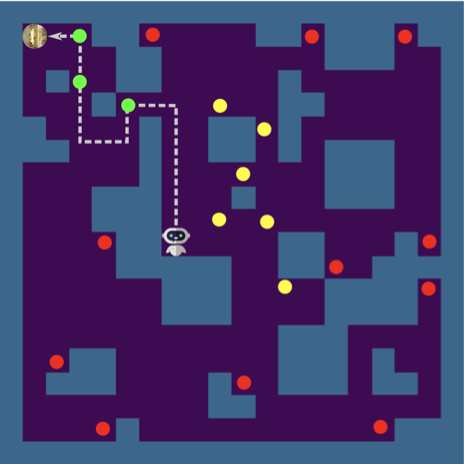}
    \caption{Blockmaze with bugs (red, green and yellow dots).}
    \label{fig:blockmazewithbugs}
\end{figure}

\textbf{1)} A Blockmaze game, Figure \ref{fig:blockmazewithbugs}, from Zheng et al. \cite{zheng2019wuji}, is selected for the evaluation of the generalist's agents. 
In the Blockmaze game, the agent's objective is to reach the goal coin, and it has 4 possible actions to choose from: north, south, east, and west. Every action leads to moving into a neighbor cell in the grid in the corresponding direction, except that a collision on a block (shown by dark green in Figure \ref{fig:blockmazewithbugs}) results in no movement. To evaluate the effectiveness of the fine-tuned generalist agents, 25 bugs are artificially injected into the Blockmaze, and randomly distributed within the task environment. A bug is a position in the Blockmaze that is triggered if the robot (agent) reaches its location in the map, as shown in Figure \ref{fig:blockmazewithbugs} with dots with colours green, red and yellow. A bug has no direct impact on the game but can be located in invalid locations of the game environment such as the Blockmaze obstacles or outside of the Blockmaze observation space. Invalid locations, on the other hand, cause the end of the game. Therefore, in this study, we consider two types of bugs: Type 1 that refers to exploratory bugs that measure the exploration capabilities of the agent, and Type 2 that refers to bugs at invalid locations of the Blockmaze.

\begin{figure}
    \centering
    \includegraphics[width=0.4\textwidth]{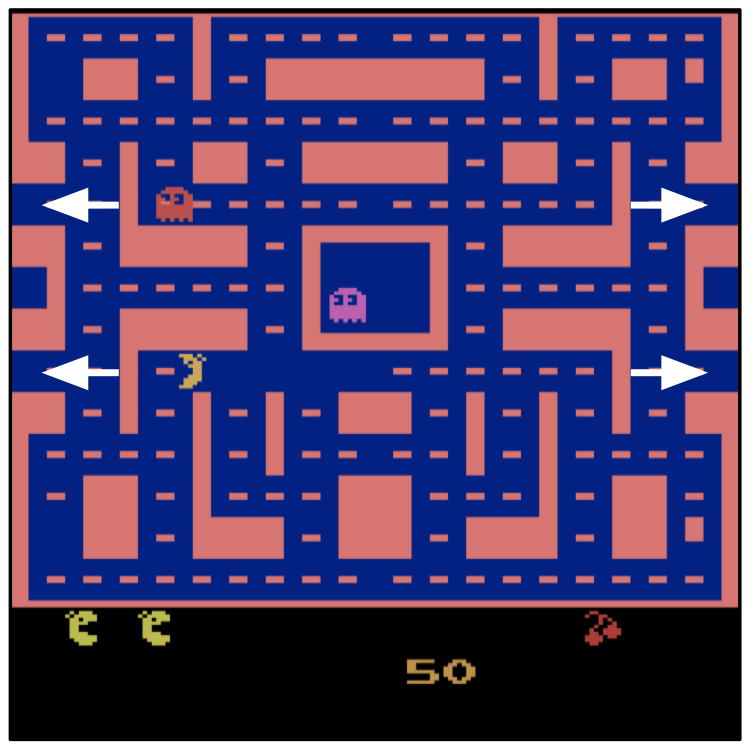}
    \caption{MsPacman game with bugs (white arrows).}
    \label{fig:pacmanwithbugs}
\end{figure}

\textbf{2)} The objective of the MsPacman \cite{tufano2022using} game is to eat all the dots without touching the ghosts. The possible actions to choose from are: north, south, east, west, and none. Four performance bugs are artificially injected into the MsPacman game, at four gates within the task environment (see white arrows in Figure \ref{fig:pacmanwithbugs}). The fine-tuned generalist agents are leveraged to look for these bugs. Specifically, a bug is a designated area within the environment, and whenever this area is reached, the agent will be rewarded by additional +50 points. For the MsPacman game, we consider 4 types of bugs representing the four gates (Figure \ref{fig:pacmanwithbugs}) where bugs can be spotted.

\textbf{3)} Finally, we consider a PDR-based scheduling \cite{zhang2020learning} technique to solve the JSSP. The objective of the task is to dispatch a set of operations of jobs while minimizing the makespan. The instances we considered $((6 \times 6),(30 \times 20) )$ are generated following the Taillard's method \cite{taillard1993benchmarks}. We used the same generated instances that have been used in the baseline work \cite{zhang2020learning} for training and testing.

\subsubsection{Experimental setup}\label{sec:Experimental setup}
We fine-tuned our pre-trained generalist agents with 0\% (zero-shot), 1\%, and 2\% of data budgets w.r.t the data budget employed to train our baselines specialist agents. Specifically, we fine-tuned the pre-trained generalist agents on the Blockmaze game with zero-shot fine-tuning steps, for 432 seconds and 864 seconds. Similarly, we fine-tuned the pre-trained generalist agents on the MsPacman game for 0, 10 and 20 episodes. Finally, we fine-tuned the pre-trained generalist agents on the task-based scheduling for 0, 360, and 720 steps for the ($6 \times 6$) instance and for 0, 60,000, and 120,000 steps for the ($30 \times 20$) instance.

We evaluated our fine-tuned generalists for 300,000 on the Blockmaze game, for 1000 episodes on the MsPacman game, and on 100 generated instances for each given size of instance ($6 \times 6$) or ($30 \times 20$). To counter the randomness effect during testing, we repeat each run 5 times to average the results. The experiments are run on the Cedar cluster servers provided by the Digital Research Alliance of Canada (the Alliance)\footnote{\url{https://docs.alliancecan.ca/wiki/Cedar}}. Each server has 32 cores at 2.2GHz with 125GB of main memory and 32GB of GPU memory. 

Zheng et al. \cite{zheng2019wuji} reported Wuji's results for 12 hours of training time, so we did the same and trained it for the same time. We evaluated Wuji for 300,000 steps. 

We trained the agent from Tufano et al. \cite{tufano2022using}, another one of our baselines, for 1,000 episodes and evaluated it for 1,000 episodes during a second phase of training. 

Finally, regarding Zhang et al.'s \cite{zhang2020learning} work, another one of our baseline papers, we trained the specialist agent's policy network for 10,000 iterations and evaluated their approach on 100 generated instances. 

As for the hyperparameters, we did not perform hyperparameters tuning, as in this paper we only study whether or not pre-trained generalist agents can be leveraged on the task at hand with little effort for fine-tuning. Table \ref{tab:The common hyperparameters that are used to fine-tune the MGDT}, \ref{tab:The common hyperparameters that are used to fine-tune IMPALA} \ref{tab:The hyperparameters that we use to fine-tune the MGDT for each online algorithm}, and \ref{tab:The hyperparameters that we use to fine-tune IMPALA for each online algorithm} summarize the hyperparameters we used on each generalist agents, as well as the ones we used for each online training algorithm considered in this paper (DQN, PPO, MAENT, and V\_TRACE).

\begin{table*}[t]
\caption{The common hyperparameters that are used to fine-tune the MGDT.}
\label{tab:The common hyperparameters that are used to fine-tune the MGDT}
\centering
\begin{tabular}{|c|c|}
\hline
Hyperparameters           & Value        \\ \hline
Number of layers          & 10           \\ \hline
Number of attention heads & 20           \\ \hline
Embedding dimension       & 1280         \\ \hline
Nonlinearity function     & TanH         \\ \hline
Batch size                & 32           \\ \hline
Updates between rollouts  & 300          \\ \hline
Target entropy            & -dim(Action) \\ \hline
Buffer size               & 10000        \\ \hline
Weight decay              & 0.0005       \\ \hline
Learning rate              & 0.0001         \\ \hline
\end{tabular}
\end{table*}

\begin{table*}[t]
\caption{The common hyperparameters that are used to fine-tune IMPALA }
\label{tab:The common hyperparameters that are used to fine-tune IMPALA}
\centering
\begin{tabular}{|c|c|}
\hline
Hyperparameters            & Value \\ \hline
Number of actors           & 48    \\ \hline
Nonlinearity function      & ReLu  \\ \hline
Batch size                 & 32    \\ \hline
Use of lstm                & True  \\ \hline
RMSProp smoothing constant & 0.99  \\ \hline
RMSProp epsilon            & 0.01  \\ \hline
entropy cost                &0.0006 \\ \hline
Learning rate              & 0.00048         \\ \hline
\end{tabular}
\end{table*}

\begin{table*}[t]
\caption{The hyperparameters that we use to fine-tune the MGDT for each online algorithm}
\label{tab:The hyperparameters that we use to fine-tune the MGDT for each online algorithm}
\centering
\resizebox{\textwidth}{!}{
\begin{tabular}{|c|c|c|c|c|}
\hline
Online algorithms & epsilon start & epsilon end & epsilon clip & surrogate loss epochs                    \\ \hline
MAENT       & NA            & NA          & NA           & NA                                      \\ \hline
PPO               & NA            & NA          & 0.2          & 15 (Blockmaze), 15 (PDR), 3 (MsPacman) \\ \hline
DQN               & 0.99          & 0.05        & NA           & NA                                      \\ \hline
\end{tabular}
}
\end{table*}

\begin{table*}[t]
\caption{The hyperparameters that we use to fine-tune IMPALA for each online algorithm}
\label{tab:The hyperparameters that we use to fine-tune IMPALA for each online algorithm}
\centering
\resizebox{\textwidth}{!}{
\begin{tabular}{|c|c|c|c|}
\hline
Online algorithms & epsilon clip & surrogate loss epochs                   & baseline cost \\ \hline
V-TRACE           & NA           & NA                                      & 0.5           \\ \hline
PPO               & 0.2          & 15 (Blockmaze), 15 (PDR), 3 (MsPacman) & NA            \\ \hline
\end{tabular}
}
\end{table*}

\subsubsection{Evaluation metrics}\label{sec:Evaluation metrics}
Depending on the task at hand, we use the following metrics to evaluate the performance of the fine-tuned generalist agents against the baselines specialist agents:
\begin{itemize}
    \item \textbf{Number of bugs detected}: the average number of bugs detected by the fine-tuned generalist agents as well as the specialist agents. We collect this metric for the Blockmaze and MsPacman games.
    \item \textbf{The average cumulative reward:} We report the reward obtained by the fine-tuned generalist agents as well as the specialist agents. We collect this metric for all our tasks.
  \item \textbf{Makespan:} We report the makespan obtained by the fine-tuned generalist agents as well as the specialist agents. We collect this metric for only the PDR-based scheduling.
 
    \item \textbf{Training time:} We collect the time consumed by the pre-trained generalist agents to fine-tune their policies on the task at hand. On the Blockmaze game, The training time lasted for 432 seconds (1\% of data budget) and 864 seconds (2\% of data budget) as the baseline specialist agent Wuji \cite{zheng2019wuji} is trained for 12 hours. On the MsPacman game, the training time was 10 episode (1\% of data budget) and 20 episodes (2\% of data budget) as the baseline specialist agent \cite{tufano2022using} is trained for 1000 episodes. On the ($6 \times 6$) instance of the PDR-based scheduling, the training time lasted for 360 (1\% of data budget) and 720 steps (2\% of data budget) as the baseline specialist agent \cite{zhang2020learning} is trained for 36,000 steps. Finally, on the ($30 \times 20$) instance of the PDR-based scheduling, we collected the training time for 60,000 and 120,000 steps (1\% and 2\% of data budget respectively) as the baseline specialist agent \cite{zhang2020learning} is trained for 6M steps.

\item \textbf{Testing time:} During the testing, we assess the performance of fine-tuned generalist agents on the task at hand. We collect the testing time for 300,000 steps for the Blockmaze game and 100 instances for the PDR-based scheduling task. Regarding the MsPacman game, once the generalist agents have been fine-tuned (for 10 and 20 episodes), we collected the time consumed by each of them to play the game for an additional 1000 episodes, similarly to the baseline \cite{tufano2022using}.

\end{itemize}
\subsubsection{Analysis method}\label{sec:Analysis method}
We proceeded as follows to answer our RQs. In \textbf{RQ1}, we fine-tuned the pre-trained generalist agents on different data budgets and collected the number of bugs detected on the Blockmaze and MsPacman games. We also collected the average makespan obtained by the fine-tuned generalist agents on the PDR-based scheduling tasks. Finally, we collected the average cumulative reward and the training and testing times of the pre-trained generalist agents for all the studied tasks. 
\\
In \textbf{RQ2}, based on the results of \textbf{RQ1}, we compared the pre-trained generalist agent that performed best against the baselines, based on the fine-tuning data budgets w.r.t to the evaluation metrics for each task. We calculate Common Language Effect Size (CLES) \cite{mcgraw1992common}, \cite{arcuri2014hitchhiker}, between the best configuration and the baseline to assess the effect size of differences. CLES estimates the probability that a randomly sampled value from one population is greater than a randomly sampled value from another population. \\
In \textbf{RQ3}, we compare the model-free DRL algorithms used to fine-tune the pre-trained generalist agents against each other. We use Tukey's post-hoc test \cite{brown2005new} to indicate the best online method. A difference with p-value $<= 0.05$ is considered significant in our assessments. When the variances in our results are not equal, we employ the Welch’s ANOVA and Games-Howell post-hoc test \cite{welch1947generalization}, \cite{games1976pairwise} instead of Tukey's test,  as Welch’s ANOVA do not assume equal variance. While Welch’s ANOVA test checks for significant differences between the generalist agent configurations, the Games-Howell post-hoc test compares each pair of configurations.  \\
\subsection{Data Availability}
The source code of our implementation and the results of experiments are publicly available \cite{replication-package}.

\section{Experimental results}\label{sec:Experimental results}
We now report the results of our experiments. We report the results  associated to \textbf{RQ1, RQ2} and \textbf{RQ3} in Sections \ref{sec:rq1}, \ref{sec:rq2} and \ref{sec:rq3} respectively. 

\subsection{\textbf{RQ1: How can we leverage generalist DRL agents for SE tasks?}} \label{sec:rq1}
To investigate whether DRL generalist agents can be leveraged for SE tasks, 
we experiment with the pre-trained models of IMPALA and MGDT as explained in Section \ref{sec:impala} and Section \ref{sec:Multi-Game Decision Transformers}. We fine-tuned these pre-trained generalist agents via model-free DRL algorithms on three different data budgets (zero-shot, 1\%, 2\%) to solve the following selected SE tasks: 1) the detection of bugs in the Blockmaze and MsPacman games, and 2) the use of PDR heuristic on the $6\times6$ and $30\times20$ instances to solve the JSSP. The details of the fine-tuning approaches we used are presented in Section \ref{alg:Online fine-tuning of generalist agents} and Section \ref{alg:Online fine-tuning of the IMPALA agent} for each pre-trained generalist agent. 

\begin{table}
\centering
\caption{Performance of the fine-tuned generalist agents on the MsPacman game on 2\% data budget and the baseline in terms of bugs detected  (in bold are the values of generalist agent configurations with greater performance).}

\resizebox{\textwidth}{!}{%
\begin{tabular}{|ccc|cccc|}
\hline
\multicolumn{3}{|c|}{Environment}                                                                      & \multicolumn{4}{c|}{MsPacman}                                                                                                            \\ \hline
\multicolumn{3}{|c|}{Metrics}                                                                           & \multicolumn{1}{c|}{Type 1}           & \multicolumn{1}{c|}{Type 2}           & \multicolumn{1}{c|}{Type 3}           & Type 4           \\ \hline
\multicolumn{2}{|c|}{\multirow{3}{*}{\textbf{Baseline agent}}}                                 & mean   & \multicolumn{1}{c|}{\textbf{7.85e+2}} & \multicolumn{1}{c|}{\textbf{7.72e+2}} & \multicolumn{1}{c|}{\textbf{6.89e+2}} & \textbf{6.75e+2} \\ \cline{3-7} 
\multicolumn{2}{|c|}{}                                                                         & std    & \multicolumn{1}{c|}{\textbf{3.41e+1}}          & \multicolumn{1}{c|}{\textbf{3.35e+1}}          & \multicolumn{1}{c|}{\textbf{4.36e+1}}          & \textbf{4.68e+1}          \\ \cline{3-7} 
\multicolumn{2}{|c|}{}                                                                         & median & \multicolumn{1}{c|}{\textbf{7.87e+2}}          & \multicolumn{1}{c|}{\textbf{7.66e+2}}          & \multicolumn{1}{c|}{\textbf{7.09e+2}}          & \textbf{6.95e+2}          \\ \hline
\multicolumn{1}{|c|}{\multirow{9}{*}{MGDT}}   & \multicolumn{1}{c|}{\multirow{3}{*}{MAENT}}    & mean   & \multicolumn{1}{c|}{0.0}          & \multicolumn{1}{c|}{0.0}          & \multicolumn{1}{c|}{1.47e+2}          & 5.86e+1          \\ \cline{3-7} 
\multicolumn{1}{|c|}{}                        & \multicolumn{1}{c|}{}                          & std    & \multicolumn{1}{c|}{0.0}          & \multicolumn{1}{c|}{0.0}          & \multicolumn{1}{c|}{3.75e+1}          & 1.41e+1          \\ \cline{3-7} 
\multicolumn{1}{|c|}{}                        & \multicolumn{1}{c|}{}                          & median & \multicolumn{1}{c|}{0.0}          & \multicolumn{1}{c|}{0.0}          & \multicolumn{1}{c|}{1.62e+2}          & 6.40e+1          \\ \cline{2-7} 
\multicolumn{1}{|c|}{}                        & \multicolumn{1}{c|}{\multirow{3}{*}{DQN}}      & mean   & \multicolumn{1}{c|}{0.0}          & \multicolumn{1}{c|}{0.0}          & \multicolumn{1}{c|}{2.04e+1}          & 1.18e+1          \\ \cline{3-7} 
\multicolumn{1}{|c|}{}                        & \multicolumn{1}{c|}{}                          & std    & \multicolumn{1}{c|}{0.0}          & \multicolumn{1}{c|}{0.0}          & \multicolumn{1}{c|}{4.76e+0}          & 2.79e+0          \\ \cline{3-7} 
\multicolumn{1}{|c|}{}                        & \multicolumn{1}{c|}{}                          & median & \multicolumn{1}{c|}{0.0}          & \multicolumn{1}{c|}{0.0}          & \multicolumn{1}{c|}{2.10e+1}          & 1.30e+1          \\ \cline{2-7} 
\multicolumn{1}{|c|}{}                        & \multicolumn{1}{c|}{\multirow{3}{*}{PPO}}      & mean   & \multicolumn{1}{c|}{0.0}          & \multicolumn{1}{c|}{0.0}          & \multicolumn{1}{c|}{0.0}          & 0.0          \\ \cline{3-7} 
\multicolumn{1}{|c|}{}                        & \multicolumn{1}{c|}{}                          & std    & \multicolumn{1}{c|}{0.0}          & \multicolumn{1}{c|}{0.0}          & \multicolumn{1}{c|}{0.0}          & 0.0          \\ \cline{3-7} 
\multicolumn{1}{|c|}{}                        & \multicolumn{1}{c|}{}                          & median & \multicolumn{1}{c|}{0.0}          & \multicolumn{1}{c|}{0.0}          & \multicolumn{1}{c|}{0.0}          & 0.0          \\ \hline
\multicolumn{1}{|c|}{\multirow{6}{*}{IMPALA}} & \multicolumn{1}{c|}{\multirow{3}{*}{V\_TRACE}} & mean   & \multicolumn{1}{c|}{2.32e+2}          & \multicolumn{1}{c|}{2.17e+2}          & \multicolumn{1}{c|}{1.22e+2}          & 1.12e+2          \\ \cline{3-7} 
\multicolumn{1}{|c|}{}                        & \multicolumn{1}{c|}{}                          & std    & \multicolumn{1}{c|}{9.13e+0}          & \multicolumn{1}{c|}{1.06e+1}          & \multicolumn{1}{c|}{1.39e+1}          & 1.77e+1          \\ \cline{3-7} 
\multicolumn{1}{|c|}{}                        & \multicolumn{1}{c|}{}                          & median & \multicolumn{1}{c|}{2.28e+2}          & \multicolumn{1}{c|}{2.20e+2}          & \multicolumn{1}{c|}{1.24e+2}          & 1.16e+2          \\ \cline{2-7} 
\multicolumn{1}{|c|}{}                        & \multicolumn{1}{c|}{\multirow{3}{*}{PPO}}      & mean   & \multicolumn{1}{c|}{2.38e+2}          & \multicolumn{1}{c|}{2.27e+2}          & \multicolumn{1}{c|}{1.19e+2}          & 1.10e+2          \\ \cline{3-7} 
\multicolumn{1}{|c|}{}                        & \multicolumn{1}{c|}{}                          & std    & \multicolumn{1}{c|}{9.85e+0}          & \multicolumn{1}{c|}{1.17e+1}          & \multicolumn{1}{c|}{1.00e+1}          & 7.81e+0          \\ \cline{3-7} 
\multicolumn{1}{|c|}{}                        & \multicolumn{1}{c|}{}                          & median & \multicolumn{1}{c|}{2.35e+2}          & \multicolumn{1}{c|}{2.24e+2}          & \multicolumn{1}{c|}{1.21e+2}          & 1.16e+2          \\ \hline
\end{tabular}
\label{tab:bugmakespan}%
}
\end{table}
\begin{table}
\centering
\caption{Performance of the fine-tuned generalist agents on the Blockmaze game on 2\% data budget and the baseline in terms of bugs detected  (in bold are the values of generalist agent configurations with greater performance).}

\resizebox{0.8\textwidth}{!}{%
\begin{tabular}{|ccc|cc|}
\hline
\multicolumn{3}{|c|}{Environment}                                                                          & \multicolumn{2}{c|}{Blockmaze}                           \\ \hline
\multicolumn{3}{|c|}{Metrics}                                                                               & \multicolumn{1}{c|}{Type 1}           & Type 2           \\ \hline
\multicolumn{2}{|c|}{\multirow{3}{*}{Baseline agent}}                                              & mean   & \multicolumn{1}{c|}{1.80e+0}          & 2.00e+0          \\ \cline{3-5} 
\multicolumn{2}{|c|}{}                                                                             & std    & \multicolumn{1}{c|}{4.00e-1}          & 0.0          \\ \cline{3-5} 
\multicolumn{2}{|c|}{}                                                                             & median & \multicolumn{1}{c|}{2.00e+0}          & 2.00e+0          \\ \hline
\multicolumn{1}{|c|}{\multirow{9}{*}{MGDT}}   & \multicolumn{1}{c|}{\multirow{3}{*}{MAENT}}        & mean   & \multicolumn{1}{c|}{0.0}          & 1.00e+0          \\ \cline{3-5} 
\multicolumn{1}{|c|}{}                        & \multicolumn{1}{c|}{}                              & std    & \multicolumn{1}{c|}{0.0}          & 0.0          \\ \cline{3-5} 
\multicolumn{1}{|c|}{}                        & \multicolumn{1}{c|}{}                              & median & \multicolumn{1}{c|}{0.0}          & 1.00e+0          \\ \cline{2-5} 
\multicolumn{1}{|c|}{}                        & \multicolumn{1}{c|}{\multirow{3}{*}{\textbf{DQN}}} & mean   & \multicolumn{1}{c|}{\textbf{3.00e+0}} & \textbf{5.00e+0} \\ \cline{3-5} 
\multicolumn{1}{|c|}{}                        & \multicolumn{1}{c|}{}                              & std    & \multicolumn{1}{c|}{\textbf{0.0}}          & \textbf{0.0}          \\ \cline{3-5} 
\multicolumn{1}{|c|}{}                        & \multicolumn{1}{c|}{}                              & median & \multicolumn{1}{c|}{\textbf{3.00e+0}}          & \textbf{5.00e+0}          \\ \cline{2-5} 
\multicolumn{1}{|c|}{}                        & \multicolumn{1}{c|}{\multirow{3}{*}{PPO}}          & mean   & \multicolumn{1}{c|}{2.00e+0}          & 2.00e+0          \\ \cline{3-5} 
\multicolumn{1}{|c|}{}                        & \multicolumn{1}{c|}{}                              & std    & \multicolumn{1}{c|}{0.0}          & 0.0          \\ \cline{3-5} 
\multicolumn{1}{|c|}{}                        & \multicolumn{1}{c|}{}                              & median & \multicolumn{1}{c|}{2.00e+0}          & 2.00e+0          \\ \hline
\multicolumn{1}{|c|}{\multirow{6}{*}{IMPALA}} & \multicolumn{1}{c|}{\multirow{3}{*}{V\_TRACE}}     & mean   & \multicolumn{1}{c|}{0.0}          & 0.0          \\ \cline{3-5} 
\multicolumn{1}{|c|}{}                        & \multicolumn{1}{c|}{}                              & std    & \multicolumn{1}{c|}{0.0}          & 0.0          \\ \cline{3-5} 
\multicolumn{1}{|c|}{}                        & \multicolumn{1}{c|}{}                              & median & \multicolumn{1}{c|}{0.0}          & 0.0          \\ \cline{2-5} 
\multicolumn{1}{|c|}{}                        & \multicolumn{1}{c|}{\multirow{3}{*}{PPO}}          & mean   & \multicolumn{1}{c|}{0.0}          & 0.0          \\ \cline{3-5} 
\multicolumn{1}{|c|}{}                        & \multicolumn{1}{c|}{}                              & std    & \multicolumn{1}{c|}{0.0}          & 0.0          \\ \cline{3-5} 
\multicolumn{1}{|c|}{}                        & \multicolumn{1}{c|}{}                              & median & \multicolumn{1}{c|}{0.0}          & 0.0          \\ \hline
\end{tabular}
\label{tab:bugmakespanblockmaze}%
}
\end{table}
\begin{table}
\centering
\caption{Performance of the fine-tuned generalist agents on the PDR-based scheduling on 2\% data budget and the baseline in terms of  makespan time (in bold are the values of generalist agent configurations with greater performance).}

\resizebox{0.8\textwidth}{!}{%
\begin{tabular}{|ccc|cc|}
\hline
\multicolumn{3}{|c|}{Environment}                                                                            & \multicolumn{2}{c|}{PDR}                                 \\ \hline
\multicolumn{3}{|c|}{Metrics}                                                                                 & \multicolumn{1}{c|}{(6 x 6)}          & (30 x 20)        \\ \hline
\multicolumn{2}{|c|}{\multirow{3}{*}{Baseline agent}}                                                & mean   & \multicolumn{1}{c|}{5.73e+2}          & 2.48e+3          \\ \cline{3-5} 
\multicolumn{2}{|c|}{}                                                                               & std    & \multicolumn{1}{c|}{3.27e+0}          & 1.07e+0          \\ \cline{3-5} 
\multicolumn{2}{|c|}{}                                                                               & median & \multicolumn{1}{c|}{5.74e+2}          & 2.47e+3          \\ \hline
\multicolumn{1}{|c|}{\multirow{9}{*}{MGDT}}   & \multicolumn{1}{c|}{\multirow{3}{*}{\textbf{MAENT}}} & mean   & \multicolumn{1}{c|}{\textbf{5.00e+2}} & 2.01e+3 \\ \cline{3-5} 
\multicolumn{1}{|c|}{}                        & \multicolumn{1}{c|}{}                                & std    & \multicolumn{1}{c|}{\textbf{0.0}}          & 5.93e+0          \\ \cline{3-5} 
\multicolumn{1}{|c|}{}                        & \multicolumn{1}{c|}{}                                & median & \multicolumn{1}{c|}{\textbf{5.00e+2}}          & 2.01e+3          \\ \cline{2-5} 
\multicolumn{1}{|c|}{}                        & \multicolumn{1}{c|}{\multirow{3}{*}{DQN}}            & mean   & \multicolumn{1}{c|}{5.02e+2}          & 2.01e+3          \\ \cline{3-5} 
\multicolumn{1}{|c|}{}                        & \multicolumn{1}{c|}{}                                & std    & \multicolumn{1}{c|}{5.68e-14}         & 5.06e+0          \\ \cline{3-5} 
\multicolumn{1}{|c|}{}                        & \multicolumn{1}{c|}{}                                & median & \multicolumn{1}{c|}{5.02e+2}          & 2.01e+3          \\ \cline{2-5} 
\multicolumn{1}{|c|}{}                        & \multicolumn{1}{c|}{\multirow{3}{*}{PPO}}            & mean   & \multicolumn{1}{c|}{5.01e+2}          & 2.01e+3          \\ \cline{3-5} 
\multicolumn{1}{|c|}{}                        & \multicolumn{1}{c|}{}                                & std    & \multicolumn{1}{c|}{0.0}          & 4.23e+0          \\ \cline{3-5} 
\multicolumn{1}{|c|}{}                        & \multicolumn{1}{c|}{}                                & median & \multicolumn{1}{c|}{5.01e+2}          & 2.01e+3          \\ \hline
\multicolumn{1}{|c|}{\multirow{6}{*}{IMPALA}} & \multicolumn{1}{c|}{\multirow{3}{*}{V\_TRACE}}       & mean   & \multicolumn{1}{c|}{5.01e+2}          & 2.01e+3          \\ \cline{3-5} 
\multicolumn{1}{|c|}{}                        & \multicolumn{1}{c|}{}                                & std    & \multicolumn{1}{c|}{3.73e+0}          & 1.70e+0          \\ \cline{3-5} 
\multicolumn{1}{|c|}{}                        & \multicolumn{1}{c|}{}                                & median & \multicolumn{1}{c|}{5.04e+2}          & 2.01e+3          \\ \cline{2-5} 
\multicolumn{1}{|c|}{}                        & \multicolumn{1}{c|}{\multirow{3}{*}{PPO}}            & mean   & \multicolumn{1}{c|}{5.03e+2}          & \textbf{2.01e+3}          \\ \cline{3-5} 
\multicolumn{1}{|c|}{}                        & \multicolumn{1}{c|}{}                                & std    & \multicolumn{1}{c|}{4.98e+0}          & \textbf{1.11e+0}          \\ \cline{3-5} 
\multicolumn{1}{|c|}{}                        & \multicolumn{1}{c|}{}                                & median & \multicolumn{1}{c|}{5.02e+2}          & \textbf{2.01e+3}          \\ \hline
\end{tabular}
\label{tab:bugmakespanpdr}%
}
\end{table}

Regarding the Blockmaze game, as described in our study design (Section \ref{studydesign}) we fine-tuned the pre-trained generalist agents using zero-shot, for 432 and 864 seconds. We collected the number of bugs detected, the training time, the testing time, and the average cumulative reward earned, which we report in Table \ref{tab:bugmakespanblockmaze}, \ref{tab:timesrewardblockmaze}, \ref{tab:bugmakespan0fblockmaze}, \ref{tab:timesreward0fblockmaze},\ref{tab:bugmakespan100fblockmaze} and \ref{tab:timesreward100fblockmaze}. 

\begin{figure}
    \centering
    \includegraphics[width=0.7\textwidth]{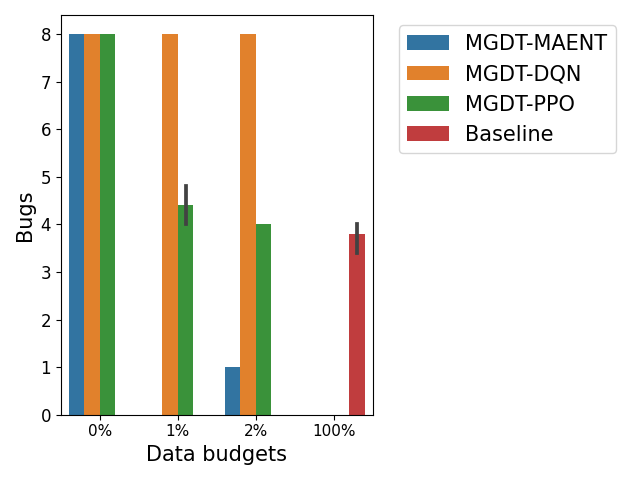}
    \caption{Number of bugs detected by the MGDT agent configurations and the baseline Wuji }
    \label{fig:Number of bugs detected by MGDT generalist agent and its configuration}
\end{figure}

In terms of bugs detected, IMPALA generalist agents (IMPALA-PPO and IMPALA-V\_TRACE) are not able to detect any bugs regardless of the data budgets allowed during fine-tuning. Nonetheless, as shown in Figure \ref{fig:Number of bugs detected by MGDT generalist agent and its configuration}, MGDT generalist agents (MGDT-PPO, MGDT-DQN, and MGDT-MAENT) are able to detect bugs across all three data budgets allowed during fine-tuning. 
\begin{table}
\centering
\caption{Cumulative reward, training, and testing times of the fine-tuned generalist agents on MsPacman game on 2\% data budget and the baseline (in bold are the values of generalist agents configurations with greater performance).}
\resizebox{0.9\textwidth}{!}{%
\begin{tabular}{|ccc|ccc|}
\hline
\multicolumn{3}{|c|}{\multirow{2}{*}{Environment}}                                                           & \multicolumn{3}{c|}{\multirow{2}{*}{MsPacman}}                                                    \\
\multicolumn{3}{|c|}{}                        & \multicolumn{3}{c|}{}                 \\ \hline
\multicolumn{3}{|c|}{Metrics}                                                                                 & \multicolumn{1}{c|}{TA time}          & \multicolumn{1}{c|}{TE time}          & Cumulative reward \\ \hline
\multicolumn{2}{|c|}{\multirow{3}{*}{\textbf{Baseline agent}}}                                       & mean   & \multicolumn{1}{c|}{\textbf{8.14e+3}} & \multicolumn{1}{c|}{\textbf{2.57e+4}} & NA                \\ \cline{3-6} 
\multicolumn{2}{|c|}{}                                                                               & std    & \multicolumn{1}{c|}{\textbf{3.57e+2}} & \multicolumn{1}{c|}{\textbf{1.08e+3}} & NA                \\ \cline{3-6} 
\multicolumn{2}{|c|}{}                                                                               & median & \multicolumn{1}{c|}{\textbf{7.99e+3}} & \multicolumn{1}{c|}{\textbf{2.55e+4}} & NA                \\ \hline
\multicolumn{1}{|c|}{\multirow{9}{*}{MGDT}}   & \multicolumn{1}{c|}{\multirow{3}{*}{\textbf{MAENT}}} & mean   & \multicolumn{1}{c|}{8.70e+3}          & \multicolumn{1}{c|}{3.92e+5}          & \textbf{2.00e+2}  \\ \cline{3-6} 
\multicolumn{1}{|c|}{}                        & \multicolumn{1}{c|}{}                                & std    & \multicolumn{1}{c|}{3.50e+3}          & \multicolumn{1}{c|}{5.52e+4}          & \textbf{4.33e+1}  \\ \cline{3-6} 
\multicolumn{1}{|c|}{}                        & \multicolumn{1}{c|}{}                                & median & \multicolumn{1}{c|}{9.19e+3}          & \multicolumn{1}{c|}{3.69e+5}          & \textbf{1.88e+2}  \\ \cline{2-6} 
\multicolumn{1}{|c|}{}                        & \multicolumn{1}{c|}{\multirow{3}{*}{DQN}}            & mean   & \multicolumn{1}{c|}{9.09e+3}          & \multicolumn{1}{c|}{3.66e+5}          & 1.50e+2           \\ \cline{3-6} 
\multicolumn{1}{|c|}{}                        & \multicolumn{1}{c|}{}                                & std    & \multicolumn{1}{c|}{2.47e+3}          & \multicolumn{1}{c|}{1.62e+4}          & 1.91e+0           \\ \cline{3-6} 
\multicolumn{1}{|c|}{}                        & \multicolumn{1}{c|}{}                                & median & \multicolumn{1}{c|}{1.10e+4}          & \multicolumn{1}{c|}{3.62e+5}          & 1.51e+2           \\ \cline{2-6} 
\multicolumn{1}{|c|}{}                        & \multicolumn{1}{c|}{\multirow{3}{*}{PPO}}            & mean   & \multicolumn{1}{c|}{1.27e+4}          & \multicolumn{1}{c|}{5.99e+5}          & 4.50e+1           \\ \cline{3-6} 
\multicolumn{1}{|c|}{}                        & \multicolumn{1}{c|}{}                                & std    & \multicolumn{1}{c|}{3.29e+3}          & \multicolumn{1}{c|}{1.54e+4}          & 0.0           \\ \cline{3-6} 
\multicolumn{1}{|c|}{}                        & \multicolumn{1}{c|}{}                                & median & \multicolumn{1}{c|}{1.11e+4}          & \multicolumn{1}{c|}{6.05e+5}          & 4.50e+1           \\ \hline
\multicolumn{1}{|c|}{\multirow{6}{*}{IMPALA}} & \multicolumn{1}{c|}{\multirow{3}{*}{V\_TRACE}}       & mean   & \multicolumn{1}{c|}{9.92e+3}          & \multicolumn{1}{c|}{3.21e+4}          & 8.90e+1           \\ \cline{3-6} 
\multicolumn{1}{|c|}{}                        & \multicolumn{1}{c|}{}                                & std    & \multicolumn{1}{c|}{4.16e+2}          & \multicolumn{1}{c|}{3.92e+3}          & 1.86e+0           \\ \cline{3-6} 
\multicolumn{1}{|c|}{}                        & \multicolumn{1}{c|}{}                                & median & \multicolumn{1}{c|}{1.00e+4}          & \multicolumn{1}{c|}{3.36e+4}          & 8.88e+1           \\ \cline{2-6} 
\multicolumn{1}{|c|}{}                        & \multicolumn{1}{c|}{\multirow{3}{*}{PPO}}            & mean   & \multicolumn{1}{c|}{1.02e+4}          & \multicolumn{1}{c|}{3.24e+4}          & 1.82e+2           \\ \cline{3-6} 
\multicolumn{1}{|c|}{}                        & \multicolumn{1}{c|}{}                                & std    & \multicolumn{1}{c|}{7.44e+2}          & \multicolumn{1}{c|}{2.36e+3}          & 1.76e+1           \\ \cline{3-6} 
\multicolumn{1}{|c|}{}                        & \multicolumn{1}{c|}{}                                & median & \multicolumn{1}{c|}{1.00e+4}          & \multicolumn{1}{c|}{3.19e+4}          & 1.73e+2           \\ \hline
\end{tabular}
\label{tab:timesreward}%
}
\begin{tablenotes}
     \item TA time and TE time refer to training time and testing time respectively.
 \end{tablenotes}
\end{table}

\begin{table}
\centering
\caption{Cumulative reward, training, and testing times of the fine-tuned generalist agents on the Blockmaze game on 2\% data budget and the baseline (in bold are the values of generalist agents configurations with greater performance).}
\resizebox{0.9\textwidth}{!}{%
\begin{tabular}{|ccc|ccc|}
\hline
\multicolumn{3}{|c|}{\multirow{2}{*}{Environment}}                                                              & \multicolumn{3}{c|}{\multirow{2}{*}{Blockmaze}}                                                   \\
\multicolumn{3}{|c|}{}                        & \multicolumn{3}{c|}{}                 \\ \hline
\multicolumn{3}{|c|}{Metrics}                                                                                    & \multicolumn{1}{c|}{TA time}          & \multicolumn{1}{c|}{TE time}          & Cumulative reward \\ \hline
\multicolumn{2}{|c|}{\multirow{3}{*}{\textbf{Baseline agent}}}                                          & mean   & \multicolumn{1}{c|}{4.32e+4}          & \multicolumn{1}{c|}{\textbf{8.82e+2}} & NA                \\ \cline{3-6} 
\multicolumn{2}{|c|}{}                                                                                  & std    & \multicolumn{1}{c|}{0.0}          & \multicolumn{1}{c|}{\textbf{2.76e+2}} & NA                \\ \cline{3-6} 
\multicolumn{2}{|c|}{}                                                                                  & median & \multicolumn{1}{c|}{4.32e+4}          & \multicolumn{1}{c|}{\textbf{9.60e+2}} & NA                \\ \hline
\multicolumn{1}{|c|}{\multirow{9}{*}{MGDT}}   & \multicolumn{1}{c|}{\multirow{3}{*}{\textbf{MAENT}}}    & mean   & \multicolumn{1}{c|}{\textbf{8.64e+2}} & \multicolumn{1}{c|}{7.91e+3}          & -2.97e+2          \\ \cline{3-6} 
\multicolumn{1}{|c|}{}                        & \multicolumn{1}{c|}{}                                   & std    & \multicolumn{1}{c|}{\textbf{0.0}} & \multicolumn{1}{c|}{2.13e+2}          & 0.0           \\ \cline{3-6} 
\multicolumn{1}{|c|}{}                        & \multicolumn{1}{c|}{}                                   & median & \multicolumn{1}{c|}{\textbf{8.64e+2}} & \multicolumn{1}{c|}{7.99e+3}          & -2.97e+2          \\ \cline{2-6} 
\multicolumn{1}{|c|}{}                        & \multicolumn{1}{c|}{\multirow{3}{*}{\textbf{DQN}}}      & mean   & \multicolumn{1}{c|}{\textbf{8.64e+2}} & \multicolumn{1}{c|}{2.19e+3}          & -1.49e+2          \\ \cline{3-6} 
\multicolumn{1}{|c|}{}                        & \multicolumn{1}{c|}{}                                   & std    & \multicolumn{1}{c|}{\textbf{0.0}} & \multicolumn{1}{c|}{8.95e+1}          & 0.0           \\ \cline{3-6} 
\multicolumn{1}{|c|}{}                        & \multicolumn{1}{c|}{}                                   & median & \multicolumn{1}{c|}{\textbf{8.64e+2}} & \multicolumn{1}{c|}{2.17e+3}          & -1.49e+2          \\ \cline{2-6} 
\multicolumn{1}{|c|}{}                        & \multicolumn{1}{c|}{\multirow{3}{*}{\textbf{PPO}}}      & mean   & \multicolumn{1}{c|}{\textbf{8.64e+2}} & \multicolumn{1}{c|}{8.68e+3}          & \textbf{-1.22e+2} \\ \cline{3-6} 
\multicolumn{1}{|c|}{}                        & \multicolumn{1}{c|}{}                                   & std    & \multicolumn{1}{c|}{\textbf{0.0}} & \multicolumn{1}{c|}{1.59e+2}          & \textbf{1.42e-14} \\ \cline{3-6} 
\multicolumn{1}{|c|}{}                        & \multicolumn{1}{c|}{}                                   & median & \multicolumn{1}{c|}{\textbf{8.64e+2}} & \multicolumn{1}{c|}{8.60e+3}          & \textbf{-1.22e+2} \\ \hline
\multicolumn{1}{|c|}{\multirow{6}{*}{IMPALA}} & \multicolumn{1}{c|}{\multirow{3}{*}{\textbf{V\_TRACE}}} & mean   & \multicolumn{1}{c|}{\textbf{8.64e+2}} & \multicolumn{1}{c|}{6.62e+3}          & -3.96e+2          \\ \cline{3-6} 
\multicolumn{1}{|c|}{}                        & \multicolumn{1}{c|}{}                                   & std    & \multicolumn{1}{c|}{\textbf{0.0}} & \multicolumn{1}{c|}{1.48e+2}          & 3.99e+0           \\ \cline{3-6} 
\multicolumn{1}{|c|}{}                        & \multicolumn{1}{c|}{}                                   & median & \multicolumn{1}{c|}{\textbf{8.64e+2}} & \multicolumn{1}{c|}{6.59e+3}          & -3.96e+2          \\ \cline{2-6} 
\multicolumn{1}{|c|}{}                        & \multicolumn{1}{c|}{\multirow{3}{*}{\textbf{PPO}}}      & mean   & \multicolumn{1}{c|}{\textbf{8.64e+2}} & \multicolumn{1}{c|}{7.38e+3}          & -4.00e+2          \\ \cline{3-6} 
\multicolumn{1}{|c|}{}                        & \multicolumn{1}{c|}{}                                   & std    & \multicolumn{1}{c|}{\textbf{0.0}} & \multicolumn{1}{c|}{2.87e+2}          & 0.0           \\ \cline{3-6} 
\multicolumn{1}{|c|}{}                        & \multicolumn{1}{c|}{}                                   & median & \multicolumn{1}{c|}{\textbf{8.64e+2}} & \multicolumn{1}{c|}{7.38e+3}          & -4.00e+2          \\ \hline
\end{tabular}
\label{tab:timesrewardblockmaze}%
}
\begin{tablenotes}
     \item TA time and TE time refer to training time and testing time respectively.
 \end{tablenotes}
\end{table}

\begin{sidewaystable}
\centering
\caption{Cumulative reward, training, and testing times of the fine-tuned generalist agents on the PDR-based scheduling on 2\% data budget and the baseline (in bold are the values of generalist agents configurations with greater performance).}
\resizebox{0.9\textwidth}{!}{%
\begin{tabular}{|ccc|cccccc|}
\hline
\multicolumn{3}{|c|}{\multirow{2}{*}{Environment}}                                                              & \multicolumn{6}{c|}{PDR}                                                                                                                                                                                                   \\ \cline{4-9} 
\multicolumn{3}{|c|}{}                        & \multicolumn{3}{c|}{(6 x 6)}                                                                                           & \multicolumn{3}{c|}{(30 x 20)}                                                                    \\ \hline
\multicolumn{3}{|c|}{Metrics}                                                                                    & \multicolumn{1}{c|}{TA time}          & \multicolumn{1}{c|}{TE time}          & \multicolumn{1}{c|}{Cumulative reward} & \multicolumn{1}{c|}{TA time}          & \multicolumn{1}{c|}{TE time}          & Cumulative reward \\ \hline
\multicolumn{2}{|c|}{\multirow{3}{*}{\textbf{Baseline agent}}}                                          & mean   & \multicolumn{1}{c|}{3.96e+3}          & \multicolumn{1}{c|}{\textbf{1.05e+1}} & \multicolumn{1}{c|}{-5.73e+2}          & \multicolumn{1}{c|}{7.42e+4}          & \multicolumn{1}{c|}{\textbf{1.73e+2}} & -2.48e+3          \\ \cline{3-9} 
\multicolumn{2}{|c|}{}                                                                                  & std    & \multicolumn{1}{c|}{1.37e+3}          & \multicolumn{1}{c|}{\textbf{1.65e+0}} & \multicolumn{1}{c|}{3.27e+0}           & \multicolumn{1}{c|}{2.42e+4}          & \multicolumn{1}{c|}{\textbf{4.70e+0}} & 1.07e+0           \\ \cline{3-9} 
\multicolumn{2}{|c|}{}                                                                                  & median & \multicolumn{1}{c|}{4.58e+3}          & \multicolumn{1}{c|}{\textbf{9.64e+0}} & \multicolumn{1}{c|}{-5.74e+2}          & \multicolumn{1}{c|}{8.79e+4}          & \multicolumn{1}{c|}{\textbf{1.73e+2}} & -2.47e+3          \\ \hline
\multicolumn{1}{|c|}{\multirow{9}{*}{MGDT}}   & \multicolumn{1}{c|}{\multirow{3}{*}{\textbf{MAENT}}}    & mean   & \multicolumn{1}{c|}{1.30e+3}          & \multicolumn{1}{c|}{1.44e+2}          & \multicolumn{1}{c|}{\textbf{-3.91e+2}} & \multicolumn{1}{c|}{4.20e+4}          & \multicolumn{1}{c|}{2.57e+3}          & \textbf{-1.26e+3} \\ \cline{3-9} 
\multicolumn{1}{|c|}{}                        & \multicolumn{1}{c|}{}                                   & std    & \multicolumn{1}{c|}{1.30e+1}          & \multicolumn{1}{c|}{6.52e+0}          & \multicolumn{1}{c|}{\textbf{0.0}}  & \multicolumn{1}{c|}{1.31e+4}          & \multicolumn{1}{c|}{8.56e+1}          & \textbf{1.20e+0}  \\ \cline{3-9} 
\multicolumn{1}{|c|}{}                        & \multicolumn{1}{c|}{}                                   & median & \multicolumn{1}{c|}{1.31e+3}          & \multicolumn{1}{c|}{1.40e+2}          & \multicolumn{1}{c|}{\textbf{-3.91e+2}} & \multicolumn{1}{c|}{4.77e+4}          & \multicolumn{1}{c|}{2.53e+3}          & \textbf{-1.26e+3} \\ \cline{2-9} 
\multicolumn{1}{|c|}{}                        & \multicolumn{1}{c|}{\multirow{3}{*}{\textbf{DQN}}}      & mean   & \multicolumn{1}{c|}{4.57e+3}          & \multicolumn{1}{c|}{1.41e+2}          & \multicolumn{1}{c|}{-3.92e+2}          & \multicolumn{1}{c|}{1.43e+5}          & \multicolumn{1}{c|}{2.50e+3}          & -1.26e+3 \\ \cline{3-9} 
\multicolumn{1}{|c|}{}                        & \multicolumn{1}{c|}{}                                   & std    & \multicolumn{1}{c|}{3.63e+1}          & \multicolumn{1}{c|}{2.12e+0}          & \multicolumn{1}{c|}{0.0}           & \multicolumn{1}{c|}{2.04e+4}          & \multicolumn{1}{c|}{5.64e+1}          & 1.71e-1  \\ \cline{3-9} 
\multicolumn{1}{|c|}{}                        & \multicolumn{1}{c|}{}                                   & median & \multicolumn{1}{c|}{4.56e+3}          & \multicolumn{1}{c|}{1.42e+2}          & \multicolumn{1}{c|}{-3.92e+2}          & \multicolumn{1}{c|}{1.40e+5}          & \multicolumn{1}{c|}{2.53e+3}          & -1.26e+3 \\ \cline{2-9} 
\multicolumn{1}{|c|}{}                        & \multicolumn{1}{c|}{\multirow{3}{*}{\textbf{PPO}}}      & mean   & \multicolumn{1}{c|}{1.56e+4}          & \multicolumn{1}{c|}{1.46e+2}          & \multicolumn{1}{c|}{-3.93e+2}          & \multicolumn{1}{c|}{2.77e+5}          & \multicolumn{1}{c|}{2.06e+3}          & -1.26e+3 \\ \cline{3-9} 
\multicolumn{1}{|c|}{}                        & \multicolumn{1}{c|}{}                                   & std    & \multicolumn{1}{c|}{7.15e+1}          & \multicolumn{1}{c|}{6.49e-1}          & \multicolumn{1}{c|}{0.0}           & \multicolumn{1}{c|}{8.26e+4}          & \multicolumn{1}{c|}{5.30e+2}          & 3.47e-1  \\ \cline{3-9} 
\multicolumn{1}{|c|}{}                        & \multicolumn{1}{c|}{}                                   & median & \multicolumn{1}{c|}{1.56e+4}          & \multicolumn{1}{c|}{1.46e+2}          & \multicolumn{1}{c|}{-3.93e+2}          & \multicolumn{1}{c|}{3.35e+5}          & \multicolumn{1}{c|}{2.46e+3}          & -1.26e+3 \\ \hline
\multicolumn{1}{|c|}{\multirow{6}{*}{IMPALA}} & \multicolumn{1}{c|}{\multirow{3}{*}{\textbf{V\_TRACE}}} & mean   & \multicolumn{1}{c|}{1.20e+2}          & \multicolumn{1}{c|}{4.94e+1}          & \multicolumn{1}{c|}{-3.92e+2}          & \multicolumn{1}{c|}{\textbf{2.60e+3}} & \multicolumn{1}{c|}{9.73e+3}          & -1.43e+3          \\ \cline{3-9} 
\multicolumn{1}{|c|}{}                        & \multicolumn{1}{c|}{}                                   & std    & \multicolumn{1}{c|}{6.07e+0}          & \multicolumn{1}{c|}{9.60e-1}          & \multicolumn{1}{c|}{6.27e-1}           & \multicolumn{1}{c|}{\textbf{9.99e+2}} & \multicolumn{1}{c|}{3.71e+2}          & 3.00e+2           \\ \cline{3-9} 
\multicolumn{1}{|c|}{}                        & \multicolumn{1}{c|}{}                                   & median & \multicolumn{1}{c|}{1.22e+2}          & \multicolumn{1}{c|}{4.99e+1}          & \multicolumn{1}{c|}{-3.92e+2}          & \multicolumn{1}{c|}{\textbf{2.05e+3}} & \multicolumn{1}{c|}{9.75e+3}          & -1.26e+3          \\ \cline{2-9} 
\multicolumn{1}{|c|}{}                        & \multicolumn{1}{c|}{\multirow{3}{*}{\textbf{PPO}}}      & mean   & \multicolumn{1}{c|}{\textbf{1.10e+2}} & \multicolumn{1}{c|}{4.85e+1}          & \multicolumn{1}{c|}{-3.92e+2}          & \multicolumn{1}{c|}{2.87e+3}          & \multicolumn{1}{c|}{9.50e+3}          & -1.26e+3          \\ \cline{3-9} 
\multicolumn{1}{|c|}{}                        & \multicolumn{1}{c|}{}                                   & std    & \multicolumn{1}{c|}{\textbf{5.67e+0}} & \multicolumn{1}{c|}{1.15e+0}          & \multicolumn{1}{c|}{4.26e-1}           & \multicolumn{1}{c|}{2.49e+2}          & \multicolumn{1}{c|}{4.45e+2}          & 2.55e-1           \\ \cline{3-9} 
\multicolumn{1}{|c|}{}                        & \multicolumn{1}{c|}{}                                   & median & \multicolumn{1}{c|}{\textbf{1.07e+2}} & \multicolumn{1}{c|}{4.88e+1}          & \multicolumn{1}{c|}{-3.92e+2}          & \multicolumn{1}{c|}{2.76e+3}          & \multicolumn{1}{c|}{9.75e+3}          & -1.26e+3          \\ \hline
\end{tabular}
\label{tab:timesrewardpdr}%
}
\begin{tablenotes}
     \item TA time and TE time refer to training time and testing time respectively.
 \end{tablenotes}
\end{sidewaystable}

Regarding the (6 $\times$ 6) instance of the PDR task, we fine-tuned the pre-trained generalist agents using zero-shot, for 320 and 720 steps. As for the (30 $\times$ 20) instance of the PDR task, we fine-tuned the generalist agents in zero-shot setting, for 60,000 and 120,000 steps. For both instances, we collected the makespan, the training time, the testing time, as well as the average cumulative reward earned. We reported the results of our experiments in Table \ref{tab:bugmakespanpdr}, \ref{tab:timesrewardpdr}, \ref{tab:bugmakespan0fpdr}, \ref{tab:timesreward0fpdr},\ref{tab:bugmakespan100fpdr} and \ref{tab:timesreward100fpdr}. 

\begin{figure}
    \centering
    \includegraphics[width=0.7\textwidth]{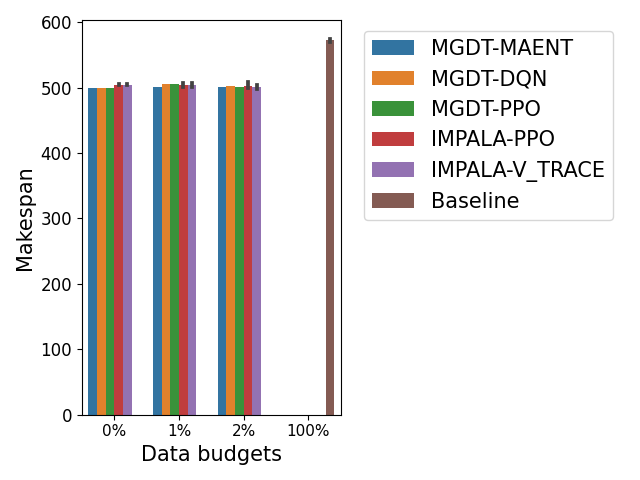}
    \caption{Makespan performance by the baseline \cite{zhang2020learning}, the MGDT and IMPALA agents  configurations on the 6 $\times$ 6 instance}
    \label{fig:Makespan performance by both MGDT and IMPALA generalist agents and their configurations66}
\end{figure}

\begin{figure}
    \centering
    \includegraphics[width=0.7\textwidth]{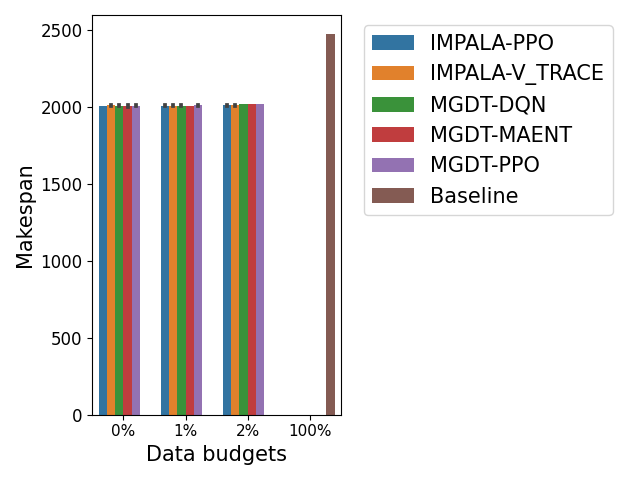}
    \caption{Makespan performance by the baseline \cite{zhang2020learning}, the MGDT and IMPALA agents  configurations on the 30 $\times$ 20 instance}
    \label{fig:Makespan performance by both MGDT and IMPALA generalist agents and their configurations3020}
\end{figure}

In terms of makespan performance, the generalist agent configurations have similar average performance across all three fine-tuning data budgets as shown in Figure \ref{fig:Makespan performance by both MGDT and IMPALA generalist agents and their configurations66} and \ref{fig:Makespan performance by both MGDT and IMPALA generalist agents and their configurations3020}.

\begin{figure}
    \centering
    \includegraphics[width=0.7\textwidth]{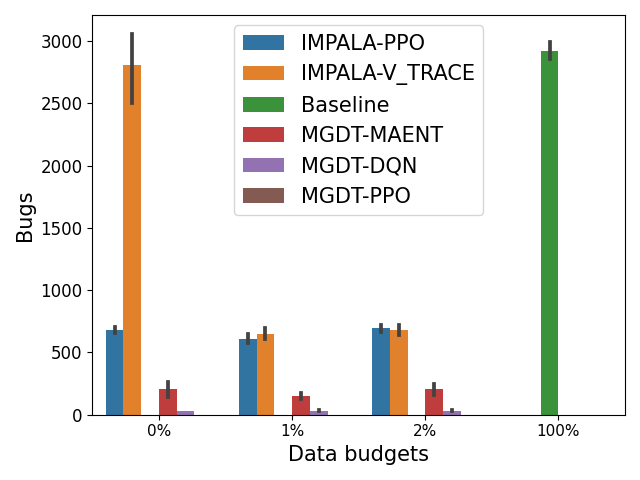}
    \caption{Number of bugs detected by the MGDT and IMPALA  agents configurations and the baseline on the MsPacman game }
    \label{fig:Number of bugs detected by the MGDT agent configurations and the baseline on the MsPacman game}
\end{figure}

Regarding the MsPacman game, we fine-tuned the pre-trained generalist agents using zero-shot, for 10 and 20 episodes as mentioned in Section \ref{studydesign}. Similarly to the previous tasks, we collected the number of bugs detected, the average cumulative reward, the training time, and the time taken to test generalist agents after training. We report the results of our experiments in Table \ref{tab:bugmakespan}, \ref{tab:timesreward}, \ref{tab:bugmakespan0f}, \ref{tab:timesreward0f},\ref{tab:bugmakespan100f} and \ref{tab:timesreward100f}. In terms of bugs detected, IMPALA-V\_TRACE configuration performs best among the generalist agent configurations during the zero-shot fine-tuning data budgets as shown in Figure \ref{fig:Number of bugs detected by the MGDT agent configurations and the baseline on the MsPacman game}.

In the next subsections, we discuss in detail the performance of the fine-tuned generalist agents in terms of the metrics mentioned in Section \ref{sec:Evaluation metrics}.

\begin{tcolorbox}
    \textbf{Finding 1: 
    Generalist agents show promising performance on the studied SE tasks, specifically the detection of bugs in two games and the minimizing makespan of PDR-based scheduling across all fine-tuning data budgets. To leverage generalist agents on the studied SE tasks we use different model-free DRL algorithms, whose performance varies based on the architecture and the pre-training settings of the generalist agents. Our results also show that at zero-shot, the performance of the generalist agents can be sufficient for achieving desirable performance.}
\end{tcolorbox}
\subsection{\textbf{RQ2: How do different pre-trained DRL generalist agents perform on SE tasks compared to DRL specialist agents?}} \label{sec:rq2}
In Tables \ref{tab:bugmakespan},\ref{tab:bugmakespanblockmaze},\ref{tab:bugmakespanpdr} \ref{tab:bugmakespan0f},\ref{tab:bugmakespan0fblockmaze},\ref{tab:bugmakespan0fpdr},\ref{tab:bugmakespan100f},\ref{tab:bugmakespan100fblockmaze} and \ref{tab:bugmakespan100fpdr}, we report, over 5 independent runs, the average number of bugs detected on MsPacman and the Blockmaze games by the fine-tuned generalist agents, as well as their makespan performance on the PDR-based scheduling task.

Regarding the Blockmaze game, at the zero-shot fine-tuning data budget, we observe that the MGDT generalist agents perform better than Wuji \cite{zheng2019wuji} on average in terms of number of bugs detected. The MGDT generalist agents detected 11\% more bugs of type 1 and 86\% more bugs of type 2. The type 2 bugs are mostly located on the lower side of the maze, suggesting a good exploration capability of the MGDT agent configurations in that area.
\begin{table}
\centering
\caption{Performance of the pre-trained generalist agents at zero-shot fine-tuning data budget on MsPacman game and the baseline in terms of bugs detected  (in bold are the values of generalist agents configurations with greater performance).}
\resizebox{0.9\textwidth}{!}{%
\begin{tabular}{|ccc|cccc|}
\hline
\multicolumn{3}{|c|}{Environment}                                                                               & \multicolumn{4}{c|}{MsPacman}                                                                                                            \\ \hline
\multicolumn{3}{|c|}{Metrics}                                                                                    & \multicolumn{1}{c|}{Type 1}           & \multicolumn{1}{c|}{Type 2}           & \multicolumn{1}{c|}{Type 3}           & Type 4           \\ \hline
\multicolumn{2}{|c|}{\multirow{3}{*}{\textbf{Baseline agent}}}                                          & mean   & \multicolumn{1}{c|}{\textbf{7.85e+2}} & \multicolumn{1}{c|}{\textbf{7.72e+2}} & \multicolumn{1}{c|}{6.89e+2}          & 6.75e+2          \\ \cline{3-7} 
\multicolumn{2}{|c|}{}                                                                                  & std    & \multicolumn{1}{c|}{\textbf{3.41e+1}} & \multicolumn{1}{c|}{\textbf{3.35e+1}} & \multicolumn{1}{c|}{4.36e+1}          & 4.68e+1          \\ \cline{3-7} 
\multicolumn{2}{|c|}{}                                                                                  & median & \multicolumn{1}{c|}{\textbf{7.87e+2}} & \multicolumn{1}{c|}{\textbf{7.66e+2}} & \multicolumn{1}{c|}{7.09e+2}          & 6.95e+2          \\ \hline
\multicolumn{1}{|c|}{\multirow{9}{*}{MGDT}}   & \multicolumn{1}{c|}{\multirow{3}{*}{MAENT}}             & mean   & \multicolumn{1}{c|}{0.0}          & \multicolumn{1}{c|}{0.0}          & \multicolumn{1}{c|}{1.47e+2}          & 5.60e+1          \\ \cline{3-7} 
\multicolumn{1}{|c|}{}                        & \multicolumn{1}{c|}{}                                   & std    & \multicolumn{1}{c|}{0.0}          & \multicolumn{1}{c|}{0.0}          & \multicolumn{1}{c|}{5.04e+1}          & 1.87e+1          \\
\hline
\end{tabular}
\label{tab:bugmakespan0f}%
}
\end{table}
\begin{table}
\centering
\caption{Performance of the pre-trained generalist agents at zero-shot fine-tuning data budget on the Blockmaze game and the baseline in terms of bugs detected (in bold are the values of generalist agents configurations with greater performance).}
\resizebox{0.8\textwidth}{!}{%
\begin{tabular}{|ccc|cc|}
\hline
\multicolumn{3}{|c|}{Environment}                                                                            & \multicolumn{2}{c|}{Blockmaze}                           \\ \hline
\multicolumn{3}{|c|}{Metrics}                                                                                 & \multicolumn{1}{c|}{Type 1}           & Type 2           \\ \hline
\multicolumn{2}{|c|}{\multirow{3}{*}{Baseline agent}}                                                & mean   & \multicolumn{1}{c|}{1.80e+0}          & 2.00e+0          \\ \cline{3-5} 
\multicolumn{2}{|c|}{}                                                                               & std    & \multicolumn{1}{c|}{4.00e-1}          & 0.0          \\ \cline{3-5} 
\multicolumn{2}{|c|}{}                                                                               & median & \multicolumn{1}{c|}{2.00e+0}          & 2.00e+0          \\ \hline
\multicolumn{1}{|c|}{\multirow{9}{*}{MGDT}}   & \multicolumn{1}{c|}{\multirow{3}{*}{\textbf{MAENT}}} & mean   & \multicolumn{1}{c|}{\textbf{3.00e+0}} & \textbf{5.00e+0} \\ \cline{3-5} 
\multicolumn{1}{|c|}{}                        & \multicolumn{1}{c|}{}                                & std    & \multicolumn{1}{c|}{\textbf{0.0}} & \textbf{0.0} \\ \cline{3-5} 
\multicolumn{1}{|c|}{}                        & \multicolumn{1}{c|}{}                                & median & \multicolumn{1}{c|}{\textbf{3.00e+0}} & \textbf{5.00e+0} \\
\hline
\end{tabular}
\label{tab:bugmakespan0fblockmaze}%
}
\end{table}
\begin{table}
\centering
\caption{Performance of the pre-trained generalist agents at zero-shot fine-tuning data budget on the PDR-based scheduling and the baseline in terms of makespan time (in bold are the values of generalist agents configurations with greater performance).}
\resizebox{0.8\textwidth}{!}{%
\begin{tabular}{|ccc|cc|}
\hline
\multicolumn{3}{|c|}{Environment}                                                                                & \multicolumn{2}{c|}{PDR}                                  \\ \hline
\multicolumn{3}{|c|}{Metrics}                                                                                    & \multicolumn{1}{c|}{(6 x 6)}           & (30 x 20)        \\ \hline
\multicolumn{2}{|c|}{\multirow{3}{*}{Baseline agent}}                                                   & mean   & \multicolumn{1}{c|}{5.73e+2}           & 2.48e+3          \\ \cline{3-5} 
\multicolumn{2}{|c|}{}                                                                                  & std    & \multicolumn{1}{c|}{3.27e+0}           & 1.07e+0          \\ \cline{3-5} 
\multicolumn{2}{|c|}{}                                                                                  & median & \multicolumn{1}{c|}{5.74e+2}           & 2.47e+3          \\ \hline
\multicolumn{1}{|c|}{\multirow{9}{*}{MGDT}}   & \multicolumn{1}{c|}{\multirow{3}{*}{MAENT}}             & mean   & \multicolumn{1}{c|}{\textbf{5.00e+2}}  & 2.02e+3          \\ \cline{3-5} 
\multicolumn{1}{|c|}{}                        & \multicolumn{1}{c|}{}                                   & std    & \multicolumn{1}{c|}{\textbf{5.68e-14}} & 0.0          \\ \cline{3-5} 
\multicolumn{1}{|c|}{}                        & \multicolumn{1}{c|}{}                                   & median & \multicolumn{1}{c|}{\textbf{5.00e+2}}  & 2.02e+3          \\
\hline
\end{tabular}
\label{tab:bugmakespan0fpdr}%
}
\end{table}
On the 1\% and 2\% fine-tuning data budgets, we observe that MGDT-DQN performs better than wuji\cite{zheng2019wuji} on average in terms of bugs detected.
\begin{table}
\centering
\caption{Cumulative reward, training and testing times performance of the pre-trained generalist agents at zero-shot fine-tuning data budget on MsPacman game and the baseline (in bold are the values of generalist agent configurations with greater performance).}
\resizebox{0.9\textwidth}{!}{%
\begin{tabular}{|ccc|ccc|}
\hline
\multicolumn{3}{|c|}{\multirow{2}{*}{Environment}}                                                              & \multicolumn{3}{c|}{\multirow{2}{*}{MsPacman}}                                           \\
\multicolumn{3}{|c|}{}                        & \multicolumn{3}{c|}{}        \\ \hline
\multicolumn{3}{|c|}{Metrics}                                                                                    & \multicolumn{1}{c|}{TA time} & \multicolumn{1}{c|}{TE time}          & Cumulative reward \\ \hline
\multicolumn{2}{|c|}{\multirow{3}{*}{\textbf{Baseline agent}}}                                          & mean   & \multicolumn{1}{c|}{8.14e+3} & \multicolumn{1}{c|}{\textbf{2.57e+4}} & NA                \\ \cline{3-6} 
\multicolumn{2}{|c|}{}                                                                                  & std    & \multicolumn{1}{c|}{3.57e+2} & \multicolumn{1}{c|}{\textbf{1.08e+3}} & NA                \\ \cline{3-6} 
\multicolumn{2}{|c|}{}                                                                                  & median & \multicolumn{1}{c|}{7.99e+3} & \multicolumn{1}{c|}{\textbf{2.55e+4}} & NA                \\ \hline
\multicolumn{1}{|c|}{\multirow{9}{*}{MGDT}}   & \multicolumn{1}{c|}{\multirow{3}{*}{MAENT}}             & mean   & \multicolumn{1}{c|}{NA}      & \multicolumn{1}{c|}{3.51e+5}          & 1.93e+2           \\ \cline{3-6} 
\multicolumn{1}{|c|}{}                        & \multicolumn{1}{c|}{}                                   & std    & \multicolumn{1}{c|}{NA}      & \multicolumn{1}{c|}{1.84e+4}          & 4.21e+1           \\
\hline
\end{tabular}
\label{tab:timesreward0f}%
}
\begin{tablenotes}
     \item  TA time and TE time refer to training time and testing time respectively.
 \end{tablenotes}
\end{table}
\begin{table}
\centering
\caption{Cumulative reward, training and testing times performance of the pre-trained generalist agents at zero-shot fine-tuning data budget on the Blockmaze game and the baseline (in bold are the values of generalist agent configurations with greater performance).}
\resizebox{0.9\textwidth}{!}{%
\begin{tabular}{|ccc|ccc|}
\hline
\multicolumn{3}{|c|}{\multirow{2}{*}{Environment}}                                                            & \multicolumn{3}{c|}{\multirow{2}{*}{Blockmaze}}                                          \\
\multicolumn{3}{|c|}{}                        & \multicolumn{3}{c|}{}        \\ \hline
\multicolumn{3}{|c|}{Metrics}                                                                                 & \multicolumn{1}{c|}{TA time} & \multicolumn{1}{c|}{TE time}          & Cumulative reward \\ \hline
\multicolumn{2}{|c|}{\multirow{3}{*}{\textbf{Baseline agent}}}                                       & mean   & \multicolumn{1}{c|}{4.32e+4} & \multicolumn{1}{c|}{\textbf{8.82e+2}} & NA                \\ \cline{3-6} 
\multicolumn{2}{|c|}{}                                                                               & std    & \multicolumn{1}{c|}{0.0} & \multicolumn{1}{c|}{\textbf{2.76e+2}} & NA                \\ \cline{3-6} 
\multicolumn{2}{|c|}{}                                                                               & median & \multicolumn{1}{c|}{4.32e+4} & \multicolumn{1}{c|}{\textbf{9.60e+2}} & NA                \\ \hline
\multicolumn{1}{|c|}{\multirow{9}{*}{MGDT}}   & \multicolumn{1}{c|}{\multirow{3}{*}{\textbf{MAENT}}} & mean   & \multicolumn{1}{c|}{NA}      & \multicolumn{1}{c|}{2.15e+3}          & \textbf{-1.49e+2} \\ \cline{3-6} 
\multicolumn{1}{|c|}{}                        & \multicolumn{1}{c|}{}                                & std    & \multicolumn{1}{c|}{NA}      & \multicolumn{1}{c|}{7.78e+1}          & \textbf{0.0}  \\
\hline
\end{tabular}
\label{tab:timesreward0fblockmaze}%
}
\begin{tablenotes}
     \item  TA time and TE time refer to training time and testing time respectively.
 \end{tablenotes}
\end{table}
\begin{sidewaystable}
\centering
\caption{Cumulative reward, training and testing times performance of the pre-trained generalist agents at zero-shot fine-tuning data budget on the PDR-based scheduling and the baseline (in bold are the values of generalist agent configurations with greater performance).}
\resizebox{0.9\textwidth}{!}{%
\begin{tabular}{|ccc|cccccc|}
\hline
\multicolumn{3}{|c|}{\multirow{2}{*}{Environment}}                                                            & \multicolumn{6}{c|}{PDR}                                                                                                                                                                                 \\ \cline{4-9} 
\multicolumn{3}{|c|}{}                        & \multicolumn{3}{c|}{(6 x 6)}                                                                                  & \multicolumn{3}{c|}{(30 x 20)}                                                           \\ \hline
\multicolumn{3}{|c|}{Metrics}                                                                                 & \multicolumn{1}{c|}{TA time} & \multicolumn{1}{c|}{TE time}          & \multicolumn{1}{c|}{Cumulative reward} & \multicolumn{1}{c|}{TA time} & \multicolumn{1}{c|}{TE time}          & Cumulative reward       \\ \hline
\multicolumn{2}{|c|}{\multirow{3}{*}{\textbf{Baseline agent}}}                                       & mean   & \multicolumn{1}{c|}{3.96e+3} & \multicolumn{1}{c|}{\textbf{1.05e+1}} & \multicolumn{1}{c|}{-5.73e+2}          & \multicolumn{1}{c|}{7.42e+4} & \multicolumn{1}{c|}{\textbf{1.73e+2}} & -2.48e+3          \\ \cline{3-9} 
\multicolumn{2}{|c|}{}                                                                               & std    & \multicolumn{1}{c|}{1.37e+3} & \multicolumn{1}{c|}{\textbf{1.65e+0}} & \multicolumn{1}{c|}{3.27e+0}           & \multicolumn{1}{c|}{2.42e+4} & \multicolumn{1}{c|}{\textbf{4.70e+0}} & 1.07e+0           \\ \cline{3-9} 
\multicolumn{2}{|c|}{}                                                                               & median & \multicolumn{1}{c|}{4.58e+3} & \multicolumn{1}{c|}{\textbf{9.64e+0}} & \multicolumn{1}{c|}{-5.74e+2}          & \multicolumn{1}{c|}{8.79e+4} & \multicolumn{1}{c|}{\textbf{1.73e+2}} & -2.47e+3          \\
\hline
\end{tabular}
\label{tab:timesreward0fpdr}%
}
\begin{tablenotes}
     \item  TA time and TE time refer to training time and testing time respectively.
 \end{tablenotes}
\end{sidewaystable}
In terms of training time, reported in Table  \ref{tab:timesrewardblockmaze}, \ref{tab:timesreward0fblockmaze}, and \ref{tab:timesreward100fblockmaze}), MGDT and IMPALA were fine-tuned on the Blockmaze game, at zero-shot fine-tuning data budget, for 432 seconds and 864 seconds which are 1\% and 2\% of the training time performed by the wuji respectively. As mentioned in Section \ref{sec:Baselines studies}, we do not compare the generalist agents against the wuji in terms of cumulative reward earned as wuji's authors did not report them \cite{zheng2019wuji}. In terms of testing time, the baseline wuji performs best.\\
\begin{table}
\centering
\caption{Performance of the fine-tuned generalist agents on MsPacman game on 1\% data budget and the baseline in terms of bugs detected  (in bold are the values of generalist agent configurations with greater performance).}
\resizebox{\textwidth}{!}{%
\begin{tabular}{|ccc|cccc|}
\hline
\multicolumn{3}{|c|}{Environments}                                                                      & \multicolumn{4}{c|}{MsPacman}                                                                                                            \\ \hline
\multicolumn{3}{|c|}{Metrics}                                                                           & \multicolumn{1}{c|}{Type 1}           & \multicolumn{1}{c|}{Type 2}           & \multicolumn{1}{c|}{Type 3}           & Type 4           \\ \hline
\multicolumn{2}{|c|}{\multirow{3}{*}{\textbf{Baseline agent}}}                                 & mean   & \multicolumn{1}{c|}{\textbf{7.85e+2}} & \multicolumn{1}{c|}{\textbf{7.72e+2}} & \multicolumn{1}{c|}{\textbf{6.89e+2}} & \textbf{6.75e+2} \\ \cline{3-7} 
\multicolumn{2}{|c|}{}                                                                         & std    & \multicolumn{1}{c|}{\textbf{3.41e+1}} & \multicolumn{1}{c|}{\textbf{3.35e+1}} & \multicolumn{1}{c|}{\textbf{4.36e+1}} & \textbf{4.68e+1} \\ \cline{3-7} 
\multicolumn{2}{|c|}{}                                                                         & median & \multicolumn{1}{c|}{\textbf{7.87e+2}} & \multicolumn{1}{c|}{\textbf{7.66e+2}} & \multicolumn{1}{c|}{\textbf{7.09e+2}} & \textbf{6.95e+2} \\ \hline
\multicolumn{1}{|c|}{\multirow{9}{*}{MGDT}}   & \multicolumn{1}{c|}{\multirow{3}{*}{MAENT}}    & mean   & \multicolumn{1}{c|}{0.0}          & \multicolumn{1}{c|}{0.0}          & \multicolumn{1}{c|}{1.05e+2}          & 4.38e+1          \\ \cline{3-7} 
\multicolumn{1}{|c|}{}                        & \multicolumn{1}{c|}{}                          & std    & \multicolumn{1}{c|}{0.0}          & \multicolumn{1}{c|}{0.0}          & \multicolumn{1}{c|}{2.13e+1}          & 5.63e+0          \\
\hline
\end{tabular}
\label{tab:bugmakespan100f}%
}
\end{table}
\begin{table}
\centering
\caption{Performance (in terms of bugs detected) of the fine-tuned generalist agents on the Blockmaze game on 1\% data budget and the baseline (in bold are the values of generalist agent configurations with greater performance).}
\resizebox{0.8\textwidth}{!}{%
\begin{tabular}{|ccc|cc|}
\hline
\multicolumn{3}{|c|}{Environments}                                                                          & \multicolumn{2}{c|}{Blockmaze}                           \\ \hline
\multicolumn{3}{|c|}{Metrics}                                                                               & \multicolumn{1}{c|}{Type 1}           & Type 2           \\ \hline
\multicolumn{2}{|c|}{\multirow{3}{*}{Baseline agent}}                                              & mean   & \multicolumn{1}{c|}{1.80e+0}          & 2.00e+0          \\ \cline{3-5} 
\multicolumn{2}{|c|}{}                                                                             & std    & \multicolumn{1}{c|}{4.00e-1}          & 0.0          \\ \cline{3-5} 
\multicolumn{2}{|c|}{}                                                                             & median & \multicolumn{1}{c|}{2.00e+0}          & 2.00e+0          \\ \hline
\multicolumn{1}{|c|}{\multirow{9}{*}{MGDT}}   & \multicolumn{1}{c|}{\multirow{3}{*}{MAENT}}        & mean   & \multicolumn{1}{c|}{0.0}          & 0.0          \\ \cline{3-5} 
\multicolumn{1}{|c|}{}                        & \multicolumn{1}{c|}{}                              & std    & \multicolumn{1}{c|}{0.0}          & 0.0          \\ \cline{3-5} 
\multicolumn{1}{|c|}{}                        & \multicolumn{1}{c|}{}                              & median & \multicolumn{1}{c|}{0.0}          & 0.0          \\
\hline
\end{tabular}
\label{tab:bugmakespan100fblockmaze}%
}
\end{table}
\begin{table}
\centering
\caption{Performance of the fine-tuned  generalist agents on the PDR-based scheduling on 1\% data budget and the baseline in terms of  makespan time (in bold are the values of generalist agent configurations with greater performance).}
\resizebox{0.8\textwidth}{!}{%
\begin{tabular}{|ccc|cc|}
\hline
\multicolumn{3}{|c|}{Environment}                                                                             & \multicolumn{2}{c|}{PDR}                                 \\ \hline
\multicolumn{3}{|c|}{Metrics}                                                                                 & \multicolumn{1}{c|}{(6 x 6)}          & (30 x 20)        \\ \hline
\multicolumn{2}{|c|}{\multirow{3}{*}{Baseline agent}}                                                & mean   & \multicolumn{1}{c|}{5.73e+2}          & 2.48e+3          \\ \cline{3-5} 
\multicolumn{2}{|c|}{}                                                                               & std    & \multicolumn{1}{c|}{3.27e+0}          & 1.07e+0          \\ \cline{3-5} 
\multicolumn{2}{|c|}{}                                                                               & median & \multicolumn{1}{c|}{5.74e+2}          & 2.47e+3          \\ \hline
\multicolumn{1}{|c|}{\multirow{9}{*}{MGDT}}   & \multicolumn{1}{c|}{\multirow{3}{*}{\textbf{MAENT}}} & mean   & \multicolumn{1}{c|}{\textbf{5.00e+2}} & \textbf{2.01e+3} \\ \cline{3-5} 
\multicolumn{1}{|c|}{}                        & \multicolumn{1}{c|}{}                                & std    & \multicolumn{1}{c|}{\textbf{0.0}} & \textbf{0.0} \\ \cline{3-5} 
\multicolumn{1}{|c|}{}                        & \multicolumn{1}{c|}{}                                & median & \multicolumn{1}{c|}{\textbf{5.00e+2}} & \textbf{2.01e+3} \\
\hline
\end{tabular}
\label{tab:bugmakespan100fpdr}%
}
\end{table}
\textbf{Statistical analysis:} Statistically, the difference between the performance of some of the MGDT agents and wuji, in terms of bugs detected, is significant with regard to our experiments.
\begin{table}
\centering
\caption{Cumulative reward, training, and testing times performance of the fine-tuned generalist agents on MsPacman game on 1\% data budget and the baseline (in bold are the values of generalist agent configurations with greater performance).}
\resizebox{\textwidth}{!}{%
\begin{tabular}{|ccc|ccc|}
\hline
\multicolumn{3}{|c|}{\multirow{2}{*}{Environment}}                                                            & \multicolumn{3}{c|}{\multirow{2}{*}{MsPacman}}                                                    \\
\multicolumn{3}{|c|}{}                        & \multicolumn{3}{c|}{}                 \\ \hline
\multicolumn{3}{|c|}{Metrics}                                                                                 & \multicolumn{1}{c|}{TA time}          & \multicolumn{1}{c|}{TE time}          & Cumulative reward \\ \hline
\multicolumn{2}{|c|}{\multirow{3}{*}{\textbf{Baseline agent}}}                                       & mean   & \multicolumn{1}{c|}{8.14e+3}          & \multicolumn{1}{c|}{\textbf{2.57e+4}} & NA                \\ \cline{3-6} 
\multicolumn{2}{|c|}{}                                                                               & std    & \multicolumn{1}{c|}{3.57e+2}          & \multicolumn{1}{c|}{\textbf{1.08e+3}} & NA                \\ \cline{3-6} 
\multicolumn{2}{|c|}{}                                                                               & median & \multicolumn{1}{c|}{7.99e+3}          & \multicolumn{1}{c|}{\textbf{2.55e+4}} & NA                \\ \hline
\multicolumn{1}{|c|}{\multirow{9}{*}{MGDT}}   & \multicolumn{1}{c|}{\multirow{3}{*}{\textbf{MAENT}}} & mean   & \multicolumn{1}{c|}{4.87e+3}          & \multicolumn{1}{c|}{4.23e+5}          & \textbf{1.69e+2}  \\ \cline{3-6} 
\multicolumn{1}{|c|}{}                        & \multicolumn{1}{c|}{}                                & std    & \multicolumn{1}{c|}{2.67e+2}          & \multicolumn{1}{c|}{6.37e+4}          & \textbf{2.93e+1}  \\
\hline
\end{tabular}

\label{tab:timesreward100f}%
}
\begin{tablenotes}
     \item TA time and TE time refer to training time and testing time respectively.
 \end{tablenotes}
\end{table}
\begin{table}
\centering
\caption{Cumulative reward, training and testing times performance of the fine-tuned generalist agents on the Blockmaze game on 1\% data budget and the baseline (in bold are the values of generalist agent configurations with greater performance).}
\resizebox{0.9\textwidth}{!}{%
\begin{tabular}{|ccc|ccc|}
\hline
\multicolumn{3}{|c|}{\multirow{2}{*}{Environment}}                                                               & \multicolumn{3}{c|}{\multirow{2}{*}{Blockmaze}}                                                   \\
\multicolumn{3}{|c|}{}                        & \multicolumn{3}{c|}{}                 \\ \hline
\multicolumn{3}{|c|}{Metrics}                                                                                    & \multicolumn{1}{c|}{TA time}          & \multicolumn{1}{c|}{TE time}          & Cumulative reward \\ \hline
\multicolumn{2}{|c|}{\multirow{3}{*}{\textbf{Baseline agent}}}                                          & mean   & \multicolumn{1}{c|}{4.32e+4}          & \multicolumn{1}{c|}{\textbf{8.82e+2}} & NA                \\ \cline{3-6} 
\multicolumn{2}{|c|}{}                                                                                  & std    & \multicolumn{1}{c|}{0.0}          & \multicolumn{1}{c|}{\textbf{2.76e+2}} & \textbf{NA}       \\ \cline{3-6} 
\multicolumn{2}{|c|}{}                                                                                  & median & \multicolumn{1}{c|}{4.32e+4}          & \multicolumn{1}{c|}{\textbf{9.60e+2}} & NA                \\ \hline
\multicolumn{1}{|c|}{\multirow{9}{*}{MGDT}}   & \multicolumn{1}{c|}{\multirow{3}{*}{\textbf{MAENT}}}    & mean   & \multicolumn{1}{c|}{\textbf{4.32e+2}} & \multicolumn{1}{c|}{8.11e+3}          & -1.42e+2          \\ \cline{3-6} 
\multicolumn{1}{|c|}{}                        & \multicolumn{1}{c|}{}                                   & std    & \multicolumn{1}{c|}{\textbf{0.0}} & \multicolumn{1}{c|}{2.20e+2}          & 7.76e+1           \\
\hline
\end{tabular}

\label{tab:timesreward100fblockmaze}%
}
\begin{tablenotes}
     \item TA time and TE time refer to training time and testing time respectively.
 \end{tablenotes}
\end{table}
\begin{sidewaystable}
\centering
\caption{Cumulative reward, training and testing times performance of the fine-tuned generalist agents on the PDR-based scheduling on 1\% data budget and the baseline (in bold are the values of generalist agent configurations with greater performance).}
\resizebox{\textwidth}{!}{%
\begin{tabular}{|ccc|cccccc|}
\hline
\multicolumn{3}{|c|}{\multirow{2}{*}{Environment}}                                                               & \multicolumn{6}{c|}{PDR}                                                                                                                                                                                                   \\ \cline{4-9} 
\multicolumn{3}{|c|}{}                        & \multicolumn{3}{c|}{(6 x 6)}                                                                                           & \multicolumn{3}{c|}{(30 x 20)}                                                                    \\ \hline
\multicolumn{3}{|c|}{Metrics}                                                                                    & \multicolumn{1}{c|}{TA time}          & \multicolumn{1}{c|}{TE time}          & \multicolumn{1}{c|}{Cumulative reward} & \multicolumn{1}{c|}{TA time}          & \multicolumn{1}{c|}{TE time}          & Cumulative reward \\ \hline
\multicolumn{2}{|c|}{\multirow{3}{*}{\textbf{Baseline agent}}}                                          & mean   & \multicolumn{1}{c|}{3.96e+3}          & \multicolumn{1}{c|}{\textbf{1.05e+1}} & \multicolumn{1}{c|}{-5.73e+2}          & \multicolumn{1}{c|}{7.42e+4}          & \multicolumn{1}{c|}{\textbf{1.73e+2}} & -2.48e+3          \\ \cline{3-9} 
\multicolumn{2}{|c|}{}                                                                                  & std    & \multicolumn{1}{c|}{1.37e+3}          & \multicolumn{1}{c|}{\textbf{1.65e+0}} & \multicolumn{1}{c|}{3.27e+0}           & \multicolumn{1}{c|}{2.42e+4}          & \multicolumn{1}{c|}{\textbf{4.70e+0}} & 1.07e+0           \\
\hline
\end{tabular}
\label{tab:timesreward100fpdr}%
}
\begin{tablenotes}
     \item TA time and TE time refer to training time and testing time respectively.
 \end{tablenotes}
\end{sidewaystable}
Table \ref{tab:Results of  post-hoc tests analysis of number of bugs detected by wuji(Baseline) and MGDT generalist agent} shows the results of post-hoc statistical test on wuji and the MGDT generalist agents in terms of bugs detected. Statistically, MGDT-DQN configuration performs best across the different percentage data budgets with CLES values close to 1 compared to the baseline. 
\begin{sidewaystable}
\caption{Results of post-hoc test analysis of the number of bugs detected by wuji and MGDT generalist agent for different fine-tuning data budgets (in bold are DRL configurations where the p-value is $<$ 0.05 and have greater performance w.r.t the effect size).}
\label{tab:Results of  post-hoc tests analysis of number of bugs detected by wuji(Baseline) and MGDT generalist agent}
\resizebox{\textwidth}{!}{
\centering
\begin{tabular}{|c|c|c|c|c|c|c|}
\hline
Data budgets & A                 & B                   & mean(A) & mean(B) & p-val    & CLES     \\ \hline
\multirow{6}{*}{zero-shot}      & Baseline          & \textbf{MGDT-DQN}   & 3.80e+0 & \textbf{8.00e+0} & \textbf{1.10e-14} & \textbf{1.48e-40} \\ \cline{2-7} 
                        & Baseline          & \textbf{MGDT-MAENT} & 3.80e+0 & \textbf{8.00e+0} & \textbf{1.10e-14} & \textbf{1.48e-40} \\ \cline{2-7} 
                        & Baseline          & \textbf{MGDT-PPO}   & 3.80e+0 & \textbf{8.00e+0} & \textbf{1.10e-14} & \textbf{1.48e-40} \\ \cline{2-7} 
                        & MGDT-DQN          & MGDT-MAENT          & 8.00e+0 & 8.00e+0 & 1.00e+0  & 5.00e-1  \\ \cline{2-7} 
                        & MGDT-DQN          & MGDT-PPO            & 8.00e+0 & 8.00e+0 & 1.00e+0  & 5.00e-1  \\ \cline{2-7} 
                        & MGDT-MAENT        & MGDT-PPO            & 8.00e+0 & 8.00e+0 & 1.00e+0  & 5.00e-1  \\ \hline
\multirow{6}{*}{1\%}      & Baseline          & \textbf{MGDT-DQN}   & 3.80e+0 & \textbf{8.00e+0} & 1.42e-11 & \textbf{2.23e-17} \\ \cline{2-7} 
                        & \textbf{Baseline} & MGDT-MAENT          & \textbf{3.80e+0} & 0.0 & \textbf{6.53e-11} & \textbf{1.00e+0}  \\ \cline{2-7} 
                        & Baseline          & MGDT-PPO            & 3.80e+0 & 4.40e+0 & 7.00e-2  & 1.15e-1  \\ \cline{2-7} 
                        & \textbf{MGDT-DQN} & MGDT-MAENT          & \textbf{8.00e+0} & 0.0 & \textbf{3.33e-16} & \textbf{1.00e+0}  \\ \cline{2-7} 
                        & \textbf{MGDT-DQN} & MGDT-PPO            & \textbf{8.00e+0} & 4.40e+0 & \textbf{1.48e-10} & \textbf{1.00e+0}  \\ \cline{2-7} 
                        & MGDT-MAENT        & \textbf{MGDT-PPO}   & 0.0 & \textbf{4.40e+0} & \textbf{6.93e-12} & \textbf{6.84e-19} \\ \hline
\multirow{6}{*}{2\%}      & Baseline          & \textbf{MGDT-DQN}   & 3.80e+0 & \textbf{8.00e+0} & \textbf{1.10e-14} & \textbf{1.48e-40} \\ \cline{2-7} 
                        & \textbf{Baseline} & MGDT-MAENT          & \textbf{3.80e+0} & 1.00e+0 & \textbf{6.31e-12} & \textbf{1.00e+0}  \\ \cline{2-7} 
                        & Baseline          & MGDT-PPO            & 3.80e+0 & 4.00e+0 & 5.09e-1  & 2.64e-1  \\ \cline{2-7} 
                        & \textbf{MGDT-DQN} & MGDT-MAENT          & \textbf{8.00e+0} & 1.00e+0 & \textbf{0.0}  & \textbf{1.00e+0}  \\ \cline{2-7} 
                        & \textbf{MGDT-DQN} & MGDT-PPO            & \textbf{8.00e+0} & 4.00e+0 & \textbf{2.41e-14} & \textbf{1.00e+0}  \\ \cline{2-7} 
                        & MGDT-MAENT        & \textbf{MGDT-PPO}   & 1.00e+0 & \textbf{4.00e+0} & \textbf{2.17e-12} & \textbf{1.19e-21} \\ \hline
\end{tabular}
}
\begin{tablenotes}
     \item mean(A) and mean(B) refer to the number of bugs detected.
 \end{tablenotes}
\end{sidewaystable}
Table \ref{tab:Results of post-hoc tests analysis of testing time performance by wuji(Baseline) and the MGDT generalist agent} shows the results of post-hoc tests for testing time of the baseline and the MGDT generalist agent. The baseline has lesser testing time with CLES values close to 1 regardless of the data budgets (time budgets in this case) that were allowed for fine-tuning the MGDT generalist agent. A possible explanation is the size of the MGDT model which has 197M parameters in comparison to the baseline model which has 68K parameters. Testing a bigger model takes longer than testing a smaller one.
\begin{tcolorbox}
    \textbf{Finding 2: Fine-tuning with the MGDT agent gives the best results in the Blockmaze game. The fact that MGDT outperforms Wuji at zero-shot, suggests that MGDT possesses pre-training abilities that can be applied effectively to new tasks without extensive fine-tuning. IMPALA agent underperforms MGDT agent on the Blockmaze game, suggesting that a scalable architecture is not sufficient for desirable performance transfer.}
\end{tcolorbox}
\begin{sidewaystable}
\caption{Results of post-hoc tests analysis of testing time performance by Wuji(Baseline) and the MGDT generalist agent (in bold are DRL configurations where p-value is $<$ 0.05 and have greater performance w.r.t the effect size).}
\label{tab:Results of post-hoc tests analysis of testing time performance by wuji(Baseline) and the MGDT generalist agent}
\centering
\resizebox{\textwidth}{!}{
\begin{tabular}{|c|c|c|c|c|c|c|}
\hline
Data budgets    & A                   & B          & mean(A)          & mean(B) & pval              & CLES               \\ \hline
\multirow{6}{*}{zero-shot} & \textbf{Baseline}   & MGDT-DQN   & \textbf{8.82e+2} & 2.15e+3 & \textbf{1.67e-3}  & \textbf{3.79e-5}   \\ \cline{2-7} 
                           & \textbf{Baseline}   & MGDT-MAENT & \textbf{8.82e+2} & 2.15e+3 & \textbf{1.67e-3}  & \textbf{3.79e-5}   \\ \cline{2-7} 
                           & \textbf{Baseline}   & MGDT-PPO   & \textbf{8.82e+2} & 2.15e+3 & \textbf{1.67e-3}  & \textbf{3.79e-5}   \\ \cline{2-7} 
                           & MGDT-DQN            & MGDT-MAENT & 2.15e+3          & 2.15e+3 & 1.00e+0           & 5.00e-1            \\ \cline{2-7} 
                           & MGDT-DQN            & MGDT-PPO   & 2.15e+3          & 2.15e+3 & 1.00e+0           & 5.00e-1            \\ \cline{2-7} 
                           & MGDT-MAENT          & MGDT-PPO   & 2.15e+3          & 2.15e+3 & 1.00e+0           & 5.00e-1            \\ \hline
\multirow{6}{*}{1\%}         & \textbf{Baseline}   & MGDT-DQN   & \textbf{8.82e+2} & 2.17e+3 & \textbf{2.11e-3}  & \textbf{1.74e-4}   \\ \cline{2-7} 
                           & \textbf{Baseline}   & MGDT-MAENT & \textbf{8.82e+2} & 8.11e+3 & \textbf{1.53e-9}  & \textbf{2.13e-75}  \\ \cline{2-7} 
                           & \textbf{Baseline}   & MGDT-PPO   & \textbf{8.82e+2} & 9.13e+3 & \textbf{1.20e-5}  & \textbf{3.87e-22}  \\ \cline{2-7} 
                           & \textbf{MGDT-DQN}   & MGDT-MAENT & \textbf{2.17e+3} & 8.11e+3 & \textbf{1.02e-6}  & \textbf{8.60e-94}  \\ \cline{2-7} 
                           & \textbf{MGDT-DQN}   & MGDT-PPO   & \textbf{2.17e+3} & 9.13e+3 & \textbf{1.40e-4}  & \textbf{3.17e-14}  \\ \cline{2-7} 
                           & \textbf{MGDT-MAENT} & MGDT-PPO   & \textbf{8.11e+3} & 9.13e+3 & \textbf{1.43e-1}  & \textbf{1.12e-1}   \\ \hline
\multirow{6}{*}{2\%}         & \textbf{Baseline}   & MGDT-DQN   & \textbf{8.82e+2} & 2.19e+3 & \textbf{1.27e-3}  & \textbf{2.77e-5}   \\ \cline{2-7} 
                           & \textbf{Baseline}   & MGDT-MAENT & \textbf{8.82e+2} & 7.91e+3 & \textbf{2.17e-9}  & \textbf{6.68e-73}  \\ \cline{2-7} 
                           & \textbf{Baseline}   & MGDT-PPO   & \textbf{8.82e+2} & 8.68e+3 & \textbf{7.76e-9}  & \textbf{1.97e-106} \\ \cline{2-7} 
                           & \textbf{MGDT-DQN}   & MGDT-MAENT & \textbf{2.19e+3} & 7.91e+3 & \textbf{9.61e-8}  & \textbf{8.53e-109} \\ \cline{2-7} 
                           & \textbf{MGDT-DQN}   & MGDT-PPO   & \textbf{2.19e+3} & 8.68e+3 & \textbf{9.43e-10} & \textbf{9.86e-222} \\ \cline{2-7} 
                           & \textbf{MGDT-MAENT} & MGDT-PPO   & \textbf{7.91e+3} & 8.68e+3 & \textbf{2.37e-3}  & \textbf{4.78e-3}   \\ \hline
\end{tabular}
}
\begin{tablenotes}
     \item mean(A) and mean(B) refer to  testing time values.
 \end{tablenotes}
\end{sidewaystable}

Regarding the PDR task, we observe that the IMPALA generalist agent outperforms the baseline \cite{zhang2020learning} based on the average of makespan and cumulative reward earned on the $6 \times 6$ instance across all fine-tuning data budgets. IMPALA agent configurations minimize the makespan by percentage values between [12.7\%, 13.3\%] for the $6 \times 6$ instance compared to the baseline across all fine-tuning data budgets. Similarly, the MGDT generalist agents outperform the baseline on average over several runs by minimizing the makespan by percentage values between [12.3\%, 13.5\%] for the $6 \times 6$ instance across all fine-tuning data budgets. In terms of training time, as expected, the IMPALA agents still have the best performance compared to the baseline across the 1\% and 2\% fine-tuning data budgets. However, the testing time of both the generalist agent configurations is longer than the baseline. Similar to the Blockmaze game, the baseline specialist agent takes less time to test due to the smaller size of its model (19K parameters), in comparison to the generalist agents with 5M and 197M parameters. \\
\textbf{Statistical analysis:} Tables \ref{tab:Results of  post-hoc tests analysis of makespan performance by the Baseline and generalist agents66}, \ref{tab:Results of  post-hoc tests analysis of training time performance by the Baseline and generalist agents on PDR task66}, and \ref{tab:Results of  post-hoc tests analysis of cumulative reward  performance by the Baseline and generalist agents on PDR task66} show the results of the post-hoc test analysis for various generalist agents and the baseline on the $6 \times 6$ instance of the PDR task.
\begin{table}[t]
\caption{Results of  post-hoc tests analysis of makespan performance by the baseline and generalist agents on the $6 \times 6$ instance (in bold are DRL configurations where p-value is $<$ 0.05 and have greater performance w.r.t the effect size).}
\label{tab:Results of  post-hoc tests analysis of makespan performance by the Baseline and generalist agents66}
\resizebox{\textwidth}{!}{
\centering
\begin{tabular}{|c|c|c|c|c|c|c|}
\hline
Data budgets     & A               & B                        & mean(A) & mean(B)          & pval             & CLES             \\ \hline
\multirow{15}{*}{zero-shot} & Baseline        & \textbf{IMPALA-PPO}      & 5.73e+2 & \textbf{5.04e+2} & \textbf{0.0} & \textbf{1.00e+0} \\ \cline{2-7} 
                            & Baseline        & \textbf{IMPALA-V\_TRACE} & 5.73e+2 & \textbf{5.04e+2} & \textbf{0.0} & \textbf{1.00e+0} \\ \cline{2-7} 
                            & Baseline        & \textbf{MGDT-DQN}        & 5.73e+2 & \textbf{5.00e+2} & \textbf{0.0} & \textbf{1.00e+0} \\ \cline{2-7} 
                            & Baseline        & \textbf{MGDT-MAENT}      & 5.73e+2 & \textbf{5.00e+2} & \textbf{0.0} & \textbf{1.00e+0} \\ \cline{2-7} 
                            & Baseline        & \textbf{MGDT-PPO}        & 5.73e+2 & \textbf{5.00e+2} & \textbf{0.0} & \textbf{1.00e+0} \\ \cline{2-7} 
                            & IMPALA-PPO      & IMPALA-V\_TRACE          & 5.04e+2 & 5.04e+2          & 1.00e+0          & 5.00e-1          \\ \cline{2-7} 
                            & IMPALA-PPO      & \textbf{MGDT-DQN}        & 5.04e+2 & \textbf{5.00e+2} & \textbf{1.95e-3} & \textbf{1.00e+0} \\ \cline{2-7} 
                            & IMPALA-PPO      & \textbf{MGDT-MAENT}      & 5.04e+2 & \textbf{5.00e+2} & \textbf{1.95e-3} & \textbf{1.00e+0} \\ \cline{2-7} 
                            & IMPALA-PPO      & \textbf{MGDT-PPO}        & 5.04e+2 & \textbf{5.00e+2} & \textbf{1.95e-3} & \textbf{1.00e+0} \\ \cline{2-7} 
                            & IMPALA-V\_TRACE & \textbf{MGDT-DQN}        & 5.04e+2 & \textbf{5.00e+2} & \textbf{1.95e-3} & \textbf{1.00e+0} \\ \cline{2-7} 
                            & IMPALA-V\_TRACE & \textbf{MGDT-MAENT}      & 5.04e+2 & \textbf{5.00e+2} & \textbf{1.95e-3} & \textbf{1.00e+0} \\ \cline{2-7} 
                            & IMPALA-V\_TRACE & \textbf{MGDT-PPO}        & 5.04e+2 & \textbf{5.00e+2} & \textbf{1.95e-3} & \textbf{1.00e+0} \\ \cline{2-7} 
                            & MGDT-DQN        & MGDT-MAENT               & 5.00e+2 & 5.00e+2          & 1.00e+0          & 5.00e-1          \\ \cline{2-7} 
                            & MGDT-DQN        & MGDT-PPO                 & 5.00e+2 & 5.00e+2          & 1.00e+0          & 5.00e-1          \\ \cline{2-7} 
                            & MGDT-MAENT      & MGDT-PPO                 & 5.00e+2 & 5.00e+2          & 1.00e+0          & 5.00e-1          \\ \hline
\multirow{15}{*}{1\%}         & Baseline        & \textbf{IMPALA-PPO}      & 5.73e+2 & \textbf{5.04e+2} & \textbf{0.0} & \textbf{1.00e+0} \\ \cline{2-7} 
                            & Baseline        & \textbf{IMPALA-V\_TRACE} & 5.73e+2 & \textbf{5.04e+2} & \textbf{0.0} & \textbf{1.00e+0} \\ \cline{2-7} 
                            & Baseline        & \textbf{MGDT-DQN}        & 5.73e+2 & \textbf{5.06e+2} & \textbf{0.0} & \textbf{1.00e+0} \\ \cline{2-7} 
                            & Baseline        & \textbf{MGDT-MAENT}      & 5.73e+2 & \textbf{5.00e+2} & \textbf{0.0} & \textbf{1.00e+0} \\ \cline{2-7} 
                            & Baseline        & \textbf{MGDT-PPO}        & 5.73e+2 & \textbf{5.05e+2} & \textbf{0.0} & \textbf{1.00e+0} \\ \cline{2-7} 
                            & IMPALA-PPO      & IMPALA-V\_TRACE          & 5.04e+2 & 5.04e+2          & 1.00e+0          & 4.80e-1          \\ \cline{2-7} 
                            & IMPALA-PPO      & MGDT-DQN                 & 5.04e+2 & 5.06e+2          & 8.12e-1          & 2.00e-1          \\ \cline{2-7} 
                            & IMPALA-PPO      & MGDT-MAENT               & 5.04e+2 & 5.00e+2          & 2.39e-1          & 1.00e+0          \\ \cline{2-7} 
                            & IMPALA-PPO      & MGDT-PPO                 & 5.04e+2 & 5.05e+2          & 9.96e-1          & 4.00e-1          \\ \cline{2-7} 
                            & IMPALA-V\_TRACE & MGDT-DQN                 & 5.04e+2 & 5.06e+2          & 9.04e-1          & 2.00e-1          \\ \cline{2-7} 
                            & IMPALA-V\_TRACE & MGDT-MAENT               & 5.04e+2 & 5.00e+2          & 1.64e-1          & 8.00e-1          \\ \cline{2-7} 
                            & IMPALA-V\_TRACE & MGDT-PPO                 & 5.04e+2 & 5.05e+2          & 1.00e+0          & 4.00e-1          \\ \cline{2-7} 
                            & MGDT-DQN        & \textbf{MGDT-MAENT}      & 5.06e+2 & \textbf{5.00e+2} & \textbf{1.90e-2} & \textbf{1.00e+0} \\ \cline{2-7} 
                            & MGDT-DQN        & MGDT-PPO                 & 5.06e+2 & 5.05e+2          & 9.74e-1          & 1.00e+0          \\ \cline{2-7} 
                            & MGDT-MAENT      & MGDT-PPO                 & 5.00e+2 & 5.05e+2          & 9.60e-2          & 0.0          \\ \hline
\multirow{15}{*}{2\%}         & Baseline        & \textbf{IMPALA-PPO}      & 5.73e+2 & \textbf{5.03e+2} & \textbf{0.0} & \textbf{1.00e+0} \\ \cline{2-7} 
                            & Baseline        & \textbf{IMPALA-V\_TRACE} & 5.73e+2 & \textbf{5.01e+2} & \textbf{0.0} & \textbf{1.00e+0} \\ \cline{2-7} 
                            & Baseline        & \textbf{MGDT-DQN}        & 5.73e+2 & \textbf{5.02e+2} & \textbf{0.0} & \textbf{1.00e+0} \\ \cline{2-7} 
                            & Baseline        & \textbf{MGDT-MAENT}      & 5.73e+2 & \textbf{5.00e+2} & \textbf{0.0} & \textbf{1.00e+0} \\ \cline{2-7} 
                            & Baseline        & \textbf{MGDT-PPO}        & 5.73e+2 & \textbf{5.01e+2} & \textbf{0.0} & \textbf{1.00e+0} \\ \cline{2-7} 
                            & IMPALA-PPO      & IMPALA-V\_TRACE          & 5.03e+2 & 5.01e+2          & 9.37e-1          & 5.60e-1          \\ \cline{2-7} 
                            & IMPALA-PPO      & MGDT-DQN                 & 5.03e+2 & 5.02e+2          & 9.97e-1          & 4.00e-1          \\ \cline{2-7} 
                            & IMPALA-PPO      & MGDT-MAENT               & 5.03e+2 & 5.00e+2          & 6.95e-1          & 6.00e-1          \\ \cline{2-7} 
                            & IMPALA-PPO      & MGDT-PPO                 & 5.03e+2 & 5.01e+2          & 8.47e-1          & 6.00e-1          \\ \cline{2-7} 
                            & IMPALA-V\_TRACE & MGDT-DQN                 & 5.01e+2 & 5.02e+2          & 9.97e-1          & 6.00e-1          \\ \cline{2-7} 
                            & IMPALA-V\_TRACE & MGDT-MAENT               & 5.01e+2 & 5.00e+2          & 9.94e-1          & 6.00e-1          \\ \cline{2-7} 
                            & IMPALA-V\_TRACE & MGDT-PPO                 & 5.01e+2 & 5.01e+2          & 1.00e+0          & 6.00e-1          \\ \cline{2-7} 
                            & MGDT-DQN        & MGDT-MAENT               & 5.02e+2 & 5.00e+2          & 9.13e-1          & 1.00e+0          \\ \cline{2-7} 
                            & MGDT-DQN        & MGDT-PPO                 & 5.02e+2 & 5.01e+2          & 9.78e-1          & 1.00e+0          \\ \cline{2-7} 
                            & MGDT-MAENT      & MGDT-PPO                 & 5.00e+2 & 5.01e+2          & 1.00e+0          & 0.0          \\ \hline
\end{tabular}
}
\begin{tablenotes}
     \item mean(A) and mean(B) refer to makespan values.
     
 \end{tablenotes}
\end{table}
Statistically, both generalist agent configurations significantly outperform the baseline on the $6 \times 6$ instance of the PDR task, in terms of makespan, and cumulative reward performance across all fine-tuning data budgets with CLES values equal to 0. 
\begin{table}[t]
\caption{Results of  post-hoc tests analysis of training time performance by the baseline and generalist agents on PDR task on the $6 \times 6$ instance(in bold are DRL configurations where p-value is $<$ 0.05 and have greater performance w.r.t the effect size).}
\label{tab:Results of  post-hoc tests analysis of training time performance by the Baseline and generalist agents on PDR task66}
\centering
\resizebox{\textwidth}{!}{
\begin{tabular}{|c|c|c|c|c|c|c|}
\hline
Data budgets & A                        & B                        & mean(A)          & mean(B)          & pval              & CLES             \\ \hline
\multirow{15}{*}{1\%}     & Baseline                 & \textbf{IMPALA-PPO}      & 3.96e+3          & \textbf{1.12e+2} & \textbf{2.77e-2}  & \textbf{1.00e+0} \\ \cline{2-7} 
                        & Baseline                 & \textbf{IMPALA-V\_TRACE} & 3.96e+3          & \textbf{1.11e+2} & \textbf{2.77e-2}  & \textbf{1.00e+0} \\ \cline{2-7} 
                        & Baseline                 & MGDT-DQN                 & 3.96e+3          & 2.28e+3          & 3.13e-1           & 8.00e-1          \\ \cline{2-7} 
                        & Baseline                 & \textbf{MGDT-MAENT}      & 3.96e+3          & \textbf{6.42e+2} & \textbf{4.59e-2}  & \textbf{1.00e+0} \\ \cline{2-7} 
                        & Baseline                 & \textbf{MGDT-PPO}        & 3.96e+3          & \textbf{7.74e+3} & \textbf{2.95e-2}  & \textbf{0.0} \\ \cline{2-7} 
                        & IMPALA-PPO               & IMPALA-V\_TRACE          & 1.12e+2          & 1.11e+2          & 9.70e-1           & 5.60e-1          \\ \cline{2-7} 
                        & \textbf{IMPALA-PPO}      & MGDT-DQN                 & \textbf{1.12e+2} & 2.28e+3          & \textbf{2.22e-14} & \textbf{0.0} \\ \cline{2-7} 
                        & \textbf{IMPALA-PPO}      & MGDT-MAENT               & \textbf{1.12e+2} & 6.42e+2          & \textbf{8.33e-10} & \textbf{0.0} \\ \cline{2-7} 
                        & \textbf{IMPALA-PPO}      & MGDT-PPO                 & \textbf{1.12e+2} & 7.74e+3          & \textbf{2.62e-9}  & \textbf{0.0} \\ \cline{2-7} 
                        & \textbf{IMPALA-V\_TRACE} & MGDT-DQN                 & \textbf{1.11e+2} & 2.28e+3          & \textbf{1.68e-11} & \textbf{0.0} \\ \cline{2-7} 
                        & \textbf{IMPALA-V\_TRACE} & MGDT-MAENT               & \textbf{1.11e+2} & 6.42e+2          & \textbf{3.66e-8}  & \textbf{0.0} \\ \cline{2-7} 
                        & \textbf{IMPALA-V\_TRACE} & MGDT-PPO                 & \textbf{1.11e+2} & 7.74e+3          & \textbf{3.22e-9}  & \textbf{0.0} \\ \cline{2-7} 
                        & MGDT-DQN                 & \textbf{MGDT-MAENT}      & 2.28e+3          & \textbf{6.42e+2} & \textbf{5.55e-16} & \textbf{1.00e+0} \\ \cline{2-7} 
                        & \textbf{MGDT-DQN}        & MGDT-PPO                 & \textbf{2.28e+3} & 7.74e+3          & \textbf{5.64e-9}  & \textbf{0.0} \\ \cline{2-7} 
                        & \textbf{MGDT-MAENT}      & MGDT-PPO                 & \textbf{6.42e+2} & 7.74e+3          & \textbf{1.41e-9}  & \textbf{0.0} \\ \hline
\multirow{15}{*}{2\%}     & Baseline                 & \textbf{IMPALA-PPO}      & 3.96e+3          & \textbf{1.10e+2} & \textbf{2.76e-2}  & \textbf{1.00e+0} \\ \cline{2-7} 
                        & Baseline                 & \textbf{IMPALA-V\_TRACE} & 3.96e+3          & \textbf{1.20e+2} & \textbf{2.79e-2}  & \textbf{1.00e+0} \\ \cline{2-7} 
                        & Baseline                 & MGDT-DQN                 & 3.96e+3          & 4.57e+3          & 9.29e-1           & 5.20e-1          \\ \cline{2-7} 
                        & Baseline                 & MGDT-MAENT               & 3.96e+3          & 1.30e+3          & 9.34e-2           & 8.00e-1          \\ \cline{2-7} 
                        & \textbf{Baseline}        & MGDT-PPO                 & \textbf{3.96e+3} & 1.56e+4          & \textbf{4.05e-4}  & \textbf{0.0} \\ \cline{2-7} 
                        & IMPALA-PPO               & IMPALA-V\_TRACE          & 1.10e+2          & 1.20e+2          & 2.12e-1           & 8.00e-2          \\ \cline{2-7} 
                        & \textbf{IMPALA-PPO}      & MGDT-DQN                 & \textbf{1.10e+2} & 4.57e+3          & \textbf{4.62e-9}  & \textbf{0.0} \\ \cline{2-7} 
                        & \textbf{IMPALA-PPO}      & MGDT-MAENT               & \textbf{1.10e+2} & 1.30e+3          & \textbf{1.86e-10} & \textbf{0.0} \\ \cline{2-7} 
                        & \textbf{IMPALA-PPO}      & MGDT-PPO                 & \textbf{1.10e+2} & 1.56e+4          & \textbf{8.23e-10} & \textbf{0.0} \\ \cline{2-7} 
                        & \textbf{IMPALA-V\_TRACE} & MGDT-DQN                 & \textbf{1.20e+2} & 4.57e+3          & \textbf{4.17e-9}  & \textbf{0.0} \\ \cline{2-7} 
                        & \textbf{IMPALA-V\_TRACE} & MGDT-MAENT               & \textbf{1.20e+2} & 1.30e+3          & \textbf{1.00e-10} & \textbf{0.0} \\ \cline{2-7} 
                        & \textbf{IMPALA-V\_TRACE} & MGDT-PPO                 & \textbf{1.20e+2} & 1.56e+4          & \textbf{7.98e-10} & \textbf{0.0} \\ \cline{2-7} 
                        & MGDT-DQN                 & \textbf{MGDT-MAENT}      & 4.57e+3          & \textbf{1.30e+3} & \textbf{9.18e-10} & \textbf{1.00e+0} \\ \cline{2-7} 
                        & \textbf{MGDT-DQN}        & MGDT-PPO                 & \textbf{4.57e+3} & 1.56e+4          & \textbf{1.56e-12} & \textbf{0.0} \\ \cline{2-7} 
                        & \textbf{MGDT-MAENT}      & MGDT-PPO                 & \textbf{1.30e+3} & 1.56e+4          & \textbf{4.37e-10} & \textbf{0.0} \\ \hline
\end{tabular}
}
\begin{tablenotes}
     \item mean(A) and mean(B) refer to training  time values.
     
 \end{tablenotes}
\end{table}
The CLES values indicate that with a 0\% chance, the baseline will have a higher makespan and lower cumulative reward earned than the generalist agents. 
\begin{table}[t]
\caption{Results of  post-hoc tests analysis of cumulative reward performance by the baseline and the fine-tuned generalist agents on the PDR task on the $6 \times 6$ instance(in bold are DRL configurations where p-value is $<$ 0.05 and have greater performance w.r.t the effect size).}
\label{tab:Results of  post-hoc tests analysis of cumulative reward  performance by the Baseline and generalist agents on PDR task66}
\centering
\resizebox{\textwidth}{!}{
\begin{tabular}{|c|c|c|c|c|c|c|}
\hline
Data budgets     & A               & B                        & mean(A)  & mean(B)           & pval             & CLES             \\ \hline
\multirow{15}{*}{zero-shot} & Baseline        & \textbf{IMPALA-PPO}      & -5.73e+2 & \textbf{-3.92e+2} & \textbf{0.0} & \textbf{0.0} \\ \cline{2-7} 
                            & Baseline        & \textbf{IMPALA-V\_TRACE} & -5.73e+2 & \textbf{-3.92e+2} & \textbf{0.0} & \textbf{0.0} \\ \cline{2-7} 
                            & Baseline        & \textbf{MGDT-DQN}        & -5.73e+2 & \textbf{-3.91e+2} & \textbf{0.0} & \textbf{0.0} \\ \cline{2-7} 
                            & Baseline        & \textbf{MGDT-MAENT}      & -5.73e+2 & \textbf{-3.91e+2} & \textbf{0.0} & \textbf{0.0} \\ \cline{2-7} 
                            & Baseline        & \textbf{MGDT-PPO}        & -5.73e+2 & \textbf{-3.91e+2} & \textbf{0.0} & \textbf{0.0} \\ \cline{2-7} 
                            & IMPALA-PPO      & IMPALA-V\_TRACE          & -3.92e+2 & -3.92e+2          & 1.00e+0          & 5.00e-1          \\ \cline{2-7} 
                            & IMPALA-PPO      & MGDT-DQN                 & -3.92e+2 & -3.91e+2          & 1.00e+0          & 6.00e-1          \\ \cline{2-7} 
                            & IMPALA-PPO      & MGDT-MAENT               & -3.92e+2 & -3.91e+2          & 1.00e+0          & 6.00e-1          \\ \cline{2-7} 
                            & IMPALA-PPO      & MGDT-PPO                 & -3.92e+2 & -3.91e+2          & 1.00e+0          & 6.00e-1          \\ \cline{2-7} 
                            & IMPALA-V\_TRACE & MGDT-DQN                 & -3.92e+2 & -3.91e+2          & 1.00e+0          & 6.00e-1          \\ \cline{2-7} 
                            & IMPALA-V\_TRACE & MGDT-MAENT               & -3.92e+2 & -3.91e+2          & 1.00e+0          & 6.00e-1          \\ \cline{2-7} 
                            & IMPALA-V\_TRACE & MGDT-PPO                 & -3.92e+2 & -3.91e+2          & 1.00e+0          & 6.00e-1          \\ \cline{2-7} 
                            & MGDT-DQN        & MGDT-MAENT               & -3.91e+2 & -3.91e+2          & 1.00e+0          & 5.00e-1          \\ \cline{2-7} 
                            & MGDT-DQN        & MGDT-PPO                 & -3.91e+2 & -3.91e+2          & 1.00e+0          & 5.00e-1          \\ \cline{2-7} 
                            & MGDT-MAENT      & MGDT-PPO                 & -3.91e+2 & -3.91e+2          & 1.00e+0          & 5.00e-1          \\ \hline
\multirow{15}{*}{1\%}         & Baseline        & \textbf{IMPALA-PPO}      & -5.73e+2 & \textbf{-3.92e+2} & \textbf{0.0} & \textbf{0.0} \\ \cline{2-7} 
                            & Baseline        & \textbf{IMPALA-V\_TRACE} & -5.73e+2 & \textbf{-3.91e+2} & \textbf{0.0} & \textbf{0.0} \\ \cline{2-7} 
                            & Baseline        & \textbf{MGDT-DQN}        & -5.73e+2 & \textbf{-3.91e+2} & \textbf{0.0} & \textbf{0.0} \\ \cline{2-7} 
                            & Baseline        & \textbf{MGDT-MAENT}      & -5.73e+2 & \textbf{-3.91e+2} & \textbf{0.0} & \textbf{0.0} \\ \cline{2-7} 
                            & Baseline        & \textbf{MGDT-PPO}        & -5.73e+2 & \textbf{-3.93e+2} & \textbf{0.0} & \textbf{0.0} \\ \cline{2-7} 
                            & IMPALA-PPO      & IMPALA-V\_TRACE          & -3.92e+2 & -3.91e+2          & 9.96e-1          & 2.00e-1          \\ \cline{2-7} 
                            & IMPALA-PPO      & MGDT-DQN                 & -3.92e+2 & -3.91e+2          & 8.48e-1          & 0.0          \\ \cline{2-7} 
                            & IMPALA-PPO      & MGDT-MAENT               & -3.92e+2 & -3.91e+2          & 9.99e-1          & 0.0          \\ \cline{2-7} 
                            & IMPALA-PPO      & MGDT-PPO                 & -3.92e+2 & -3.93e+2          & 7.68e-1          & 1.00e+0          \\ \cline{2-7} 
                            & IMPALA-V\_TRACE & MGDT-DQN                 & -3.91e+2 & -3.91e+2          & 9.84e-1          & 0.0          \\ \cline{2-7} 
                            & IMPALA-V\_TRACE & MGDT-MAENT               & -3.91e+2 & -3.91e+2          & 1.00e+0          & 8.00e-1          \\ \cline{2-7} 
                            & IMPALA-V\_TRACE & MGDT-PPO                 & -3.91e+2 & -3.93e+2          & 4.68e-1          & 1.00e+0          \\ \cline{2-7} 
                            & MGDT-DQN        & MGDT-MAENT               & -3.91e+2 & -3.91e+2          & 9.60e-1          & 1.00e+0          \\ \cline{2-7} 
                            & MGDT-DQN        & MGDT-PPO                 & -3.91e+2 & -3.93e+2          & 1.66e-1          & 1.00e+0          \\ \cline{2-7} 
                            & MGDT-MAENT      & MGDT-PPO                 & -3.91e+2 & -3.93e+2          & 5.64e-1          & 1.00e+0          \\ \hline
\multirow{15}{*}{2\%}         & Baseline        & \textbf{IMPALA-PPO}      & -5.73e+2 & \textbf{-3.92e+2} & \textbf{0.0} & \textbf{0.0} \\ \cline{2-7} 
                            & Baseline        & \textbf{IMPALA-V\_TRACE} & -5.73e+2 & \textbf{-3.92e+2} & \textbf{0.0} & \textbf{0.0} \\ \cline{2-7} 
                            & Baseline        & \textbf{MGDT-DQN}        & -5.73e+2 & \textbf{-3.92e+2} & \textbf{0.0} & \textbf{0.0} \\ \cline{2-7} 
                            & Baseline        & \textbf{MGDT-MAENT}      & -5.73e+2 & \textbf{-3.91e+2} & \textbf{0.0} & \textbf{0.0} \\ \cline{2-7} 
                            & Baseline        & \textbf{MGDT-PPO}        & -5.73e+2 & \textbf{-3.93e+2} & \textbf{0.0} & \textbf{0.0} \\ \cline{2-7} 
                            & IMPALA-PPO      & IMPALA-V\_TRACE          & -3.92e+2 & -3.92e+2          & 9.99e-1          & 6.00e-1          \\ \cline{2-7} 
                            & IMPALA-PPO      & MGDT-DQN                 & -3.92e+2 & -3.92e+2          & 1.00e+0          & 7.00e-1          \\ \cline{2-7} 
                            & IMPALA-PPO      & MGDT-MAENT               & -3.92e+2 & -3.91e+2          & 1.00e+0          & 4.00e-1          \\ \cline{2-7} 
                            & IMPALA-PPO      & MGDT-PPO                 & -3.92e+2 & -3.93e+2          & 5.43e-1          & 1.00e+0          \\ \cline{2-7} 
                            & IMPALA-V\_TRACE & MGDT-DQN                 & -3.92e+2 & -3.92e+2          & 1.00e+0          & 6.00e-1          \\ \cline{2-7} 
                            & IMPALA-V\_TRACE & MGDT-MAENT               & -3.92e+2 & -3.91e+2          & 9.99e-1          & 4.00e-1          \\ \cline{2-7} 
                            & IMPALA-V\_TRACE & MGDT-PPO                 & -3.92e+2 & -3.93e+2          & 7.43e-1          & 1.00e+0          \\ \cline{2-7} 
                            & MGDT-DQN        & MGDT-MAENT               & -3.92e+2 & -3.91e+2          & 1.00e+0          & 0.0          \\ \cline{2-7} 
                            & MGDT-DQN        & MGDT-PPO                 & -3.92e+2 & -3.93e+2          & 6.75e-1          & 1.00e+0          \\ \cline{2-7} 
                            & MGDT-MAENT      & MGDT-PPO                 & -3.91e+2 & -3.93e+2          & 5.17e-1          & 1.00e+0          \\ \hline
\end{tabular}
}
\begin{tablenotes}
     \item mean(A) and mean(B) refer to cumulative reward values.
     
 \end{tablenotes}
\end{table}
In terms of training time across the 1\% and 2\% fine-tuning data budgets, the IMPALA agents perform significantly better. The results also show that IMPALA and MGDT configurations are not significantly different in terms of makespan, and cumulative reward earned.  In terms of testing time, the baseline on the $6 \times 6$ instance of the PDR task performs best, as reported in Table \ref{tab:Results of  post-hoc tests analysis of testing time performance by the Baseline and generalist agents on PDR task66}. As previously mentioned, the larger size of both generalist agent's models negatively impacted their testing time performance, in comparison to the baseline specialist agents.
\begin{table}[t]
\caption{Results of post-hoc tests analysis of testing time performance by the baseline and the fine-tuned generalist agents on PDR task on the $6 \times 6$ instance (in bold are DRL configurations where p-value is $<$0.05 and have greater performance w.r.t the effect size).}
\label{tab:Results of  post-hoc tests analysis of testing time performance by the Baseline and generalist agents on PDR task66}
\centering
\resizebox{\textwidth}{!}{
\begin{tabular}{|c|c|c|c|c|c|c|}
\hline
Data budgets     & A                        & B               & mean(A)          & mean(B) & pval              & CLES               \\ \hline
\multirow{15}{*}{zero-shot} & \textbf{Baseline}        & IMPALA-PPO      & \textbf{1.05e+1} & 5.10e+1 & \textbf{2.50e-7}  & \textbf{5.63e-34}  \\ \cline{2-7} 
                            & \textbf{Baseline}        & IMPALA-V\_TRACE & \textbf{1.05e+1} & 5.10e+1 & \textbf{2.50e-7}  & \textbf{5.63e-34}  \\ \cline{2-7} 
                            & \textbf{Baseline}        & MGDT-DQN        & \textbf{1.05e+1} & 1.59e+2 & \textbf{2.67e-6}  & \textbf{5.33e-87}  \\ \cline{2-7} 
                            & \textbf{Baseline}        & MGDT-MAENT      & \textbf{1.05e+1} & 1.59e+2 & \textbf{2.67e-6}  & \textbf{5.33e-87}  \\ \cline{2-7} 
                            & \textbf{Baseline}        & MGDT-PPO        & \textbf{1.05e+1} & 1.59e+2 & \textbf{2.67e-6}  & \textbf{5.33e-87}  \\ \cline{2-7} 
                            & IMPALA-PPO               & IMPALA-V\_TRACE & 5.10e+1          & 5.10e+1 & 1.00e+0           & 5.00e-1            \\ \cline{2-7} 
                            & \textbf{IMPALA-PPO}      & MGDT-DQN        & \textbf{5.10e+1} & 1.59e+2 & \textbf{3.48e-6}  & \textbf{1.34e-43}  \\ \cline{2-7} 
                            & \textbf{IMPALA-PPO}      & MGDT-MAENT      & \textbf{5.10e+1} & 1.59e+2 & \textbf{3.48e-6}  & \textbf{1.34e-43}  \\ \cline{2-7} 
                            & \textbf{IMPALA-PPO}      & MGDT-PPO        & \textbf{5.10e+1} & 1.59e+2 & \textbf{3.48e-6}  & \textbf{1.34e-43}  \\ \cline{2-7} 
                            & \textbf{IMPALA-V\_TRACE} & MGDT-DQN        & \textbf{5.10e+1} & 1.59e+2 & \textbf{3.48e-6}  & \textbf{1.34e-43}  \\ \cline{2-7} 
                            & \textbf{IMPALA-V\_TRACE} & MGDT-MAENT      & \textbf{5.10e+1} & 1.59e+2 & \textbf{3.48e-6}  & \textbf{1.34e-43}  \\ \cline{2-7} 
                            & \textbf{IMPALA-V\_TRACE} & MGDT-PPO        & \textbf{5.10e+1} & 1.59e+2 & \textbf{3.48e-6}  & \textbf{1.34e-43}  \\ \cline{2-7} 
                            & MGDT-DQN                 & MGDT-MAENT      & 1.59e+2          & 1.59e+2 & 1.00e+0           & 5.00e-1            \\ \cline{2-7} 
                            & MGDT-DQN                 & MGDT-PPO        & 1.59e+2          & 1.59e+2 & 1.00e+0           & 5.00e-1            \\ \cline{2-7} 
                            & MGDT-MAENT               & MGDT-PPO        & 1.59e+2          & 1.59e+2 & 1.00e+0           & 5.00e-1            \\ \hline
\multirow{15}{*}{1\%}         & \textbf{Baseline}        & IMPALA-PPO      & \textbf{1.05e+1} & 4.76e+1 & \textbf{3.97e-8}  & \textbf{9.82e-38}  \\ \cline{2-7} 
                            & \textbf{Baseline}        & IMPALA-V\_TRACE & \textbf{1.05e+1} & 4.86e+1 & \textbf{1.90e-7}  & \textbf{1.28e-77}  \\ \cline{2-7} 
                            & \textbf{Baseline}        & MGDT-DQN        & \textbf{1.05e+1} & 1.40e+2 & \textbf{7.91e-11} & \textbf{0.0}   \\ \cline{2-7} 
                            & \textbf{Baseline}        & MGDT-MAENT      & \textbf{1.05e+1} & 1.38e+2 & \textbf{9.07e-10} & \textbf{0.0}   \\ \cline{2-7} 
                            & \textbf{Baseline}        & MGDT-PPO        & \textbf{1.05e+1} & 1.41e+2 & \textbf{1.23e-12} & \textbf{0.0}   \\ \cline{2-7} 
                            & IMPALA-PPO               & IMPALA-V\_TRACE & 4.76e+1          & 4.86e+1 & 9.15e-1           & 3.34e-1            \\ \cline{2-7} 
                            & \textbf{IMPALA-PPO}      & MGDT-DQN        & \textbf{4.76e+1} & 1.40e+2 & \textbf{2.25e-10} & \textbf{4.14e-146} \\ \cline{2-7} 
                            & \textbf{IMPALA-PPO}      & MGDT-MAENT      & \textbf{4.76e+1} & 1.38e+2 & \textbf{4.44e-8}  & \textbf{0.0}   \\ \cline{2-7} 
                            & \textbf{IMPALA-PPO}      & MGDT-PPO        & \textbf{4.76e+1} & 1.41e+2 & \textbf{2.58e-11} & \textbf{4.68e-202} \\ \cline{2-7} 
                            & \textbf{IMPALA-V\_TRACE} & MGDT-DQN        & \textbf{4.86e+1} & 1.40e+2 & \textbf{1.50e-7}  & \textbf{2.35e-212} \\ \cline{2-7} 
                            & \textbf{IMPALA-V\_TRACE} & MGDT-MAENT      & \textbf{4.86e+1} & 1.38e+2 & \textbf{4.22e-14} & \textbf{0.0}   \\ \cline{2-7} 
                            & \textbf{IMPALA-V\_TRACE} & MGDT-PPO        & \textbf{4.86e+1} & 1.41e+2 & \textbf{7.99e-9}  & \textbf{0.0}   \\ \cline{2-7} 
                            & MGDT-DQN                 & MGDT-MAENT      & 1.40e+2          & 1.38e+2 & 8.64e-1           & 6.88e-1            \\ \cline{2-7} 
                            & MGDT-DQN                 & MGDT-PPO        & 1.40e+2          & 1.41e+2 & 8.69e-1           & 3.12e-1            \\ \cline{2-7} 
                            & MGDT-MAENT               & MGDT-PPO        & 1.38e+2          & 1.41e+2 & 1.51e-1           & 8.24e-2            \\ \hline
\multirow{15}{*}{2\%}         & \textbf{Baseline}        & IMPALA-PPO      & \textbf{1.05e+1} & 4.85e+1 & \textbf{1.63e-8}  & \textbf{2.27e-64}  \\ \cline{2-7} 
                            & \textbf{Baseline}        & IMPALA-V\_TRACE & \textbf{1.05e+1} & 4.94e+1 & \textbf{4.66e-8}  & \textbf{2.91e-74}  \\ \cline{2-7} 
                            & \textbf{Baseline}        & MGDT-DQN        & \textbf{1.05e+1} & 1.41e+2 & \textbf{5.43e-12} & \textbf{0.0}   \\ \cline{2-7} 
                            & \textbf{Baseline}        & MGDT-MAENT      & \textbf{1.05e+1} & 1.44e+2 & \textbf{4.27e-6}  & \textbf{2.91e-71}  \\ \cline{2-7} 
                            & \textbf{Baseline}        & MGDT-PPO        & \textbf{1.05e+1} & 1.46e+2 & \textbf{7.90e-10} & \textbf{0.0}   \\ \cline{2-7} 
                            & IMPALA-PPO               & IMPALA-V\_TRACE & 4.85e+1          & 4.94e+1 & 8.50e-1           & 3.04e-1            \\ \cline{2-7} 
                            & \textbf{IMPALA-PPO}      & MGDT-DQN        & \textbf{4.85e+1} & 1.41e+2 & \textbf{1.74e-9}  & \textbf{3.77e-257} \\ \cline{2-7} 
                            & \textbf{IMPALA-PPO}      & MGDT-MAENT      & \textbf{4.85e+1} & 1.44e+2 & \textbf{3.07e-5}  & \textbf{1.05e-38}  \\ \cline{2-7} 
                            & \textbf{IMPALA-PPO}      & MGDT-PPO        & \textbf{4.85e+1} & 1.46e+2 & \textbf{2.12e-11} & \textbf{0.0}   \\ \cline{2-7} 
                            & \textbf{IMPALA-V\_TRACE} & MGDT-DQN        & \textbf{4.94e+1} & 1.41e+2 & \textbf{8.28e-9}  & \textbf{9.97e-271} \\ \cline{2-7} 
                            & \textbf{IMPALA-V\_TRACE} & MGDT-MAENT      & \textbf{4.94e+1} & 1.44e+2 & \textbf{3.65e-5}  & \textbf{2.20e-38}  \\ \cline{2-7} 
                            & \textbf{IMPALA-V\_TRACE} & MGDT-PPO        & \textbf{4.94e+1} & 1.46e+2 & \textbf{7.78e-13} & \textbf{0.0}   \\ \cline{2-7} 
                            & MGDT-DQN                 & MGDT-MAENT      & 1.41e+2          & 1.44e+2 & 8.79e-1           & 3.18e-1            \\ \cline{2-7} 
                            & \textbf{MGDT-DQN}        & MGDT-PPO        & \textbf{1.41e+2} & 1.46e+2 & \textbf{4.61e-2}  & \textbf{2.30e-2}   \\ \cline{2-7} 
                            & MGDT-MAENT               & MGDT-PPO        & 1.44e+2          & 1.46e+2 & 9.98e-1           & 4.29e-1            \\ \hline
\end{tabular}
}
\begin{tablenotes}
     \item mean(A) and mean(B) refer to testing time values.
     
 \end{tablenotes}
\end{table}

Regarding the $30 \times 20$ instance of the PDR task, we observe that the IMPALA generalist agents outperform the baseline \cite{zhang2020learning} on the average of makespan and cumulative reward across all fine-tuning data budgets. IMPALA agents minimize the makespan [20.6\%, 20.8\%] more than the baseline. Similarly, the MGDT generalist agents outperform the baseline on average over several runs  by minimizing the makespan [20.3\%, 20.8\%] more than baseline across all fine-tuning data budgets. In terms of training time, because of its distributed nature, the IMPALA agents still have the best performance compared to the baseline across the 1\% and 2\% fine-tuning data budgets.  However, the testing time of both the generalist agents is longer than the baseline. \\
\textbf{Statistical analysis:} Tables  \ref{tab:Results of  post-hoc tests analysis of makespan performance by the Baseline and generalist agents3020}, \ref{tab:Results of  post-hoc tests analysis of training time performance by the Baseline and generalist agents on PDR task3020}, and \ref{tab:Results of  post-hoc tests analysis of cumulative reward  performance by the Baseline and generalist agents on PDR task3020} show the results of the post-hoc test analysis for the generalist agents with different configuration and the baseline on the $30 \times 20$ instance of the PDR task.
\begin{table}[t]
\caption{Results of  post-hoc tests analysis of makespan performance by the baseline and generalist agents on the $30 \times 20$ instance (in bold are DRL configurations where p-value is $<$ 0.05 and have greater performance w.r.t the effect size).}
\label{tab:Results of  post-hoc tests analysis of makespan performance by the Baseline and generalist agents3020}
\resizebox{\textwidth}{!}{
\centering
\begin{tabular}{|c|c|c|c|c|c|c|}
\hline
Data budgets     & A                   & B                        & mean(A)          & mean(B)          & pval              & CLES             \\ \hline
\multirow{13}{*}{zero-shot} & Baseline            & \textbf{IMPALA-PPO}      & 2.48e+3          & \textbf{2.01e+3} & \textbf{7.31e-11} & \textbf{1.00e+0} \\ \cline{2-7} 
                            & Baseline            & \textbf{IMPALA-V\_TRACE} & 2.48e+3          & \textbf{2.01e+3} & \textbf{7.31e-11} & \textbf{1.00e+0} \\ \cline{2-7} 
                            & Baseline            & \textbf{MGDT-DQN}        & 2.48e+3          & \textbf{2.02e+3} & \textbf{2.51e-11} & \textbf{1.00e+0} \\ \cline{2-7} 
                            & Baseline            & \textbf{MGDT-MAENT}      & 2.48e+3          & \textbf{2.02e+3} & \textbf{2.51e-11} & \textbf{1.00e+0} \\ \cline{2-7} 
                            & Baseline            & \textbf{MGDT-PPO}        & 2.48e+3          & \textbf{2.02e+3} & \textbf{2.51e-11} & \textbf{1.00e+0} \\ \cline{2-7} 
                            & IMPALA-PPO          & IMPALA-V\_TRACE          & 2.01e+3          & 2.01e+3          & 1.00e+0           & 5.00e-1          \\ \cline{2-7} 
                            & IMPALA-PPO          & MGDT-DQN                 & 2.01e+3          & 2.02e+3          & 9.61e-2           & 5.33e-2          \\ \cline{2-7} 
                            & IMPALA-PPO          & MGDT-MAENT               & 2.01e+3          & 2.02e+3          & 9.61e-2           & 5.33e-2          \\ \cline{2-7} 
                            & IMPALA-V\_TRACE     & MGDT-DQN                 & 2.01e+3          & 2.02e+3          & 9.61e-2           & 5.33e-2          \\ \cline{2-7} 
                            & IMPALA-V\_TRACE     & MGDT-MAENT               & 2.01e+3          & 2.02e+3          & 9.61e-2           & 5.33e-2          \\ \cline{2-7} 
                            & MGDT-DQN            & MGDT-MAENT               & 2.02e+3          & 2.02e+3          & 1.00e+0           & 5.00e-1          \\ \cline{2-7} 
                            & MGDT-DQN            & MGDT-PPO                 & 2.02e+3          & 2.02e+3          & 1.00e+0           & 5.00e-1          \\ \cline{2-7} 
                            & MGDT-MAENT          & MGDT-PPO                 & 2.02e+3          & 2.02e+3          & 1.00e+0           & 5.00e-1          \\ \hline
\multirow{15}{*}{1\%}         & Baseline            & \textbf{IMPALA-PPO}      & 2.48e+3          & \textbf{2.01e+3} & \textbf{9.18e-10} & \textbf{1.00e+0} \\ \cline{2-7} 
                            & Baseline            & \textbf{IMPALA-V\_TRACE} & 2.48e+3          & \textbf{2.01e+3} & \textbf{8.12e-9}  & \textbf{1.00e+0} \\ \cline{2-7} 
                            & Baseline            & \textbf{MGDT-DQN}        & 2.48e+3          & \textbf{2.01e+3} & \textbf{7.16e-11} & \textbf{1.00e+0} \\ \cline{2-7} 
                            & Baseline            & \textbf{MGDT-MAENT}      & 2.48e+3          & \textbf{2.01e+3} & \textbf{2.64e-11} & \textbf{1.00e+0} \\ \cline{2-7} 
                            & Baseline            & \textbf{MGDT-PPO}        & 2.48e+3          & \textbf{2.01e+3} & \textbf{5.55e-16} & \textbf{1.00e+0} \\ \cline{2-7} 
                            & IMPALA-PPO          & IMPALA-V\_TRACE          & 2.01e+3          & 2.01e+3          & 1.00e+0           & 4.61e-1          \\ \cline{2-7} 
                            & IMPALA-PPO          & MGDT-DQN                 & 2.01e+3          & 2.01e+3          & 9.77e-1           & 3.78e-1          \\ \cline{2-7} 
                            & IMPALA-PPO          & MGDT-MAENT               & 2.01e+3          & 2.01e+3          & 9.20e-1           & 6.61e-1          \\ \cline{2-7} 
                            & IMPALA-PPO          & MGDT-PPO                 & 2.01e+3          & 2.01e+3          & 8.37e-1           & 3.02e-1          \\ \cline{2-7} 
                            & IMPALA-V\_TRACE     & MGDT-DQN                 & 2.01e+3          & 2.01e+3          & 9.89e-1           & 3.92e-1          \\ \cline{2-7} 
                            & IMPALA-V\_TRACE     & MGDT-MAENT               & 2.01e+3          & 2.01e+3          & 6.51e-1           & 7.75e-1          \\ \cline{2-7} 
                            & IMPALA-V\_TRACE     & MGDT-PPO                 & 2.01e+3          & 2.01e+3          & 8.17e-1           & 2.84e-1          \\ \cline{2-7} 
                            & MGDT-DQN            & MGDT-MAENT               & 2.01e+3          & 2.01e+3          & 3.61e-1           & 8.48e-1          \\ \cline{2-7} 
                            & MGDT-DQN            & MGDT-PPO                 & 2.01e+3          & 2.01e+3          & 9.99e-1           & 4.36e-1          \\ \cline{2-7} 
                            & \textbf{MGDT-MAENT} & MGDT-PPO                 & \textbf{2.01e+3} & 2.01e+3          & \textbf{1.80e-3}  & \textbf{7.19e-8} \\ \hline
\multirow{15}{*}{2\%}         & Baseline            & \textbf{IMPALA-PPO}      & 2.48e+3          & \textbf{2.01e+3} & \textbf{0.0}  & \textbf{1.00e+0} \\ \cline{2-7} 
                            & Baseline            & \textbf{IMPALA-V\_TRACE} & 2.48e+3          & \textbf{2.01e+3} & \textbf{2.71e-11} & \textbf{1.00e+0} \\ \cline{2-7} 
                            & Baseline            & \textbf{MGDT-DQN}        & 2.48e+3          & \textbf{2.01e+3} & \textbf{8.67e-9}  & \textbf{1.00e+0} \\ \cline{2-7} 
                            & Baseline            & \textbf{MGDT-MAENT}      & 2.48e+3          & \textbf{2.01e+3} & \textbf{2.52e-6}  & \textbf{1.00e+0} \\ \cline{2-7} 
                            & Baseline            & \textbf{MGDT-PPO}        & 2.48e+3          & \textbf{2.01e+3} & \textbf{2.24e-9}  & \textbf{1.00e+0} \\ \cline{2-7} 
                            & IMPALA-PPO          & IMPALA-V\_TRACE          & 2.01e+3          & 2.01e+3          & 2.18e-1           & 9.58e-2          \\ \cline{2-7} 
                            & IMPALA-PPO          & MGDT-DQN                 & 2.01e+3          & 2.01e+3          & 9.95e-1           & 4.15e-1          \\ \cline{2-7} 
                            & IMPALA-PPO          & MGDT-MAENT               & 2.01e+3          & 2.01e+3          & 9.96e-1           & 5.80e-1          \\ \cline{2-7} 
                            & IMPALA-PPO          & MGDT-PPO                 & 2.01e+3          & 2.01e+3          & 8.52e-1           & 3.07e-1          \\ \cline{2-7} 
                            & IMPALA-V\_TRACE     & MGDT-DQN                 & 2.01e+3          & 2.01e+3          & 9.76e-1           & 6.28e-1          \\ \cline{2-7} 
                            & IMPALA-V\_TRACE     & MGDT-MAENT               & 2.01e+3          & 2.01e+3          & 7.81e-1           & 7.40e-1          \\ \cline{2-7} 
                            & IMPALA-V\_TRACE     & MGDT-PPO                 & 2.01e+3          & 2.01e+3          & 1.00e+0           & 5.52e-1          \\ \cline{2-7} 
                            & MGDT-DQN            & MGDT-MAENT               & 2.01e+3          & 2.01e+3          & 9.83e-1           & 6.19e-1          \\ \cline{2-7} 
                            & MGDT-DQN            & MGDT-PPO                 & 2.01e+3          & 2.01e+3          & 9.99e-1           & 4.34e-1          \\ \cline{2-7} 
                            & MGDT-MAENT          & MGDT-PPO                 & 2.01e+3          & 2.01e+3          & 9.07e-1           & 3.21e-1          \\ \hline
\end{tabular}
}
\begin{tablenotes}
     \item mean(A) and mean(B) refer to makespan values.
     
 \end{tablenotes}
\end{table}
Statistically, both generalist agents significantly outperform the baseline on the $30 \times 20$ instance of the PDR task, in terms of makespan, and cumulative reward across all fine-tuning data budgets with CLES values close to 0. 
\begin{table}[t]
\caption{Results of post-hoc tests analysis of training time performance by the baseline and generalist agents on PDR task on the $30 \times 20$ instance (in bold are DRL configurations where p-value is $<$ 0.05 and have greater performance w.r.t the effect size).}
\label{tab:Results of  post-hoc tests analysis of training time performance by the Baseline and generalist agents on PDR task3020}
\centering
\resizebox{\textwidth}{!}{
\begin{tabular}{|c|c|c|c|c|c|c|}
\hline
Data budgets & A                        & B                        & mean(A)          & mean(B)          & pval             & CLES               \\ \hline
\multirow{15}{*}{1\%}     & Baseline                 & \textbf{IMPALA-PPO}      & 7.42e+4          & \textbf{1.37e+3} & \textbf{2.91e-2} & \textbf{9.94e-1}   \\ \cline{2-7} 
                        & Baseline                 & \textbf{IMPALA-V\_TRACE} & 7.42e+4          & \textbf{1.31e+3} & \textbf{2.90e-2} & \textbf{9.96e-1}   \\ \cline{2-7} 
                        & Baseline                 & MGDT-DQN                 & 7.42e+4          & 6.77e+4          & 9.94e-1          & 5.86e-1            \\ \cline{2-7} 
                        & Baseline                 & MGDT-MAENT               & 7.42e+4          & 1.80e+4          & 6.90e-2          & 9.72e-1            \\ \cline{2-7} 
                        & \textbf{Baseline}        & MGDT-PPO                 & \textbf{7.42e+4} & 1.59e+5          & \textbf{1.46e-2} & \textbf{2.40e-3}   \\ \cline{2-7} 
                        & IMPALA-PPO               & IMPALA-V\_TRACE          & 1.37e+3          & 1.31e+3          & 1.00e+0          & 5.43e-1            \\ \cline{2-7} 
                        & \textbf{IMPALA-PPO}      & MGDT-DQN                 & \textbf{1.37e+3} & 6.77e+4          & \textbf{1.31e-5} & \textbf{1.29e-74}  \\ \cline{2-7} 
                        & \textbf{IMPALA-PPO}      & MGDT-MAENT               & \textbf{1.37e+3} & 1.80e+4          & \textbf{2.35e-5} & \textbf{1.02e-55}  \\ \cline{2-7} 
                        & \textbf{IMPALA-PPO}      & MGDT-PPO                 & \textbf{1.37e+3} & 1.59e+5          & \textbf{3.91e-6} & \textbf{2.63e-135} \\ \cline{2-7} 
                        & \textbf{IMPALA-V\_TRACE} & MGDT-DQN                 & \textbf{1.31e+3} & 6.77e+4          & \textbf{8.78e-6} & \textbf{5.86e-82}  \\ \cline{2-7} 
                        & \textbf{IMPALA-V\_TRACE} & MGDT-MAENT               & \textbf{1.31e+3} & 1.80e+4          & \textbf{7.11e-7} & \textbf{9.24e-50}  \\ \cline{2-7} 
                        & \textbf{IMPALA-V\_TRACE} & MGDT-PPO                 & \textbf{1.31e+3} & 1.59e+5          & \textbf{3.35e-6} & \textbf{6.59e-151} \\ \cline{2-7} 
                        & MGDT-DQN                 & \textbf{MGDT-MAENT}      & 6.77e+4          & \textbf{1.80e+4} & \textbf{1.21e-5} & \textbf{1.00e+0}   \\ \cline{2-7} 
                        & \textbf{MGDT-DQN}        & MGDT-PPO                 & 6.77e+4          & 1.59e+5          & 5.99e-7          & 1.27e-35           \\ \cline{2-7} 
                        & \textbf{MGDT-MAENT}      & MGDT-PPO                 & 1.80e+4          & 1.59e+5          & 3.62e-6          & 7.92e-106          \\ \hline
\multirow{15}{*}{2\%}     & Baseline                 & \textbf{IMPALA-PPO}      & 7.42e+4          & \textbf{2.87e+3} & \textbf{3.13e-2} & \textbf{9.93e-1}   \\ \cline{2-7} 
                        & Baseline                 & \textbf{IMPALA-V\_TRACE} & 7.42e+4          & \textbf{2.60e+3} & \textbf{3.07e-2} & \textbf{9.95e-1}   \\ \cline{2-7} 
                        & \textbf{Baseline}        & MGDT-DQN                 & \textbf{7.42e+4} & 1.43e+5          & \textbf{2.85e-2} & \textbf{3.14e-2}   \\ \cline{2-7} 
                        & Baseline                 & \textbf{MGDT-MAENT}      & 7.42e+4          & \textbf{4.20e+4} & \textbf{3.56e-1} & \textbf{8.37e-1}   \\ \cline{2-7} 
                        & \textbf{Baseline}        & MGDT-PPO                 & \textbf{7.42e+4} & 2.77e+5          & \textbf{3.75e-2} & \textbf{1.81e-2}   \\ \cline{2-7} 
                        & IMPALA-PPO               & IMPALA-V\_TRACE          & 2.87e+3          & 2.60e+3          & 9.95e-1          & 5.87e-1            \\ \cline{2-7} 
                        & \textbf{IMPALA-PPO}      & MGDT-DQN                 & \textbf{2.87e+3} & 1.43e+5          & \textbf{9.71e-4} & \textbf{3.67e-10}  \\ \cline{2-7} 
                        & \textbf{IMPALA-PPO}      & MGDT-MAENT               & \textbf{2.87e+3} & 4.20e+4          & \textbf{2.30e-2} & \textbf{3.92e-3}   \\ \cline{2-7} 
                        & \textbf{IMPALA-PPO}      & MGDT-PPO                 & \textbf{2.87e+3} & 2.77e+5          & \textbf{1.55e-2} & \textbf{1.49e-3}   \\ \cline{2-7} 
                        & \textbf{IMPALA-V\_TRACE} & MGDT-DQN                 & \textbf{2.60e+3} & 1.43e+5          & \textbf{9.37e-4} & \textbf{3.18e-11}  \\ \cline{2-7} 
                        & \textbf{IMPALA-V\_TRACE} & MGDT-MAENT               & \textbf{2.60e+3} & 4.20e+4          & \textbf{2.19e-2} & \textbf{2.33e-3}   \\ \cline{2-7} 
                        & \textbf{IMPALA-V\_TRACE} & MGDT-PPO                 & \textbf{2.60e+3} & 2.77e+5          & \textbf{1.54e-2} & \textbf{8.06e-4}   \\ \cline{2-7} 
                        & MGDT-DQN                 & \textbf{MGDT-MAENT}      & 1.43e+5          & \textbf{4.20e+4} & \textbf{6.79e-4} & \textbf{1.00e+0}   \\ \cline{2-7} 
                        & MGDT-DQN                 & MGDT-PPO                 & 1.43e+5          & 2.77e+5          & 1.57e-1          & 7.95e-2            \\ \cline{2-7} 
                        & \textbf{MGDT-MAENT}      & MGDT-PPO                 & \textbf{4.20e+4} & 2.77e+5          & \textbf{2.50e-2} & \textbf{5.94e-3}   \\ \hline
\end{tabular}
}
\begin{tablenotes}
     \item mean(A) and mean(B) refer to training  time values.
     
 \end{tablenotes}
\end{table}
The CLES values indicate that with a 0\% chance, the baseline will have a higher makespan and lower cumulative reward earned than the generalist agents. 
\begin{table}[t]
\caption{Results of post-hoc tests analysis of cumulative reward performance by the baseline and generalist agents on the PDR task on the $30 \times 20$ instance (in bold are DRL configurations where p-value is $<$ 0.05 and have greater performance w.r.t the effect size).}
\label{tab:Results of  post-hoc tests analysis of cumulative reward  performance by the Baseline and generalist agents on PDR task3020}
\centering
\resizebox{\textwidth}{!}{
\begin{tabular}{|c|c|c|c|c|c|c|}
\hline
Data budgets     & A               & B                        & mean(A)  & mean(B)           & pval              & CLES             \\ \hline
\multirow{15}{*}{zero-shot} & Baseline        & \textbf{IMPALA-PPO}      & -2.48e+3 & \textbf{-1.26e+3} & \textbf{0.0}  & \textbf{0.0} \\ \cline{2-7} 
                            & Baseline        & \textbf{IMPALA-V\_TRACE} & -2.48e+3 & \textbf{-1.26e+3} & \textbf{0.0}  & \textbf{0.0} \\ \cline{2-7} 
                            & Baseline        & \textbf{MGDT-DQN}        & -2.48e+3 & \textbf{-1.26e+3} & \textbf{4.21e-12} & \textbf{0.0} \\ \cline{2-7} 
                            & Baseline        & \textbf{MGDT-MAENT}      & -2.48e+3 & \textbf{-1.26e+3} & \textbf{4.21e-12} & \textbf{0.0} \\ \cline{2-7} 
                            & Baseline        & \textbf{MGDT-PPO}        & -2.48e+3 & \textbf{-1.26e+3} & \textbf{4.21e-12} & \textbf{0.0} \\ \cline{2-7} 
                            & IMPALA-PPO      & IMPALA-V\_TRACE          & -1.26e+3 & -1.26e+3          & 1.00e+0           & 5.00e-1          \\ \cline{2-7} 
                            & IMPALA-PPO      & MGDT-DQN                 & -1.26e+3 & -1.26e+3          & 1.00e+0           & 4.68e-1          \\ \cline{2-7} 
                            & IMPALA-PPO      & MGDT-MAENT               & -1.26e+3 & -1.26e+3          & 1.00e+0           & 4.68e-1          \\ \cline{2-7} 
                            & IMPALA-PPO      & MGDT-PPO                 & -1.26e+3 & -1.26e+3          & 1.00e+0           & 4.68e-1          \\ \cline{2-7} 
                            & IMPALA-V\_TRACE & MGDT-DQN                 & -1.26e+3 & -1.26e+3          & 1.00e+0           & 4.68e-1          \\ \cline{2-7} 
                            & IMPALA-V\_TRACE & MGDT-MAENT               & -1.26e+3 & -1.26e+3          & 1.00e+0           & 4.68e-1          \\ \cline{2-7} 
                            & IMPALA-V\_TRACE & MGDT-PPO                 & -1.26e+3 & -1.26e+3          & 1.00e+0           & 4.68e-1          \\ \cline{2-7} 
                            & MGDT-DQN        & MGDT-MAENT               & -1.26e+3 & -1.26e+3          & 1.00e+0           & 5.00e-1          \\ \cline{2-7} 
                            & MGDT-DQN        & MGDT-PPO                 & -1.26e+3 & -1.26e+3          & 1.00e+0           & 5.00e-1          \\ \cline{2-7} 
                            & MGDT-MAENT      & MGDT-PPO                 & -1.26e+3 & -1.26e+3          & 1.00e+0           & 5.00e-1          \\ \hline
\multirow{15}{*}{1\%}         & Baseline        & \textbf{IMPALA-PPO}      & -2.48e+3 & \textbf{-1.26e+3} & \textbf{0.0}  & \textbf{0.0} \\ \cline{2-7} 
                            & Baseline        & \textbf{IMPALA-V\_TRACE} & -2.48e+3 & \textbf{-1.44e+3} & \textbf{4.50e-2}  & \textbf{2.53e-3} \\ \cline{2-7} 
                            & Baseline        & \textbf{MGDT-DQN}        & -2.48e+3 & \textbf{-1.26e+3} & \textbf{0.0}  & \textbf{0.0} \\ \cline{2-7} 
                            & Baseline        & \textbf{MGDT-MAENT}      & -2.48e+3 & \textbf{-1.26e+3} & \textbf{4.19e-12} & \textbf{0.0} \\ \cline{2-7} 
                            & Baseline        & \textbf{MGDT-PPO}        & -2.48e+3 & \textbf{-1.26e+3} & \textbf{0.0}  & \textbf{0.0} \\ \cline{2-7} 
                            & IMPALA-PPO      & IMPALA-V\_TRACE          & -1.26e+3 & -1.44e+3          & 8.90e-1           & 6.84e-1          \\ \cline{2-7} 
                            & IMPALA-PPO      & MGDT-DQN                 & -1.26e+3 & -1.26e+3          & 4.19e-1           & 8.15e-1          \\ \cline{2-7} 
                            & IMPALA-PPO      & MGDT-MAENT               & -1.26e+3 & -1.26e+3          & 5.62e-1           & 2.15e-1          \\ \cline{2-7} 
                            & IMPALA-PPO      & MGDT-PPO                 & -1.26e+3 & -1.26e+3          & 8.35e-1           & 7.00e-1          \\ \cline{2-7} 
                            & IMPALA-V\_TRACE & MGDT-DQN                 & -1.44e+3 & -1.26e+3          & 8.92e-1           & 3.17e-1          \\ \cline{2-7} 
                            & IMPALA-V\_TRACE & MGDT-MAENT               & -1.44e+3 & -1.26e+3          & 8.88e-1           & 3.15e-1          \\ \cline{2-7} 
                            & IMPALA-V\_TRACE & MGDT-PPO                 & -1.44e+3 & -1.26e+3          & 8.91e-1           & 3.17e-1          \\ \cline{2-7} 
                            & MGDT-DQN        & MGDT-MAENT               & -1.26e+3 & -1.26e+3          & 9.37e-2           & 4.08e-2          \\ \cline{2-7} 
                            & MGDT-DQN        & MGDT-PPO                 & -1.26e+3 & -1.26e+3          & 9.99e-1           & 4.43e-1          \\ \cline{2-7} 
                            & MGDT-MAENT      & MGDT-PPO                 & -1.26e+3 & -1.26e+3          & 3.76e-1           & 8.43e-1          \\ \hline
\multirow{15}{*}{2\%}         & Baseline        & \textbf{IMPALA-PPO}      & -2.48e+3 & -1.26e+3          & 5.88e-14          & 0.0          \\ \cline{2-7} 
                            & Baseline        & \textbf{IMPALA-V\_TRACE} & -2.48e+3 & \textbf{-1.43e+3} & \textbf{4.27e-2}  & \textbf{2.14e-3} \\ \cline{2-7} 
                            & Baseline        & \textbf{MGDT-DQN}        & -2.48e+3 & \textbf{-1.26e+3} & \textbf{3.51e-13} & \textbf{0.0} \\ \cline{2-7} 
                            & Baseline        & \textbf{MGDT-MAENT}      & -2.48e+3 & \textbf{-1.26e+3} & \textbf{0.0}  & \textbf{0.0} \\ \cline{2-7} 
                            & Baseline        & \textbf{MGDT-PPO}        & -2.48e+3 & \textbf{-1.26e+3} & \textbf{1.63e-14} & \textbf{0.0} \\ \cline{2-7} 
                            & IMPALA-PPO      & IMPALA-V\_TRACE          & -1.26e+3 & -1.43e+3          & 8.90e-1           & 6.84e-1          \\ \cline{2-7} 
                            & IMPALA-PPO      & MGDT-DQN                 & -1.26e+3 & -1.26e+3          & 5.63e-1           & 2.22e-1          \\ \cline{2-7} 
                            & IMPALA-PPO      & MGDT-MAENT               & -1.26e+3 & -1.26e+3          & 9.61e-1           & 6.39e-1          \\ \cline{2-7} 
                            & IMPALA-PPO      & MGDT-PPO                 & -1.26e+3 & -1.26e+3          & 7.70e-1           & 7.22e-1          \\ \cline{2-7} 
                            & IMPALA-V\_TRACE & MGDT-DQN                 & -1.43e+3 & -1.26e+3          & 8.90e-1           & 3.16e-1          \\ \cline{2-7} 
                            & IMPALA-V\_TRACE & MGDT-MAENT               & -1.43e+3 & -1.26e+3          & 8.91e-1           & 3.08e-1          \\ \cline{2-7} 
                            & IMPALA-V\_TRACE & MGDT-PPO                 & -1.43e+3 & -1.26e+3          & 8.91e-1           & 3.17e-1          \\ \cline{2-7} 
                            & MGDT-DQN        & MGDT-MAENT               & -1.26e+3 & -1.26e+3          & 8.45e-1           & 7.04e-1          \\ \cline{2-7} 
                            & MGDT-DQN        & MGDT-PPO                 & -1.26e+3 & -1.26e+3          & 1.80e-1           & 8.97e-1          \\ \cline{2-7} 
                            & MGDT-MAENT      & MGDT-PPO                 & -1.26e+3 & -1.26e+3          & 9.99e-1           & 4.36e-1          \\ \hline
\end{tabular}
}
\begin{tablenotes}
     \item mean(A) and mean(B) refer to cumulative reward values.
     
 \end{tablenotes}
\end{table}
In terms of training time across the 1\% and 2\% fine-tuning data budgets, the IMPALA agent configurations perform significantly better. In terms of testing time the baseline on the $30 \times 20$ instance of the PDR task performs best, as reported in Table \ref{tab:Results of  post-hoc tests analysis of testing time performance by the Baseline and generalist agents on PDR task3020}.
\begin{table}[t]
\caption{Results of post-hoc tests analysis of testing time by the baseline and generalist agents on PDR task on the $30 \times 20$ instance (in bold are DRL configurations where p-value is $<$ 0.05 and have greater performance w.r.t the effect size).}
\label{tab:Results of  post-hoc tests analysis of testing time performance by the Baseline and generalist agents on PDR task3020}
\centering
\resizebox{\textwidth}{!}{
\begin{tabular}{|c|c|c|c|c|c|c|}
\hline
Data budgets    & A                 & B                   & mean(A)          & mean(B)          & pval             & CLES               \\ \hline
\multirow{15}{*}{zero-shot} & \textbf{Baseline} & IMPALA-PPO          & \textbf{1.73e+2} & 1.05e+4          & \textbf{1.80e-4} & \textbf{1.73e-21}  \\ \cline{2-7} 
                            & \textbf{Baseline} & IMPALA-V\_TRACE     & \textbf{1.73e+2} & 1.05e+4          & \textbf{1.80e-4} & \textbf{1.73e-21}  \\ \cline{2-7} 
                            & \textbf{Baseline} & MGDT-DQN            & \textbf{1.73e+2} & 2.70e+3          & \textbf{3.46e-5} & \textbf{1.95e-46}  \\ \cline{2-7} 
                            & \textbf{Baseline} & MGDT-MAENT          & \textbf{1.73e+2} & 2.70e+3          & \textbf{3.46e-5} & \textbf{1.95e-46}  \\ \cline{2-7} 
                            & \textbf{Baseline} & MGDT-PPO            & \textbf{1.73e+2} & 2.70e+3          & \textbf{3.46e-5} & \textbf{1.95e-46}  \\ \cline{2-7} 
                            & IMPALA-PPO        & IMPALA-V\_TRACE     & 1.05e+4          & 1.05e+4          & 1.00e+0          & 5.00e-1            \\ \cline{2-7} 
                            & IMPALA-PPO        & \textbf{MGDT-DQN}   & 1.05e+4          & \textbf{2.70e+3} & \textbf{4.19e-4} & \textbf{1.00e+0}   \\ \cline{2-7} 
                            & IMPALA-PPO        & \textbf{MGDT-MAENT} & 1.05e+4          & \textbf{2.70e+3} & \textbf{4.19e-4} & \textbf{1.00e+0}   \\ \cline{2-7} 
                            & IMPALA-PPO        & \textbf{MGDT-PPO}   & 1.05e+4          & \textbf{2.70e+3} & \textbf{4.19e-4} & \textbf{1.00e+0}   \\ \cline{2-7} 
                            & IMPALA-V\_TRACE   & \textbf{MGDT-DQN}   & 1.05e+4          & \textbf{2.70e+3} & \textbf{4.19e-4} & \textbf{1.00e+0}   \\ \cline{2-7} 
                            & IMPALA-V\_TRACE   & \textbf{MGDT-MAENT} & 1.05e+4          & \textbf{2.70e+3} & \textbf{4.19e-4} & \textbf{1.00e+0}   \\ \cline{2-7} 
                            & IMPALA-V\_TRACE   & \textbf{MGDT-PPO}   & 1.05e+4          & \textbf{2.70e+3} & \textbf{4.19e-4} & \textbf{1.00e+0}   \\ \cline{2-7} 
                            & MGDT-DQN          & MGDT-MAENT          & 2.70e+3          & 2.70e+3          & 1.00e+0          & 5.00e-1            \\ \cline{2-7} 
                            & MGDT-DQN          & MGDT-PPO            & 2.70e+3          & 2.70e+3          & 1.00e+0          & 5.00e-1            \\ \cline{2-7} 
                            & MGDT-MAENT        & MGDT-PPO            & 2.70e+3          & 2.70e+3          & 1.00e+0          & 5.00e-1            \\ \hline
\multirow{15}{*}{1\%}         & \textbf{Baseline} & IMPALA-PPO          & \textbf{1.73e+2} & 9.35e+3          & \textbf{1.36e-5} & \textbf{2.46e-73}  \\ \cline{2-7} 
                            & \textbf{Baseline} & IMPALA-V\_TRACE     & \textbf{1.73e+2} & 1.02e+4          & \textbf{1.13e-4} & \textbf{5.45e-104} \\ \cline{2-7} 
                            & \textbf{Baseline} & MGDT-DQN            & \textbf{1.73e+2} & 2.47e+3          & \textbf{2.96e-7} & \textbf{0.0}   \\ \cline{2-7} 
                            & \textbf{Baseline} & MGDT-MAENT          & \textbf{1.73e+2} & 3.01e+3          & \textbf{8.64e-3} & \textbf{2.50e-4}   \\ \cline{2-7} 
                            & \textbf{Baseline} & MGDT-PPO            & \textbf{1.73e+2} & 2.47e+3          & \textbf{1.07e-9} & \textbf{0.0}   \\ \cline{2-7} 
                            & IMPALA-PPO        & IMPALA-V\_TRACE     & 9.35e+3          & 1.02e+4          & 1.75e-1          & 9.42e-2            \\ \cline{2-7} 
                            & IMPALA-PPO        & \textbf{MGDT-DQN}   & 9.35e+3          & \textbf{2.47e+3} & \textbf{3.68e-5} & \textbf{1.00e+0}   \\ \cline{2-7} 
                            & IMPALA-PPO        & \textbf{MGDT-MAENT} & 9.35e+3          & \textbf{3.01e+3} & \textbf{2.04e-5} & \textbf{1.00e+0}   \\ \cline{2-7} 
                            & IMPALA-PPO        & \textbf{MGDT-PPO}   & 9.35e+3          & \textbf{2.47e+3} & \textbf{4.21e-5} & \textbf{1.00e+0}   \\ \cline{2-7} 
                            & IMPALA-V\_TRACE   & \textbf{MGDT-DQN}   & 1.02e+4          & \textbf{2.47e+3} & \textbf{2.14e-4} & \textbf{1.00e+0}   \\ \cline{2-7} 
                            & IMPALA-V\_TRACE   & \textbf{MGDT-MAENT} & 1.02e+4          & \textbf{3.01e+3} & \textbf{1.38e-5} & \textbf{1.00e+0}   \\ \cline{2-7} 
                            & IMPALA-V\_TRACE   & \textbf{MGDT-PPO}   & 1.02e+4          & \textbf{2.47e+3} & \textbf{2.41e-4} & \textbf{1.00e+0}   \\ \cline{2-7} 
                            & MGDT-DQN          & MGDT-MAENT          & 2.47e+3          & 3.01e+3          & 7.00e-1          & 2.56e-1            \\ \cline{2-7} 
                            & MGDT-DQN          & MGDT-PPO            & 2.47e+3          & 2.47e+3          & 1.00e+0          & 5.10e-1            \\ \cline{2-7} 
                            & MGDT-MAENT        & MGDT-PPO            & 3.01e+3          & 2.47e+3          & 6.97e-1          & 7.45e-1            \\ \hline
\multirow{15}{*}{2\%}         & \textbf{Baseline} & IMPALA-PPO          & \textbf{1.73e+2} & 9.50e+3          & \textbf{1.18e-5} & \textbf{9.36e-79}  \\ \cline{2-7} 
                            & \textbf{Baseline} & IMPALA-V\_TRACE     & \textbf{1.73e+2} & 9.73e+3          & \textbf{1.20e-4} & \textbf{7.84e-100} \\ \cline{2-7} 
                            & \textbf{Baseline} & MGDT-DQN            & \textbf{1.73e+2} & 2.50e+3          & \textbf{6.70e-7} & \textbf{5.62e-297} \\ \cline{2-7} 
                            & \textbf{Baseline} & MGDT-MAENT          & \textbf{1.73e+2} & 2.57e+3          & \textbf{9.07e-5} & \textbf{2.38e-117} \\ \cline{2-7} 
                            & \textbf{Baseline} & MGDT-PPO            & \textbf{1.73e+2} & 2.06e+3          & \textbf{1.21e-2} & \textbf{7.46e-4}   \\ \cline{2-7} 
                            & IMPALA-PPO        & IMPALA-V\_TRACE     & 9.50e+3          & 9.73e+3          & 9.71e-1          & 3.64e-1            \\ \cline{2-7} 
                            & IMPALA-PPO        & \textbf{MGDT-DQN}   & 9.50e+3          & \textbf{2.50e+3} & \textbf{2.90e-5} & \textbf{1.00e+0}   \\ \cline{2-7} 
                            & IMPALA-PPO        & \textbf{MGDT-MAENT} & 9.50e+3          & \textbf{2.57e+3} & \textbf{1.85e-5} & \textbf{1.00e+0}   \\ \cline{2-7} 
                            & IMPALA-PPO        & \textbf{MGDT-PPO}   & 9.50e+3          & \textbf{2.06e+3} & \textbf{3.23e-7} & \textbf{1.00e+0}   \\ \cline{2-7} 
                            & IMPALA-V\_TRACE   & \textbf{MGDT-DQN}   & 9.73e+3          & \textbf{2.50e+3} & \textbf{2.24e-4} & \textbf{1.00e+0}   \\ \cline{2-7} 
                            & IMPALA-V\_TRACE   & \textbf{MGDT-MAENT} & 9.73e+3          & \textbf{2.57e+3} & \textbf{1.49e-4} & \textbf{1.00e+0}   \\ \cline{2-7} 
                            & IMPALA-V\_TRACE   & \textbf{MGDT-PPO}   & 9.73e+3          & \textbf{2.06e+3} & \textbf{8.25e-7} & \textbf{1.00e+0}   \\ \cline{2-7} 
                            & MGDT-DQN          & MGDT-MAENT          & 2.50e+3          & 2.57e+3          & 8.17e-1          & 2.83e-1            \\ \cline{2-7} 
                            & MGDT-DQN          & MGDT-PPO            & 2.50e+3          & 2.06e+3          & 6.00e-1          & 7.74e-1            \\ \cline{2-7} 
                            & MGDT-MAENT        & MGDT-PPO            & 2.57e+3          & 2.06e+3          & 4.94e-1          & 8.18e-1            \\ \hline
\end{tabular}
}
\begin{tablenotes}
     \item mean(A) and mean(B) refer to testing time values.
     
 \end{tablenotes}
\end{table}
\begin{tcolorbox}
    \textbf{Finding 3: On both studied instances, the fine-tuned generalist agents show statistically significantly better performance compared to the PDR task's baseline in terms of makespan and cumulative reward earned on all fine-tuning data budgets. Suggesting that generalist agents' capabilities are particularly well-suited for scheduling-based tasks.}
\end{tcolorbox}

Regarding the MsPacman game, we observe that the IMPALA generalist agent on its V\_TRACE configuration outperforms the baseline on average over 5 runs in terms of number of detected bugs of types 3 and 4  during zero-shot fine-tuning data budget. Specifically, the IMPALA agent detects on average over 5 runs, 18.5\%, and 16\% more bugs of type 3 and type 4, respectively. Even though MGDT agents perform poorly in detecting bugs in comparison to the baseline, it is important to note that it only detects bugs of types 3 and 4 located at the lower side of MsPacman's environment during all fine-tuning data budgets. Further, similarly to its performance in the Blockmaze game, its exploration capability is limited to certain areas of the game environment. In terms of training time, MGDT agents achieved the lowest training (fine-tuning) time in comparison to the baseline during the 1\% fine-tuning data budget. In terms of testing time the baseline performs best during all fine-tuning data budgets.  
\begin{table}[t]
\caption{Results of post-hoc tests analysis of the number of bugs detected by the baseline and the generalist agents on MsPacman game (in bold are DRL configurations where the p-value is $<$ 0.05 and have superior performance w.r.t the effect size).}
\label{tab:Results of  post-hoc tests analysis of the number of bugs detected by the Baseline and the generalist agents on MsPacman task}
\resizebox{\textwidth}{!}{
\centering
\begin{tabular}{|c|c|c|c|c|c|c|}
\hline
Data budgets    & A                        & B                        & mean(A)          & mean(B)          & pval              & CLES             \\ \hline
\multirow{15}{*}{zero-shot} & \textbf{BASELINE}        & IMPALA-PPO               & \textbf{2.92e+3} & 6.81e+2          & \textbf{2.12e-7}  & \textbf{1.00e+0} \\ \cline{2-7} 
                            & BASELINE                 & IMPALA-V\_TRACE          & 2.92e+3          & 2.80e+3          & 9.74e-1           & 6.21e-1          \\ \cline{2-7} 
                            & \textbf{BASELINE}        & MGDT-DQN                 & \textbf{2.92e+3} & 2.98e+1          & \textbf{1.01e-6}  & \textbf{1.00e+0} \\ \cline{2-7} 
                            & \textbf{BASELINE}        & MGDT-MAENT               & \textbf{2.92e+3} & 2.03e+2          & \textbf{1.95e-10} & \textbf{1.00e+0} \\ \cline{2-7} 
                            & \textbf{BASELINE}        & MGDT-PPO                 & \textbf{2.92e+3} & 0.0          & \textbf{1.00e-6}  & \textbf{1.00e+0} \\ \cline{2-7} 
                            & IMPALA-PPO               & \textbf{IMPALA-V\_TRACE} & 6.81e+2          & \textbf{2.80e+3} & \textbf{1.15e-3}  & \textbf{3.46e-9} \\ \cline{2-7} 
                            & \textbf{IMPALA-PPO}      & MGDT-DQN                 & \textbf{6.81e+2} & 2.98e+1          & \textbf{5.99e-6}  & \textbf{1.00e+0} \\ \cline{2-7} 
                            & \textbf{IMPALA-PPO}      & MGDT-MAENT               & \textbf{6.81e+2} & 2.03e+2          & \textbf{2.66e-4}  & \textbf{1.00e+0} \\ \cline{2-7} 
                            & \textbf{IMPALA-PPO}      & MGDT-PPO                 & \textbf{6.81e+2} & 0.0          & \textbf{6.17e-6}  & \textbf{1.00e+0} \\ \cline{2-7} 
                            & \textbf{IMPALA-V\_TRACE} & MGDT-DQN                 & \textbf{2.80e+3} & 2.98e+1          & \textbf{4.26e-4}  & \textbf{1.00e+0} \\ \cline{2-7} 
                            & \textbf{IMPALA-V\_TRACE} & MGDT-MAENT               & \textbf{2.80e+3} & 2.03e+2          & \textbf{3.55e-4}  & \textbf{1.00e+0} \\ \cline{2-7} 
                            & \textbf{IMPALA-V\_TRACE} & MGDT-PPO                 & \textbf{2.80e+3} & 0.0          & \textbf{4.09e-4}  & \textbf{1.00e+0} \\ \cline{2-7} 
                            & MGDT-DQN                 & \textbf{MGDT-MAENT}      & 2.98e+1          & \textbf{2.03e+2} & \textbf{4.12e-2}  & \textbf{1.25e-2} \\ \cline{2-7} 
                            & \textbf{MGDT-DQN}        & MGDT-PPO                 & \textbf{2.98e+1} & 0.0          & \textbf{2.08e-4}  & \textbf{1.00e+0} \\ \cline{2-7} 
                            & \textbf{MGDT-MAENT}      & MGDT-PPO                 & \textbf{2.03e+2} & 0.0          & \textbf{2.39e-2}  & \textbf{9.96e-1} \\ \hline
\multirow{15}{*}{1\%}         & \textbf{BASELINE}        & IMPALA-PPO               & \textbf{2.92e+3} & 6.09e+2          & \textbf{1.56e-8}  & \textbf{1.00e+0} \\ \cline{2-7} 
                            & \textbf{BASELINE}        & IMPALA-V\_TRACE          & \textbf{2.92e+3} & 6.47e+2          & \textbf{4.29e-9}  & \textbf{1.00e+0} \\ \cline{2-7} 
                            & \textbf{BASELINE}        & MGDT-DQN                 & \textbf{2.92e+3} & 3.22e+1          & \textbf{9.97e-7}  & \textbf{1.00e+0} \\ \cline{2-7} 
                            & \textbf{BASELINE}        & MGDT-MAENT               & \textbf{2.92e+3} & 1.49e+2          & \textbf{9.78e-8}  & \textbf{1.00e+0} \\ \cline{2-7} 
                            & \textbf{BASELINE}        & MGDT-PPO                 & \textbf{2.92e+3} & 0.0          & \textbf{1.00e-6}  & \textbf{1.00e+0} \\ \cline{2-7} 
                            & IMPALA-PPO               & IMPALA-V\_TRACE          & 6.09e+2          & 6.47e+2          & 8.30e-1           & 2.97e-1          \\ \cline{2-7} 
                            & \textbf{IMPALA-PPO}      & MGDT-DQN                 & \textbf{6.09e+2} & 3.22e+1          & \textbf{4.85e-5}  & \textbf{1.00e+0} \\ \cline{2-7} 
                            & \textbf{IMPALA-PPO}      & MGDT-MAENT               & \textbf{6.09e+2} & 1.49e+2          & \textbf{3.05e-6}  & \textbf{1.00e+0} \\ \cline{2-7} 
                            & \textbf{IMPALA-PPO}      & MGDT-PPO                 & \textbf{6.09e+2} & 0.0          & \textbf{4.37e-5}  & \textbf{1.00e+0} \\ \cline{2-7} 
                            & \textbf{IMPALA-V\_TRACE} & MGDT-DQN                 & \textbf{6.47e+2} & 3.22e+1          & \textbf{7.88e-5}  & \textbf{1.00e+0} \\ \cline{2-7} 
                            & \textbf{IMPALA-V\_TRACE} & MGDT-MAENT               & \textbf{6.47e+2} & 1.49e+2          & \textbf{1.16e-5}  & \textbf{1.00e+0} \\ \cline{2-7} 
                            & \textbf{IMPALA-V\_TRACE} & MGDT-PPO                 & \textbf{6.47e+2} & 0.0          & \textbf{6.93e-5}  & \textbf{1.00e+0} \\ \cline{2-7} 
                            & MGDT-DQN                 & \textbf{MGDT-MAENT}      & 3.22e+1          & \textbf{1.49e+2} & \textbf{4.53e-3}  & \textbf{3.41e-5} \\ \cline{2-7} 
                            & \textbf{MGDT-DQN}        & MGDT-PPO                 & \textbf{3.22e+1} & 0.0          & \textbf{3.27e-4}  & \textbf{1.00e+0} \\ \cline{2-7} 
                            & \textbf{MGDT-MAENT}      & MGDT-PPO                 & \textbf{1.49e+2} & 0.0          & \textbf{1.98e-3}  & \textbf{1.00e+0} \\ \hline
\multirow{15}{*}{2\%}         & \textbf{BASELINE}        & IMPALA-PPO               & \textbf{2.92e+3} & 6.94e+2          & \textbf{1.51e-7}  & \textbf{1.00e+0} \\ \cline{2-7} 
                            & \textbf{BASELINE}        & IMPALA-V\_TRACE          & \textbf{2.92e+3} & 6.82e+2          & \textbf{1.07e-8}  & \textbf{1.00e+0} \\ \cline{2-7} 
                            & \textbf{BASELINE}        & MGDT-DQN                 & \textbf{2.92e+3} & 3.22e+1          & \textbf{8.79e-7}  & \textbf{1.00e+0} \\ \cline{2-7} 
                            & \textbf{BASELINE}        & MGDT-MAENT               & \textbf{2.92e+3} & 2.06e+2          & \textbf{7.27e-10} & \textbf{1.00e+0} \\ \cline{2-7} 
                            & \textbf{BASELINE}        & MGDT-PPO                 & \textbf{2.92e+3} & 0.0          & \textbf{1.00e-6}  & \textbf{1.00e+0} \\ \cline{2-7} 
                            & IMPALA-PPO               & IMPALA-V\_TRACE          & 6.94e+2          & 6.82e+2          & 9.97e-1           & 5.79e-1          \\ \cline{2-7} 
                            & \textbf{IMPALA-PPO}      & MGDT-DQN                 & \textbf{6.94e+2} & 3.22e+1          & \textbf{3.78e-6}  & \textbf{1.00e+0} \\ \cline{2-7} 
                            & \textbf{IMPALA-PPO}      & MGDT-MAENT               & \textbf{6.94e+2} & 2.06e+2          & \textbf{1.54e-5}  & \textbf{1.00e+0} \\ \cline{2-7} 
                            & \textbf{IMPALA-PPO}      & MGDT-PPO                 & \textbf{6.94e+2} & 0.0          & \textbf{7.64e-6}  & \textbf{1.00e+0} \\ \cline{2-7} 
                            & \textbf{IMPALA-V\_TRACE} & MGDT-DQN                 & \textbf{6.82e+2} & 3.22e+1          & \textbf{3.19e-5}  & \textbf{1.00e+0} \\ \cline{2-7} 
                            & \textbf{IMPALA-V\_TRACE} & MGDT-MAENT               & \textbf{6.82e+2} & 2.06e+2          & \textbf{7.64e-6}  & \textbf{1.00e+0} \\ \cline{2-7} 
                            & \textbf{IMPALA-V\_TRACE} & MGDT-PPO                 & \textbf{6.82e+2} & 0.0          & \textbf{3.72e-5}  & \textbf{1.00e+0} \\ \cline{2-7} 
                            & MGDT-DQN                 & \textbf{MGDT-MAENT}      & 3.22e+1          & \textbf{2.06e+2} & \textbf{1.37e-2}  & \textbf{1.38e-3} \\ \cline{2-7} 
                            & \textbf{MGDT-DQN}        & MGDT-PPO                 & \textbf{3.22e+1} & 0.0          & \textbf{3.78e-3}  & \textbf{1.00e+0} \\ \cline{2-7} 
                            & \textbf{MGDT-MAENT}      & MGDT-PPO                 & \textbf{2.06e+2} & 0.0          & \textbf{7.80e-3}  & \textbf{1.00e+0} \\ \hline
\end{tabular}
}
\begin{tablenotes}
     \item mean(A) and mean(B) refer to the number of bugs detected.
 \end{tablenotes}
\end{table}
As mentioned in Section \ref{sec:Baselines studies}, we do not compare the generalist agents against the baseline in terms of cumulative reward earned as Tufano et al.\cite{tufano2022using} do not report them either.\\
\textbf{Statistical analysis:} Tables  \ref{tab:Results of  post-hoc tests analysis of the number of bugs detected by the Baseline and the generalist agents on MsPacman task}, \ref{tab:Results of  post-hoc tests analysis of training time performance by the baseline and the generalist agents on MsPacman task}, and \ref{tab:Results of  post-hoc tests analysis of testing time performance by the baseline and the generalist agents on MsPacman task} show the results of post-hoc test analysis for the baseline and the generalist agents on the MsPacman game involving the number of bugs detected, the training and testing time performance.
\begin{table}[t]
\caption{Results of post-hoc tests analysis of training time performance by the baseline and the generalist agents on MsPacman game (in bold are DRL configurations where p-value is $<$ 0.05 and have greater performance w.r.t the effect size).}
\label{tab:Results of  post-hoc tests analysis of training time performance by the baseline and the generalist agents on MsPacman task}
\resizebox{\textwidth}{!}{
\centering
\begin{tabular}{|c|c|c|c|c|c|c|}
\hline
Data budgets & A                 & B                   & mean(A)          & mean(B)          & pval             & CLES             \\ \hline
\multirow{15}{*}{1\%}     & \textbf{BASELINE} & IMPALA-PPO          & \textbf{8.14e+3} & 1.09e+4          & \textbf{1.81e-1} & \textbf{5.93e-2} \\ \cline{2-7} 
                        & \textbf{BASELINE} & IMPALA-V\_TRACE     & \textbf{8.14e+3} & 1.02e+4          & \textbf{2.25e-4} & \textbf{4.99e-6} \\ \cline{2-7} 
                        & BASELINE          & \textbf{MGDT-DQN}   & 8.14e+3          & \textbf{4.06e+3} & \textbf{6.32e-3} & \textbf{9.99e-1} \\ \cline{2-7} 
                        & BASELINE          & \textbf{MGDT-MAENT} & 8.14e+3          & \textbf{4.87e+3} & \textbf{9.07e-6} & \textbf{1.00e+0} \\ \cline{2-7} 
                        & BASELINE          & \textbf{MGDT-PPO }  & 8.14e+3          & \textbf{5.46e+3} & \textbf{5.20e-4} & \textbf{1.00e+0} \\ \cline{2-7} 
                        & IMPALA-PPO        & IMPALA-V\_TRACE     & 1.09e+4          & 1.02e+4          & 9.54e-1          & 6.45e-1          \\ \cline{2-7} 
                        & IMPALA-PPO        & \textbf{MGDT-DQN}   & 1.09e+4          & \textbf{4.06e+3} & \textbf{4.78e-3} & \textbf{1.00e+0} \\ \cline{2-7} 
                        & IMPALA-PPO        & \textbf{MGDT-MAENT} & 1.09e+4          & \textbf{4.87e+3} & \textbf{2.21e-2} & \textbf{1.00e+0} \\ \cline{2-7} 
                        & IMPALA-PPO        & \textbf{MGDT-PPO}   & 1.09e+4          & \textbf{5.46e+3} & \textbf{3.23e-2} & \textbf{9.99e-1} \\ \cline{2-7} 
                        & IMPALA-V\_TRACE   & \textbf{MGDT-DQN}   & 1.02e+4          & \textbf{4.06e+3} & \textbf{1.45e-3} & \textbf{1.00e+0} \\ \cline{2-7} 
                        & IMPALA-V\_TRACE   & \textbf{MGDT-MAENT} & 1.02e+4          & \textbf{4.87e+3} & \textbf{1.96e-8} & \textbf{1.00e+0} \\ \cline{2-7} 
                        & IMPALA-V\_TRACE   & \textbf{MGDT-PPO}   & 1.02e+4          & \textbf{5.46e+3} & \textbf{3.91e-6} & \textbf{1.00e+0} \\ \cline{2-7} 
                        & MGDT-DQN          & MGDT-MAENT          & 4.06e+3          & 4.87e+3          & 6.99e-1          & 2.56e-1          \\ \cline{2-7} 
                        & MGDT-DQN          & MGDT-PPO            & 4.06e+3          & 5.46e+3          & 2.81e-1          & 1.25e-1          \\ \cline{2-7} 
                        & MGDT-MAENT        & MGDT-PPO            & 4.87e+3          & 5.46e+3          & 6.08e-2          & 2.84e-2          \\ \hline
\multirow{15}{*}{2\%}     & \textbf{BASELINE} & IMPALA-PPO          & \textbf{8.14e+3} & 1.02e+4          & \textbf{1.86e-2} & \textbf{1.18e-2} \\ \cline{2-7} 
                        & \textbf{BASELINE} & IMPALA-V\_TRACE     & \textbf{8.14e+3} & 9.92e+3          & \textbf{1.90e-3} & \textbf{1.86e-3} \\ \cline{2-7} 
                        & \textbf{BASELINE} & MGDT-DQN            & \textbf{8.14e+3} & 9.09e+3          & \textbf{9.62e-1} & \textbf{3.67e-1} \\ \cline{2-7} 
                        & \textbf{BASELINE} & MGDT-MAENT          & \textbf{8.14e+3} & 8.70e+3          & \textbf{9.99e-1} & \textbf{4.43e-1} \\ \cline{2-7} 
                        & \textbf{BASELINE} & MGDT-PPO            & \textbf{8.14e+3} & 1.27e+4          & \textbf{2.38e-1} & \textbf{1.09e-1} \\ \cline{2-7} 
                        & IMPALA-PPO        & IMPALA-V\_TRACE     & 1.02e+4          & 9.92e+3          & 9.70e-1          & 6.28e-1          \\ \cline{2-7} 
                        & IMPALA-PPO        & MGDT-DQN            & 1.02e+4          & 9.09e+3          & 9.35e-1          & 6.54e-1          \\ \cline{2-7} 
                        & IMPALA-PPO        & MGDT-MAENT          & 1.02e+4          & 8.70e+3          & 9.41e-1          & 6.49e-1          \\ \cline{2-7} 
                        & IMPALA-PPO        & MGDT-PPO            & 1.02e+4          & 1.27e+4          & 6.97e-1          & 2.56e-1          \\ \cline{2-7} 
                        & IMPALA-V\_TRACE   & MGDT-DQN            & 9.92e+3          & 9.09e+3          & 9.78e-1          & 6.16e-1          \\ \cline{2-7} 
                        & IMPALA-V\_TRACE   & MGDT-MAENT          & 9.92e+3          & 8.70e+3          & 9.74e-1          & 6.21e-1          \\ \cline{2-7} 
                        & IMPALA-V\_TRACE   & MGDT-PPO            & 9.92e+3          & 1.27e+4          & 5.98e-1          & 2.26e-1          \\ \cline{2-7} 
                        & MGDT-DQN          & MGDT-MAENT          & 9.09e+3          & 8.70e+3          & 1.00e+0          & 5.32e-1          \\ \cline{2-7} 
                        & MGDT-DQN          & MGDT-PPO            & 9.09e+3          & 1.27e+4          & 5.38e-1          & 2.16e-1          \\ \cline{2-7} 
                        & MGDT-MAENT        & MGDT-PPO            & 8.70e+3          & 1.27e+4          & 5.83e-1          & 2.28e-1          \\ \hline
\end{tabular}
}
\begin{tablenotes}
     \item mean(A) and mean(B) refer to training values.
     
 \end{tablenotes}
\end{table}
The results show that the MGDT agents significantly outperform the baseline in terms of training time on the MsPacman game during the 1\% fine-tuning data budgets. At 2\% fine-tuning data budget, the baseline achieved the lowest training time. Further, the difference between the IMPALA-V\_TRACE  configuration and the baseline agent is not significant in terms of bugs detected (combined bugs of Type 1, 2, 3, and 4) at zero-shot fine-tuning data budget. At 1\% and 2\% fine-tuning data budgets, the baseline detects the highest number of bugs. In terms of testing time, the difference between the IMPALA agents and the baseline agent is not significant at zero-shot fine-tuning data budget. At 1\% and 2\% fine-tuning data budgets, the baseline achieved the lowest testing time. 
\begin{tcolorbox}
    \textbf{Finding 4: In the MsPacman game, at zero-shot, the difference between the specialist and IMPALA agents is not significant in terms of testing time, despite different training approaches. Suggesting that IMPALA agent policy is as optimized as the specialist agent with no fine-tuning effort.}
\end{tcolorbox}
\begin{table}[t]
\caption{Results of post-hoc tests analysis of testing time performance of the baseline and the generalist agents on MsPacman game (in bold are DRL configurations where the p-value is $<$ 0.05 and have superior performance w.r.t the effect size).}
\label{tab:Results of  post-hoc tests analysis of testing time performance by the baseline and the generalist agents on MsPacman task}
\resizebox{\textwidth}{!}{
\centering
\begin{tabular}{|c|c|c|c|c|c|c|}
\hline
Data budgets     & A                        & B               & mean(A)          & mean(B) & pval             & CLES               \\ \hline
\multirow{15}{*}{zero-shot} & BASELINE                 & IMPALA-PPO      & 2.57e+4          & 2.96e+4 & 4.25e-1          & 1.78e-1            \\ \cline{2-7} 
                            & BASELINE                 & IMPALA-V\_TRACE & 2.57e+4          & 3.89e+4 & 6.50e-2          & 2.86e-2            \\ \cline{2-7} 
                            & \textbf{BASELINE}        & MGDT-DQN        & \textbf{2.57e+4} & 3.90e+5 & \textbf{3.16e-6} & \textbf{3.27e-139} \\ \cline{2-7} 
                            & \textbf{BASELINE}        & MGDT-MAENT      & \textbf{2.57e+4} & 3.51e+5 & \textbf{2.21e-5} & \textbf{2.60e-56}  \\ \cline{2-7} 
                            & \textbf{BASELINE}        & MGDT-PPO        & \textbf{2.57e+4} & 6.77e+5 & \textbf{3.31e-6} & \textbf{2.26e-143} \\ \cline{2-7} 
                            & IMPALA-PPO               & IMPALA-V\_TRACE & 2.96e+4          & 3.89e+4 & 2.16e-1          & 1.21e-1            \\ \cline{2-7} 
                            & \textbf{IMPALA-PPO}      & MGDT-DQN        & \textbf{2.96e+4} & 3.90e+5 & \textbf{8.44e-7} & \textbf{1.58e-127} \\ \cline{2-7} 
                            & \textbf{IMPALA-PPO}      & MGDT-MAENT      & \textbf{2.96e+4} & 3.51e+5 & \textbf{1.33e-5} & \textbf{3.02e-53}  \\ \cline{2-7} 
                            & \textbf{IMPALA-PPO}      & MGDT-PPO        & \textbf{2.96e+4} & 6.77e+5 & \textbf{2.16e-6} & \textbf{1.27e-138} \\ \cline{2-7} 
                            & \textbf{IMPALA-V\_TRACE} & MGDT-DQN        & \textbf{3.89e+4} & 3.90e+5 & \textbf{8.46e-8} & \textbf{4.03e-107} \\ \cline{2-7} 
                            & \textbf{IMPALA-V\_TRACE} & MGDT-MAENT      & \textbf{3.89e+4} & 3.51e+5 & \textbf{5.23e-6} & \textbf{3.08e-47}  \\ \cline{2-7} 
                            & \textbf{IMPALA-V\_TRACE} & MGDT-PPO        & \textbf{3.89e+4} & 6.77e+5 & \textbf{9.42e-7} & \textbf{4.91e-129} \\ \cline{2-7} 
                            & MGDT-DQN                 & MGDT-MAENT      & 3.90e+5          & 3.51e+5 & 7.20e-2          & 9.39e-1            \\ \cline{2-7} 
                            & \textbf{MGDT-DQN}        & MGDT-PPO        & \textbf{3.90e+5} & 6.77e+5 & \textbf{2.85e-6} & \textbf{6.03e-23}  \\ \cline{2-7} 
                            & \textbf{MGDT-MAENT}      & MGDT-PPO        & \textbf{3.51e+5} & 6.77e+5 & \textbf{3.00e-7} & \textbf{1.23e-23}  \\ \hline
\multirow{15}{*}{1\%}         & \textbf{BASELINE}        & IMPALA-PPO      & \textbf{2.57e+4} & 3.01e+4 & \textbf{2.53e-2} & \textbf{2.43e-2}   \\ \cline{2-7} 
                            & \textbf{BASELINE}        & IMPALA-V\_TRACE & \textbf{2.57e+4} & 3.01e+4 & \textbf{2.47e-2} & \textbf{2.43e-2}   \\ \cline{2-7} 
                            & \textbf{BASELINE}        & MGDT-DQN        & \textbf{2.57e+4} & 3.50e+5 & \textbf{3.63e-4} & \textbf{1.47e-15}  \\ \cline{2-7} 
                            & \textbf{BASELINE}        & MGDT-MAENT      & \textbf{2.57e+4} & 4.23e+5 & \textbf{1.42e-3} & \textbf{1.16e-8}   \\ \cline{2-7} 
                            & \textbf{BASELINE}        & MGDT-PPO        & \textbf{2.57e+4} & 5.74e+5 & \textbf{9.27e-9} & \textbf{0.0}   \\ \cline{2-7} 
                            & IMPALA-PPO               & IMPALA-V\_TRACE & 3.01e+4          & 3.01e+4 & 1.00e+0          & 5.08e-1            \\ \cline{2-7} 
                            & \textbf{IMPALA-PPO}      & MGDT-DQN        & \textbf{3.01e+4} & 3.50e+5 & \textbf{3.78e-4} & \textbf{3.58e-15}  \\ \cline{2-7} 
                            & \textbf{IMPALA-PPO}      & MGDT-MAENT      & \textbf{3.01e+4} & 4.23e+5 & \textbf{1.48e-3} & \textbf{1.66e-8}   \\ \cline{2-7} 
                            & \textbf{IMPALA-PPO}      & MGDT-PPO        & \textbf{3.01e+4} & 5.74e+5 & \textbf{3.14e-9} & \textbf{0.0}   \\ \cline{2-7} 
                            & \textbf{IMPALA-V\_TRACE} & MGDT-DQN        & \textbf{3.01e+4} & 3.50e+5 & \textbf{3.78e-4} & \textbf{3.53e-15}  \\ \cline{2-7} 
                            & \textbf{IMPALA-V\_TRACE} & MGDT-MAENT      & \textbf{3.01e+4} & 4.23e+5 & \textbf{1.48e-3} & \textbf{1.66e-8}   \\ \cline{2-7} 
                            & \textbf{IMPALA-V\_TRACE} & MGDT-PPO        & \textbf{3.01e+4} & 5.74e+5 & \textbf{3.48e-9} & \textbf{0.0}   \\ \cline{2-7} 
                            & MGDT-DQN                 & MGDT-MAENT      & 3.50e+5          & 4.23e+5 & 4.37e-1          & 1.87e-1            \\ \cline{2-7} 
                            & \textbf{MGDT-DQN}        & MGDT-PPO        & \textbf{3.50e+5} & 5.74e+5 & \textbf{1.29e-3} & \textbf{4.16e-8}   \\ \cline{2-7} 
                            & \textbf{MGDT-MAENT}      & MGDT-PPO        & \textbf{4.23e+5} & 5.74e+5 & \textbf{4.99e-2} & \textbf{1.77e-2}   \\ \hline
\multirow{15}{*}{2\%}         & \textbf{BASELINE}        & IMPALA-PPO      & \textbf{2.57e+4} & 3.24e+4 & \textbf{1.87e-2} & \textbf{1.10e-2}   \\ \cline{2-7} 
                            & BASELINE                 & IMPALA-V\_TRACE & 2.57e+4          & 3.21e+4 & 1.54e-1          & 7.95e-2            \\ \cline{2-7} 
                            & \textbf{BASELINE}        & MGDT-DQN        & \textbf{2.57e+4} & 3.66e+5 & \textbf{1.09e-5} & \textbf{2.80e-78}  \\ \cline{2-7} 
                            & \textbf{BASELINE}        & MGDT-MAENT      & \textbf{2.57e+4} & 3.92e+5 & \textbf{1.12e-3} & \textbf{1.53e-9}   \\ \cline{2-7} 
                            & \textbf{BASELINE}        & MGDT-PPO        & \textbf{2.57e+4} & 5.99e+5 & \textbf{1.07e-6} & \textbf{2.55e-241} \\ \cline{2-7} 
                            & IMPALA-PPO               & IMPALA-V\_TRACE & 3.24e+4          & 3.21e+4 & 1.00e+0          & 5.19e-1            \\ \cline{2-7} 
                            & \textbf{IMPALA-PPO}      & MGDT-DQN        & \textbf{3.24e+4} & 3.66e+5 & \textbf{8.87e-6} & \textbf{3.90e-74}  \\ \cline{2-7} 
                            & \textbf{IMPALA-PPO}      & MGDT-MAENT      & \textbf{3.24e+4} & 3.92e+5 & \textbf{1.19e-3} & \textbf{3.00e-9}   \\ \cline{2-7} 
                            & \textbf{IMPALA-PPO}      & MGDT-PPO        & \textbf{3.24e+4} & 5.99e+5 & \textbf{7.50e-7} & \textbf{1.39e-231} \\ \cline{2-7} 
                            & \textbf{IMPALA-V\_TRACE} & MGDT-DQN        & \textbf{3.21e+4} & 3.66e+5 & \textbf{4.70e-6} & \textbf{1.04e-71}  \\ \cline{2-7} 
                            & \textbf{IMPALA-V\_TRACE} & MGDT-MAENT      & \textbf{3.21e+4} & 3.92e+5 & \textbf{1.16e-3} & \textbf{3.10e-9}   \\ \cline{2-7} 
                            & \textbf{IMPALA-V\_TRACE} & MGDT-PPO        & \textbf{3.21e+4} & 5.99e+5 & \textbf{3.08e-7} & \textbf{6.72e-223} \\ \cline{2-7} 
                            & MGDT-DQN                 & MGDT-MAENT      & 3.66e+5          & 3.92e+5 & 9.27e-1          & 3.42e-1            \\ \cline{2-7} 
                            & \textbf{MGDT-DQN}        & MGDT-PPO        & \textbf{3.66e+5} & 5.99e+5 & \textbf{3.04e-7} & \textbf{6.79e-21}  \\ \cline{2-7} 
                            & \textbf{MGDT-MAENT}      & MGDT-PPO        & \textbf{3.92e+5} & 5.99e+5 & \textbf{7.23e-3} & \textbf{6.35e-4}   \\ \hline
\end{tabular}
}
\begin{tablenotes}
     \item mean(A) and mean(B) refer to testing time values.
     
 \end{tablenotes}
\end{table}
\subsection{\textbf{RQ3: How do different model-free DRL algorithms affect the performance of pre-trained generalist agents on SE tasks?}} \label{sec:rq3}
Regarding the Blockmaze game, we observe the same average number of detected bugs among  MGDT-DQN, MGDT-MAENT, and MGDT-PPO in terms of bugs detected on the zero-shot fine-tuning data budgets. On the 1\% and 2\% fine-tuning data budgets, MGDT-DQN performs best. Regarding the IMPALA agent, we observe the same performance among IMPALA-V\_TRACE and IMPALA-PPO in terms of bugs detected. Nevertheless, the studied model-free algorithms applied to the MGDT generalist agents perform better than when applied to the IMPALA generalist agent in terms of the number of bugs detected. A possible explanation lies in the differences in architecture between MGDT and IMPALA. Lee et al. \cite{lee2022multi} argued that transformer architecture makes it easier to discover correlations between input and output components. With very little effort for fine-tuning (zero-shot, 1\%, 2\% fine-tuning data-budgets)  MGDT can show better performance in detecting bugs than a classic neural network architecture. In terms of testing time, the MGDT agent performs best on the zero-shot fine-tuning data budgets. On the 1\% and 2\% fine-tuning data budgets, MGDT-DQN has the best testing time performance. In terms of rewards earned, the MGDT generalist agents have the best average performance on the zero-shot fine-tuning data budgets. On the 1\% and 2\% fine-tuning data budgets, MGDT-PPO performs best.\\
\textbf{Statistical analysis:} Statistically, the difference between the model-free algorithms is significant in terms of bugs detected. As shown in Table~\ref{tab:Results of post-hoc tests analysis of testing time performance by wuji(Baseline) and the MGDT generalist agent}, on the 1\% and 2\% fine-tuning data budgets, MGDT-DQN performs significantly better than the other configurations with CLES values equal to one. The IMPALA agent configurations perform poorly with no bug detected after fine-tuning (see Table \ref{tab:bugmakespanblockmaze}, \ref{tab:bugmakespan0fblockmaze} and \ref{tab:bugmakespan100fblockmaze}). Table \ref{tab:Results of  post-hoc tests analysis of testing time performance by each model-free algorithm on the Blockmaze game} shows results of post-hoc test analysis of testing time for different model-free algorithms on the Blockmaze game.
\begin{sidewaystable}
\caption{Results of post-hoc tests analysis of testing time performance by each model-free algorithm on the Blockmaze game (in bold are DRL configurations where p-value is $<$ 0.05 and have greater performance w.r.t the effect size).}
\label{tab:Results of post-hoc tests analysis of testing time performance by each model-free algorithm on the Blockmaze game}
\resizebox{0.8\textwidth}{!}{
\centering
\begin{tabular}{|c|c|c|c|c|c|c|}
\hline
Data budgets     & A                        & B                   & mean(A)          & mean(B)          & pval             & CLES               \\ \hline
\multirow{10}{*}{zero-shot} & IMPALA-PPO               & IMPALA-V\_TRACE     & 7.77e+3          & 7.77e+3          & 1.00e+0          & 5.00e-1            \\ \cline{2-7} 
                            & IMPALA-PPO               & \textbf{MGDT-DQN}   & 7.77e+3          & \textbf{2.15e+3} & \textbf{1.22e-2} & \textbf{9.99e-1}   \\ \cline{2-7} 
                            & IMPALA-PPO               & \textbf{MGDT-MAENT} & 7.77e+3          & \textbf{2.15e+3} & \textbf{1.22e-2} & \textbf{9.99e-1}   \\ \cline{2-7} 
                            & IMPALA-PPO               & \textbf{MGDT-PPO}   & 7.77e+3          & \textbf{2.15e+3} & \textbf{1.22e-2} & \textbf{9.99e-1}   \\ \cline{2-7} 
                            & IMPALA-V\_TRACE          & \textbf{MGDT-DQN}   & 7.77e+3          & \textbf{2.15e+3} & \textbf{1.22e-2} & \textbf{9.99e-1}   \\ \cline{2-7} 
                            & IMPALA-V\_TRACE          & \textbf{MGDT-MAENT} & 7.77e+3          & \textbf{2.15e+3} & \textbf{1.22e-2} & \textbf{9.99e-1}   \\ \cline{2-7} 
                            & IMPALA-V\_TRACE          & \textbf{MGDT-PPO}   & 7.77e+3          & \textbf{2.15e+3} & \textbf{1.22e-2} & \textbf{9.99e-1}   \\ \cline{2-7} 
                            & MGDT-DQN                 & MGDT-MAENT          & 2.15e+3          & 2.15e+3          & 1.00e+0          & 5.00e-1            \\ \cline{2-7} 
                            & MGDT-DQN                 & MGDT-PPO            & 2.15e+3          & 2.15e+3          & 1.00e+0          & 5.00e-1            \\ \cline{2-7} 
                            & MGDT-MAENT               & MGDT-PPO            & 2.15e+3          & 2.15e+3          & 1.00e+0          & 5.00e-1            \\ \hline
\multirow{10}{*}{1\%}         & IMPALA-PPO               & IMPALA-V\_TRACE     & 6.66e+3          & 6.58e+3          & 9.98e-1          & 5.54e-1            \\ \cline{2-7} 
                            & IMPALA-PPO               & \textbf{MGDT-DQN}   & 6.66e+3          & \textbf{2.17e+3} & \textbf{2.33e-4} & \textbf{1.00e+0}   \\ \cline{2-7} 
                            & \textbf{IMPALA-PPO}      & MGDT-MAENT          & 6.66e+3          & 8.11e+3          & 1.20e-2          & 8.04e-3            \\ \cline{2-7} 
                            & \textbf{IMPALA-PPO}      & MGDT-PPO            & 6.66e+3          & 9.13e+3          & 4.48e-3          & 5.53e-3            \\ \cline{2-7} 
                            & IMPALA-V\_TRACE          & \textbf{MGDT-DQN}   & 6.58e+3          & \textbf{2.17e+3} & \textbf{7.52e-7} & \textbf{1.00e+0}   \\ \cline{2-7} 
                            & \textbf{IMPALA-V\_TRACE} & MGDT-MAENT          & \textbf{6.58e+3} & 8.11e+3          & \textbf{4.58e-5} & \textbf{3.03e-7}   \\ \cline{2-7} 
                            & \textbf{IMPALA-V\_TRACE} & MGDT-PPO            & \textbf{6.58e+3} & 9.13e+3          & \textbf{7.73e-3} & \textbf{9.68e-4}   \\ \cline{2-7} 
                            & \textbf{MGDT-DQN}        & MGDT-MAENT          & \textbf{2.17e+3} & 8.11e+3          & \textbf{1.42e-6} & \textbf{8.60e-94}  \\ \cline{2-7} 
                            & \textbf{MGDT-DQN}        & MGDT-PPO            & \textbf{2.17e+3} & 9.13e+3          & \textbf{1.91e-4} & \textbf{3.17e-14}  \\ \cline{2-7} 
                            & MGDT-MAENT               & MGDT-PPO            & 8.11e+3          & 9.13e+3          & 1.87e-1          & 1.12e-1            \\ \hline
\multirow{10}{*}{2\%}         & IMPALA-PPO               & IMPALA-V\_TRACE     & 7.38e+3          & 6.62e+3          & 5.43e-2          & 9.80e-1            \\ \cline{2-7} 
                            & IMPALA-PPO               & \textbf{MGDT-DQN}   & 7.38e+3          & \textbf{2.19e+3} & \textbf{1.17e-4} & \textbf{1.00e+0}   \\ \cline{2-7} 
                            & IMPALA-PPO               & MGDT-MAENT          & 7.38e+3          & 7.91e+3          & 1.82e-1          & 1.02e-1            \\ \cline{2-7} 
                            & \textbf{IMPALA-PPO}      & MGDT-PPO            & \textbf{7.38e+3} & 8.68e+3          & \textbf{7.39e-3} & \textbf{4.00e-4}   \\ \cline{2-7} 
                            & IMPALA-V\_TRACE          & \textbf{MGDT-DQN}   & 6.62e+3          & \textbf{2.19e+3} & \textbf{1.38e-6} & \textbf{1.00e+0}   \\ \cline{2-7} 
                            & \textbf{IMPALA-V\_TRACE} & MGDT-MAENT          & \textbf{6.62e+3} & 7.91e+3          & \textbf{2.09e-4} & \textbf{3.67e-6}   \\ \cline{2-7} 
                            & \textbf{IMPALA-V\_TRACE} & MGDT-PPO            & \textbf{6.62e+3} & 8.68e+3          & \textbf{4.84e-6} & \textbf{2.90e-17}  \\ \cline{2-7} 
                            & \textbf{MGDT-DQN}        & MGDT-MAENT          & \textbf{2.19e+3} & 7.91e+3          & \textbf{1.38e-7} & \textbf{8.53e-109} \\ \cline{2-7} 
                            & \textbf{MGDT-DQN}        & MGDT-PPO            & \textbf{2.19e+3} & 8.68e+3          & \textbf{1.38e-9} & \textbf{9.86e-222} \\ \cline{2-7} 
                            & \textbf{MGDT-MAENT}      & MGDT-PPO            & \textbf{7.91e+3} & 8.68e+3          & \textbf{3.48e-3} & \textbf{4.78e-3}   \\ \hline
\end{tabular}
}
\begin{tablenotes}
     \item mean(A) and mean(B) refer to testing  time values.
     
 \end{tablenotes}
\end{sidewaystable}
The results show that the MGDT agents significantly outperform the IMPALA configurations in terms of testing time on the zero-shot fine-tuning data budgets. On the 1\% and 2\% fine-tuning data budgets, MGDT-DQN performs best with CLES values close to 1 followed by IMPALA-PPO. Table \ref{tab:Results of  post-hoc tests analysis of cumulative reward performance by  each model-free algorithm on the Blockmaze game} shows the results of post-hoc test analysis involving the model-free algorithms in terms of cumulative reward performance.
\begin{sidewaystable}
\caption{Results of post-hoc tests analysis of cumulative reward performance by  each model-free algorithm on the Blockmaze game (in bold are DRL configurations where p-value is $<$ 0.05 and have greater performance w.r.t the effect size).}
\label{tab:Results of post-hoc tests analysis of cumulative reward performance by each model-free algorithm on the Blockmaze game}
\resizebox{0.9\textwidth}{!}{
\centering
\begin{tabular}{|c|c|c|c|c|c|c|}
\hline
Data budgets    & A                 & B                        & mean(A)           & mean(B)           & pval              & CLES             \\ \hline
\multirow{6}{*}{zero-shot} & IMPALA-PPO        & \textbf{MGDT-DQN}        & -4.00e+2          & \textbf{-1.49e+2} & \textbf{0.0}  & \textbf{0.0} \\ \cline{2-7} 
                           & IMPALA-PPO        & \textbf{MGDT-MAENT}      & -4.00e+2          & \textbf{-1.49e+2} & \textbf{0.0}  & \textbf{0.0} \\ \cline{2-7} 
                           & IMPALA-PPO        & \textbf{MGDT-PPO}        & -4.00e+2          & \textbf{-1.49e+2} & \textbf{0.0}  & \textbf{0.0} \\ \cline{2-7} 
                           & IMPALA-V\_TRACE   & \textbf{MGDT-DQN}        & -4.00e+2          & \textbf{-1.49e+2} & \textbf{0.0}  & \textbf{0.0} \\ \cline{2-7} 
                           & IMPALA-V\_TRACE   & \textbf{MGDT-MAENT}      & -4.00e+2          & \textbf{-1.49e+2} & \textbf{0.0}  & \textbf{0.0} \\ \cline{2-7} 
                           & IMPALA-V\_TRACE   & \textbf{MGDT-PPO}        & -4.00e+2          & \textbf{-1.49e+2} & \textbf{0.0}  & \textbf{0.0} \\ \hline
\multirow{10}{*}{1\%}        & IMPALA-PPO        & \textbf{IMPALA-V\_TRACE} & -4.00e+2          & \textbf{-3.96e+2} & \textbf{1.00e+0}  & \textbf{2.00e-1} \\ \cline{2-7} 
                           & IMPALA-PPO        & \textbf{MGDT-DQN}        & -4.00e+2          & \textbf{-1.49e+2} & \textbf{2.06e-8}  & \textbf{0.0} \\ \cline{2-7} 
                           & IMPALA-PPO        & \textbf{MGDT-MAENT}      & -4.00e+2          & \textbf{-1.42e+2} & \textbf{1.27e-8}  & \textbf{0.0} \\ \cline{2-7} 
                           & IMPALA-PPO        & \textbf{MGDT-PPO}        & -4.00e+2          & \textbf{-1.22e+2} & \textbf{3.72e-9}  & \textbf{0.0} \\ \cline{2-7} 
                           & IMPALA-V\_TRACE   & \textbf{MGDT-DQN}        & -3.96e+2          & \textbf{-1.49e+2} & \textbf{2.62e-8}  & \textbf{0.0} \\ \cline{2-7} 
                           & IMPALA-V\_TRACE   & \textbf{MGDT-MAENT}      & -3.96e+2          & \textbf{-1.42e+2} & \textbf{1.60e-8}  & \textbf{0.0} \\ \cline{2-7} 
                           & IMPALA-V\_TRACE   & \textbf{MGDT-PPO}        & -3.96e+2          & \textbf{-1.22e+2} & \textbf{4.65e-9}  & \textbf{0.0} \\ \cline{2-7} 
                           & MGDT-DQN          & MGDT-MAENT               & -1.49e+2          & -1.42e+2          & 9.98e-1           & 2.00e-1          \\ \cline{2-7} 
                           & MGDT-DQN          & MGDT-PPO                 & -1.49e+2          & -1.22e+2          & 8.18e-1           & 0.0          \\ \cline{2-7} 
                           & MGDT-MAENT        & MGDT-PPO                 & -1.42e+2          & -1.22e+2          & 9.35e-1           & 8.00e-1          \\ \hline
\multirow{10}{*}{2\%}        & IMPALA-PPO        & \textbf{IMPALA-V\_TRACE} & -4.00e+2          & \textbf{-3.96e+2} & \textbf{2.59e-2}  & \textbf{1.25e-1} \\ \cline{2-7} 
                           & IMPALA-PPO        & \textbf{MGDT-DQN}        & -4.00e+2          & \textbf{-1.49e+2} & \textbf{7.77e-16} & \textbf{0.0} \\ \cline{2-7} 
                           & IMPALA-PPO        & \textbf{MGDT-MAENT}      & -4.00e+2          & \textbf{-2.97e+2} & \textbf{7.77e-16} & \textbf{0.0} \\ \cline{2-7} 
                           & IMPALA-PPO        & \textbf{MGDT-PPO}        & -4.00e+2          & \textbf{-1.22e+2} & \textbf{7.77e-16} & \textbf{0.0} \\ \cline{2-7} 
                           & IMPALA-V\_TRACE   & \textbf{MGDT-DQN}        & -3.96e+2          & \textbf{-1.49e+2} & \textbf{7.77e-16} & \textbf{0.0} \\ \cline{2-7} 
                           & IMPALA-V\_TRACE   & \textbf{MGDT-MAENT}      & -3.96e+2          & \textbf{-2.97e+2} & \textbf{7.77e-16} & \textbf{0.0} \\ \cline{2-7} 
                           & IMPALA-V\_TRACE   & \textbf{MGDT-PPO}        & -3.96e+2          & \textbf{-1.22e+2} & \textbf{7.77e-16} & \textbf{0.0} \\ \cline{2-7} 
                           & \textbf{MGDT-DQN} & MGDT-MAENT               & \textbf{-1.49e+2} & -2.97e+2          & \textbf{7.77e-16} & \textbf{1.00e+0} \\ \cline{2-7} 
                           & MGDT-DQN          & \textbf{MGDT-PPO}        & -1.49e+2          & \textbf{-1.22e+2} & \textbf{1.58e-13} & \textbf{0.0} \\ \cline{2-7} 
                           & MGDT-MAENT        & \textbf{MGDT-PPO}        & -2.97e+2          & \textbf{-1.22e+2} & \textbf{7.77e-16} & \textbf{0.0} \\ \hline
\end{tabular}
}
\begin{tablenotes}
     \item mean(A) and mean(B) refer to the cumulative reward values.
     
 \end{tablenotes}
\end{sidewaystable}
The results show that MGDT agents perform significantly better when earning rewards.
\begin{tcolorbox}
    \textbf{Finding 5: MGDT-DQN consistently shows good performance across all fine-tuning data budgets compared to the other generalist agents on the Blockmaze game. This also suggests that the MGDT-DQN configuration possesses quick learning and adaptability in limited data scenarios. However, based on our results, the trade-off between training and testing time should be carefully weighed depending on the requirements of the task at hand.}
\end{tcolorbox}

Regarding the $6 \times 6$ instance of the PDR task, on average, all IMPALA agents performs similarly in terms of makespan, testing time and cumulative reward earned after fine-tuning. Similarly, each of the MGDT agents has the same average performance when applied to the $6 \times 6$ instance of the PDR task in terms of makespan, testing time, and cumulative reward earned. MGDT-PPO achieves the longest training time on the 1\% and 2\% fine-tuning data budgets, suggesting that the larger the datasets on top of a large model (197M parameters), the longer it takes to optimize PPO surrogate loss \cite{schulman2017proximal}. \\
\textbf{Statistical analysis:} Statistically, the difference between both generalist agents is not significant in terms of makespan values on the 1\% and 2\% data budgets (see Table \ref{tab:Results of  post-hoc tests analysis of makespan performance by the Baseline and generalist agents66}). On the zero-shot, the MGDT agents perform best with CLES values equal to 1. In terms of training time the IMPALA configurations perform significantly better than the other training methods in the ($6 \times 6$) instance (as reported in Table \ref{tab:Results of  post-hoc tests analysis of training time performance by the Baseline and generalist agents on PDR task66}) on the 1\% and 2\% data budgets. In terms of testing time, IMPALA configurations perform significantly better than MGDT configurations on the ($6 \times 6$) instance. In terms of cumulative reward, the two generalist agents do not show significantly different performance against each other on all fine-tuning data budgets.

On the $30 \times 20$ instance of the PDR task, on average, all IMPALA agents perform similarly in terms of makespan, and cumulative reward earned on all fine-tuning data budgets. Similarly, each of the MGDT agents has the same average performance when applied to the $30 \times 20$ instance of the PDR task in terms of makespan and cumulative reward earned. MGDT-PPO has the longest training time on the 1\% and 2\% fine-tuning data budgets. IMPALA-V\_TRACE and IMPALA-PPO have the longest testing time.\\
\textbf{Statistical analysis:} Statistically, the difference between both generalist agents is not significant in terms of makespan values on the zero-shot and 2\% fine-tuning (see Table \ref{tab:Results of  post-hoc tests analysis of makespan performance by the Baseline and generalist agents3020}). On the 1\% fine-tuning data budgets, in terms of makespan,  MGDT-PPO performs worse than MGDT-MAENT with CLES values close to 0. In terms of training time, the IMPALA configurations perform significantly better than the other training methods in the ($30 \times 20$) instance (Table \ref{tab:Results of  post-hoc tests analysis of training time performance by the Baseline and generalist agents on PDR task3020}) on the 1\% and 2\% data budgets. In terms of testing time, IMPALA configurations perform worse than MGDT on the ($30 \times 20$) instance. In terms of cumulative reward, the observed differences between the two generalist agents on all fine-tuning data budgets are not significant.
\begin{tcolorbox}
\textbf{Finding 6: Both generalist agents easily adapt to the PDR task on both small and medium instances. In this case, pre-training on other tasks (compared to training from scratch) indeed helps with rapid adaptation to the new task (PDR). Further, in the case of PDR tasks, the feature extractor is a GNN, suggesting that feature extraction representation is important for good performance transfer.}
\end{tcolorbox}

Regarding MsPacman game, among the model-free algorithms, IMPALA agents perform best on average in terms of bugs detected and testing time at all fine-tuning data budgets.
In terms of training time, at the 1\% fine-tuning data budget, the MGDT agent configurations performed on average the lowest fine-tuning time. In terms of reward earned, IMPALA-V\_TRACE performs best on average at the zero-shot fine-tuning data budget.\\
\begin{tcolorbox}
\textbf{Finding 7: IMPALA agent configurations achieved the lowest testing time across all fine-tuning data budgets on the MsPacman game. The IMPALA agent takes advantage of the distributed nature of its architecture as well as multiple actors stepping into the game environment increasing the chance of detecting bugs.}
\end{tcolorbox}
\textbf{Statistical analysis:} Tables~\ref{tab:Results of  post-hoc tests analysis of the number of bugs detected by the Baseline and the generalist agents on MsPacman task}, \ref{tab:Results of  post-hoc tests analysis of training time performance by the baseline and the generalist agents on MsPacman task}, \ref{tab:Results of  post-hoc tests analysis of testing time performance by the baseline and the generalist agents on MsPacman task}, and \ref{tab:Results of  post-hoc tests analysis of cumulative reward performance by each online methods on MsPacman task} show the results of the post-hoc test analysis of model-free algorithms regarding their performance on the MsPacman game.
\begin{sidewaystable}
\caption{Results of post-hoc tests analysis of cumulative reward performance by each model-free algorithm on the MsPacman game (in bold are DRL configurations where p-value is $<$ 0.05 and have greater performance w.r.t the effect size).}
\label{tab:Results of  post-hoc tests analysis of cumulative reward performance by each online methods on MsPacman task}
\resizebox{0.9\textwidth}{!}{
\centering
\begin{tabular}{|c|c|c|c|c|c|c|}
\hline
Data budgets     & A                        & B                        & mean(A)          & mean(B)          & pval              & CLES              \\ \hline
\multirow{10}{*}{zero-shot} & IMPALA-PPO               & \textbf{IMPALA-V\_TRACE} & 8.84e+1          & \textbf{1.96e+2} & \textbf{1.79e-2}  & \textbf{3.69e-3}  \\ \cline{2-7} 
                            & IMPALA-PPO               & \textbf{MGDT-DQN}        & 8.84e+1          & \textbf{1.52e+2} & \textbf{8.90e-6}  & \textbf{6.40e-56} \\ \cline{2-7} 
                            & IMPALA-PPO               & \textbf{MGDT-MAENT}      & 8.84e+1          & \textbf{1.93e+2} & \textbf{3.48e-2}  & \textbf{1.34e-2}  \\ \cline{2-7} 
                            & \textbf{IMPALA-PPO}      & MGDT-PPO                 & \textbf{8.84e+1} & 4.50e+1          & \textbf{1.32e-7}  & \textbf{1.00e+0}  \\ \cline{2-7} 
                            & IMPALA-V\_TRACE          & MGDT-DQN                 & 1.96e+2          & 1.52e+2          & 2.60e-1           & 8.65e-1           \\ \cline{2-7} 
                            & IMPALA-V\_TRACE          & MGDT-MAENT               & 1.96e+2          & 1.93e+2          & 1.00e+0           & 5.21e-1           \\ \cline{2-7} 
                            & \textbf{IMPALA-V\_TRACE} & MGDT-PPO                 & \textbf{1.96e+2} & 4.50e+1          & \textbf{5.16e-3}  & \textbf{1.00e+0}  \\ \cline{2-7} 
                            & MGDT-DQN                 & MGDT-MAENT               & 1.52e+2          & 1.93e+2          & 4.17e-1           & 1.91e-1           \\ \cline{2-7} 
                            & \textbf{MGDT-DQN}        & MGDT-PPO                 & \textbf{1.52e+2} & 4.50e+1          & \textbf{2.16e-6}  & \textbf{1.00e+0}  \\ \cline{2-7} 
                            & \textbf{MGDT-MAENT}      & MGDT-PPO                 & \textbf{1.93e+2} & 4.50e+1          & \textbf{1.01e-2}  & \textbf{9.99e-1}  \\ \hline
\multirow{10}{*}{1\%}         & IMPALA-PPO               & IMPALA-V\_TRACE          & 8.84e+1          & 8.79e+1          & 9.27e-1           & 6.38e-1           \\ \cline{2-7} 
                            & IMPALA-PPO               & \textbf{MGDT-DQN}        & 8.84e+1          & \textbf{1.51e+2} & \textbf{5.86e-7}  & \textbf{5.21e-90} \\ \cline{2-7} 
                            & IMPALA-PPO               & \textbf{MGDT-MAENT}      & 8.84e+1          & \textbf{1.69e+2} & \textbf{2.42e-2}  & \textbf{6.98e-3}  \\ \cline{2-7} 
                            & \textbf{IMPALA-PPO}      & MGDT-PPO                 & \textbf{8.84e+1} & 4.50e+1          & \textbf{3.86e-7}  & \textbf{1.00e+0}  \\ \cline{2-7} 
                            & IMPALA-V\_TRACE          & \textbf{MGDT-DQN}        & 8.79e+1          & \textbf{1.51e+2} & \textbf{4.79e-7}  & \textbf{9.54e-91} \\ \cline{2-7} 
                            & IMPALA-V\_TRACE          & \textbf{MGDT-MAENT}      & 8.79e+1          & \textbf{1.69e+2} & \textbf{2.37e-2}  & \textbf{6.67e-3}  \\ \cline{2-7} 
                            & \textbf{IMPALA-V\_TRACE} & MGDT-PPO                 & \textbf{8.79e+1} & 4.50e+1          & \textbf{4.69e-7}  & \textbf{1.00e+0}  \\ \cline{2-7} 
                            & MGDT-DQN                 & MGDT-MAENT               & 1.51e+2          & 1.69e+2          & 7.39e-1           & 2.91e-1           \\ \cline{2-7} 
                            & \textbf{MGDT-DQN}        & MGDT-PPO                 & \textbf{1.51e+2} & 4.50e+1          & \textbf{6.84e-7}  & \textbf{1.00e+0}  \\ \cline{2-7} 
                            & \textbf{MGDT-MAENT}      & MGDT-PPO                 & \textbf{1.69e+2} & 4.50e+1          & \textbf{5.05e-3}  & \textbf{1.00e+0}  \\ \hline
\multirow{10}{*}{2\%}         & \textbf{IMPALA-PPO}      & IMPALA-V\_TRACE          & \textbf{1.82e+2} & 8.90e+1          & \textbf{2.00e-3}  & \textbf{1.00e+0}  \\ \cline{2-7} 
                            & IMPALA-PPO               & MGDT-DQN                 & 1.82e+2          & 1.50e+2          & 9.48e-2           & 9.46e-1           \\ \cline{2-7} 
                            & IMPALA-PPO               & MGDT-MAENT               & 1.82e+2          & 2.00e+2          & 9.32e-1           & 3.67e-1           \\ \cline{2-7} 
                            & \textbf{IMPALA-PPO}      & MGDT-PPO                 & \textbf{1.82e+2} & 4.50e+1          & \textbf{4.77e-4}  & \textbf{1.00e+0}  \\ \cline{2-7} 
                            & IMPALA-V\_TRACE          & \textbf{MGDT-DQN}        & 8.90e+1          & \textbf{1.50e+2} & \textbf{4.14e-10} & \textbf{1.00e-93} \\ \cline{2-7} 
                            & IMPALA-V\_TRACE          & \textbf{MGDT-MAENT}      & 8.90e+1          & \textbf{2.00e+2} & \textbf{3.13e-2}  & \textbf{1.13e-2}  \\ \cline{2-7} 
                            & \textbf{IMPALA-V\_TRACE} & MGDT-PPO                 & \textbf{8.90e+1} & 4.50e+1          & \textbf{5.71e-6}  & \textbf{1.00e+0}  \\ \cline{2-7} 
                            & MGDT-DQN                 & MGDT-MAENT               & 1.50e+2          & 2.00e+2          & 3.08e-1           & 1.53e-1           \\ \cline{2-7} 
                            & \textbf{MGDT-DQN}        & MGDT-PPO                 & \textbf{1.50e+2} & 4.50e+1          & \textbf{1.98e-7}  & \textbf{1.00e+0}  \\ \cline{2-7} 
                            & \textbf{MGDT-MAENT}      & MGDT-PPO                 & \textbf{2.00e+2} & 4.50e+1          & \textbf{9.52e-3}  & \textbf{9.99e-1}  \\ \hline
\end{tabular}
}
\begin{tablenotes}
     \item mean(A) and mean(B) refer to cumulative reward values.
 \end{tablenotes}
\end{sidewaystable}
Statistically, among the model-free DRL algorithms, IMPALA agent configurations perform best with a CLE value of 100\% in terms of detected bugs. MGDT agent configurations achieve the lowest training time compared to the IMPALA agent configurations with CLE values close to 100\% at the 1\% fine-tuning data budget (Table \ref{tab:Results of  post-hoc tests analysis of training time performance by the baseline and the generalist agents on MsPacman task}). IMPALA agent configurations have the lowest testing time in comparison to MGDT agent configurations with CLE values close to 0\% at all fine-tuning data budgets (Table \ref{tab:Results of  post-hoc tests analysis of testing time performance by the baseline and the generalist agents on MsPacman task}). In terms of cumulative reward earned, among the MGDT agent configurations, MGDT-DQN and MGDT-MAENT perform best at all fine-tuning data budgets with CLE values of 100\% (Table \ref{tab:Results of  post-hoc tests analysis of cumulative reward performance by each online methods on MsPacman task}). Among the IMPALA agent configurations, IMPALA-PPO performs worst in terms of cumulative reward earned at zero-shot fine-tuning data budget with CLE value close to 0\%, nevertheless, it performs best at 2\% fine-tuning data budget. The PPO algorithm features a smaller number of random actions during training, thanks to its update rule. As a result, it leverages rewards that have already been discovered. This characteristic means that the PPO algorithm's performance increases or decreases in terms of accumulated reward throughout training, based on past data. Furthermore, this property explains its improved reward performance over IMPALA-V\_TRACE at 2\% fine-tuning data budget, given that IMPALA-PPO achieves positive rewards at zero-shot and  1\% fine-tuning data budgets.
\begin{tcolorbox}
    \textbf{Finding 8: IMPALA-V\_TRACE shows a statistically significant improved performance in comparison to the other model-free DRL algorithms in finding bugs in the MsPacman game. Our results also provide insights regarding the strengths and weaknesses of different model-free DRL algorithms on the MsPacman game. The choice of the model-free DRL algorithm depends on specific metrics such as bug detection, training efficiency or cumulative reward optimization. IMPALA-PPO outperforms IMPALA-V\_TRACE at the 2\% fine-tuning data budget in terms of cumulative reward obtained, given its ability to achieve positive rewards at zero-shot and 1\% fine-tuning data budgets.}
\end{tcolorbox}

\section{Recommendations about the selection of generalist agents}
\label{Recommendations about generalist agents selection}
In this section, we discuss our recommendations regarding selecting generalist agents by researchers and practitioners.

The exploration capability of the MGDT generalist agent has led it to detect more bugs in part of the Blockmaze game environment where IMPALA could not, across all fine-tuning data budgets.

\begin{tcolorbox}
    Recommendation 1: For exploration-intensive tasks, we recommend generalist agents that offer sequence modeling representation of data for rapid adaptation and exploration of the unknown in an environment.
\end{tcolorbox}

MGDT and IMPALA generalist agents have shown good performance on the studied PDR-based scheduling in both evaluated scheduling instances across all fine-tuning data budgets. Compared to the Blockmaze and MsPacman games, the PDR-based scheduling solves the problem of sparse rewards by often rewarding the agent according to its makespan during an episode. 

\begin{tcolorbox}
    Recommendation 2: We recommend using generalist agents on tasks where the distribution of rewards is dense.
\end{tcolorbox}

The scalability property of IMPALA has led it to outperform MGDT on the MsPacman game. Moreover, in the MsPacman game, we evaluated IMPALA bug detection capability during a continuous learning mechanism and it achieved the desired outcome.
\begin{tcolorbox}
    Recommendation 3: We recommend using scalable generalist agents in a continuous learning setting for data efficiency.
\end{tcolorbox}

Our results also indicate that the performance of the generalist agents varies among the model-free DRL algorithms used for fine-tuning. In the MsPacman game, V\_TRACE algorithm outperforms the classic model-free algorithm PPO. Similarly, MAENT outperforms DQN and PPO.
\begin{tcolorbox}
    Recommendation 4: When fine-tuning a generalist on a task, we recommend using a model-free DRL algorithm adapted to the generalist agent's architecture for sample-efficient policy optimization.
\end{tcolorbox}

Our results indicate that the generalist agents significantly outperform the specialist ones on the studied PDR-based scheduling tasks across all fine-tuning data budgets, by performing the lowest makespan time.
\begin{tcolorbox}
    Recommendation 5: We recommend using generalist agents for solving scheduling problems.
\end{tcolorbox}
\section{Related work} \label{Related work}
Recently, researchers have started using pre-trained deep learning models in SE for code-related tasks. Wang et al. \cite{wang2022bridging} exploited natural language pre-trained models in code-related tasks. Specifically, the natural language models are first augmented with semantic-preserving transformation sequences for pre-training purposes, then fine-tuned on two downstream tasks: code clone detection and code search tasks. Kanade et al. \cite{kanade2020learning} proposed cuBERT a contextual embedding of source code derived from BERT \cite{devlin2018bert} model. CuBERT is pre-trained on a massive corpus of python codes and then fine-tuned on downstream code-related tasks. The results show that by pre-training with cuBERT, the fine-tuned models outperformed the baseline model trained from scratch. Similarly, Feng et al. \cite{feng2020codebert} proposed codeBERT, a BERT transformer-based model pre-trained on code from github and fine-tuned for code search and documentation generation tasks. In this work, similar to these previous studies, we adopt BERT as the model's architecture of the MGDT agent. Moreover, we pre-train our generalist agents to leverage prior skills during fine-tuning.

DRL algorithms have been leveraged by researchers to improve the performance of SE tasks \cite{singh2013architecture,bahrpeyma2015adaptive,chen2020enhanced, vuong2018reinforcement}. Zheng et al. \cite{zheng2019wuji}, Tufano et al. \cite{tufano2022using} our baselines studies, trained specialist agents to detect game bugs. Bagherzadeh et al. \cite{bagherzadeh2021reinforcement} leveraged  DRL algorithms in continuous improvement regression testing. They investigated ranking models as DRL problems to find the optimal prioritized test cases. Zhang et al. \cite{zhang2020learning} another one of our baseline studies, trained a specialist agent to solve the JSSP on scheduling instances. These studies are similar to our work as we examine specialist agents' performance on SE tasks. However, we compare specialist agents' performance against pre-trained generalist agents' performance.

Researchers have developed DRL generalist agents, to solve multiple tasks simultaneously. Mendonca et al. \cite{mendonca2021discovering} proposed Latent Explorer Achiever (LEXA), a DRL-based agent capable of learning general skills in an environment where some states are estimated based on the uncertainty of the agent policy. LEXA is evaluated on robotic locomotion and manipulation tasks and outperforms prior studies. Similarly, Kalashnikov et al. \cite{kalashnikov2021mt} proposed a multi-robot learning system where multiple robots can learn a variety of tasks. The authors show that their proposed approach enables the robots to generalize to never-before-seen tasks and achieve high performance. Other prior works include Lee et al. \cite{lee2022multi}, Espeholt et al. \cite{espeholt2018impala}, our baseline studies where authors developed generalist agents trained on the Atari suite environments and capable of achieving performance close to or better than the ones of specialist agents. In this paper, we leverage generalist agents on SE tasks, which none of these studies had explored.
\section{Threats to validity}\label{threats}

\textbf{Conclusion validity.}
The conclusion's limitations concern the degree of accuracy of the statistical findings on the best-performing fine-tuned general agents.
We use Tukey and Welch's ANOVA post-hoc tests as statistical tests. The significance level is set to 0.05 which is standard across the literature as shown by Welch et al. \cite{welch1947generalization}, Games et al. \cite{games1976pairwise}. The non-deterministic nature of DRL algorithms can also threaten the conclusions made in this work. We address this by collecting results from 5 independent runs for all our experiments.

\textbf{Internal validity.} 
 A potential limitation is the number of pre-trained generalist agents used. We have chosen to evaluate only two generalist agents because of the availability of their source code. Moreover, the goal of our study was to show that generalist agents can be leveraged in SE tasks for resource efficiency, thus using any other generalist agents does not invalidate our findings.

\textbf{Construct validity.}
A potential threat to construct validity stems from our evaluation criteria. However, these criteria are standard in the literature. We use these measures( the number of bugs detected, the average cumulative reward, the makespan, the training and testing time) to make a fair comparison between the generalist agents in identical circumstances. We discussed how they should be interpreted in Section \ref{sec:Evaluation metrics}

\textbf{External validity.} 
Since our goal is to investigate whether or not pre-trained generalist agents can be leveraged for SE  tasks, a potential limitation is the choice of the tasks used to evaluate the pre-trained generalist agents. We address this threat by choosing bug detection and scheduling tasks that are two different types of SE tasks, in order to achieve diversity. While in scheduling we aim to minimize the makespan, in the bug detection task we aim to find bugs in a game as early as possible. 

\textbf{Reliability validity.} To allow other researchers to replicate or build on our research, we provide a detailed replication package \cite{replication-package} including the code and obtained results.
\section{Conclusion and Discussions} \label{Conclusion}
In this paper, we investigated the applicability of pre-trained generalist agents to two important SE tasks: the detection of bugs in two games and the minimization of makespan on two instances of the task scheduling problem. We compare the efficiency of pre-trained generalist agents against that of three baseline approaches 
in terms of (i) the detection of bugs, and (ii) the minimization of the makespan of scheduling instances. Our results show that the fine-tuned generalist agents can perform close to or better than the specialist agents on some SE tasks. 
Our results also highlight the difficulties of some model-free DRL algorithms in handling the complexity of generalist agents, which suggests further investigation. We formulate recommendations to help SE practitioners make an informed decision when leveraging generalist agents to develop SE tasks. Our experiments also show that generalist agents achieve good transfer performance on the PDR-based scheduling on both evaluated scheduling instances, which have the higher action space sets. Recall that the generalist agents were pre-trained on the Atari suite environments (Section \ref{sec:impala} and Section \ref{sec:Multi-Game Decision Transformers}) with the full Atari action set consisting of 18 actions. This makes us wonder whether, for good transfer learning performance, the full action set of the generalist agents has to be reached. In the future, we plan to investigate this in more detail. Moreover, we plan to expand our study to investigate larger models and more SE tasks. We also plan to use some SE tasks to pre-train the generalist agents.

\section{Conflict of interest}
The authors declared that they have no conflict of interest. 
\bibliographystyle{spbasic}
\bibliography{references.bib}
\end{document}